\documentclass{aastex61}
\usepackage{amsmath}
\usepackage{bm}
\usepackage{relsize,exscale}

\published{2018 December 6}

\shorttitle{Computation of Light Curves and the R-M Effect}
\shortauthors{Short et al.}

\begin{document}

\title{Accurate Computation of Light Curves and the Rossiter--McLaughlin Effect in Multi-Body Eclipsing Systems}

\correspondingauthor{Gur Windmiller}
\email{gwindmiller@sdsu.edu}

\author{Donald R Short}
\affil{Department of Astronomy, San Diego State University, 5500 Campanile Drive,  San Diego CA 92182, USA}

\author{Jerome A Orosz}
\affil{Department of Astronomy, San Diego State University, 5500 Campanile Drive,  San Diego CA 92182, USA}

\author{Gur Windmiller}
\affil{Department of Astronomy, San Diego State University, 5500 Campanile Drive,  San Diego CA 92182, USA}

\author{William F Welsh}
\affil{Department of Astronomy, San Diego State University, 5500 Campanile Drive,  San Diego CA 92182, USA}

\begin{abstract}
We present here an efficient method for computing the visible flux for each body during a multi-body eclipsing event for all commonly used limb darkening laws. Our approach follows the idea put forth by \citet{Pal2012} to apply Green's Theorem on the limb darkening integral, thus transforming the two-dimensional flux integral over the visible disk into a one-dimensional integral over the visible boundary.
We implement this idea through an iterative process which combines a fast method for describing the visible boundary of each body with a fast numerical integration scheme to compute the integrals. For the two-body case, our method compares well in speed with both that of \citet{Mandel2002} and that of \citet{Gimenez2006a}. The strength of the method is that it works for any number of spherical bodies, with a computational accuracy that is adjustable through the use of a tolerance parameter. Most significantly, the method offers two main advantages over previously used techniques:  (i) it can employ a multitude of limb darkening laws, including all of the commonly used ones;  (ii) it can compute the Rossiter-McLaughlin effect for rigid body rotation with an arbitrary orientation of the rotation axis, using any of these limb darkening laws. In addition, we can compute the Rossiter-McLaughlin effect for stars exhibiting differential rotation, using the quadratic limb darkening law. We provide the mathematical background for the method and explain in detail how to implement the technique with the help of several examples and codes which we make available.
\end{abstract}


\section{Introduction} \label{sec:intro}

The eminent astronomer Henry Norris Russell asserted
``... there are ways of approach to unknown territory which lead 
surprisingly far, and repay their followers richly. There is probably no 
better example of this than eclipses of heavenly bodies''  \citep{Russell1948}. 
Russell's declaration of the ``royal road'' to stellar astrophysics 
and, by extension, nearly all branches of astronomy, has indeed been 
fulfilled by the study of eclipsing binary stars. Russell's 
pioneering work, with Harlow Shapley, and later with Merrill, 
provided the means to travel down this road
(see \citealt{Russell1912}; \citealt{RussellShapely1912}; \citealt{Russell1952}).
But the Russell-Merrill method was a product of its time (decades before
the advent of computers), and although the method was employed into the 
1970s, its limitations were apparent in the 1950s, highlighted most 
notably by Zden\v{e}k Kopal (e.g.\ see \citealt{Kopal1959}, \citealt{Kopal1979}). Kopal's own 
approach was both more mathematically robust and faithful to the data
(e.g.\ using an iterative least-squares approach to matching 
observations). However it would take until the late 1960s and 1970s 
before computational resources allowed the use of light curve synthesis 
and modeling techniques (e.g.\ \citealt{Lucy1968}, and the famous code of \citealt{Wilson1971}). Since then, eclipse modeling codes have grown more 
and more sophisticated, incorporating additional astrophysics and 
data-fitting methods (for a concise review of the development of binary star research, 
see \citealt{Southworth2012} and references therein). \\

The discovery of exoplanets near the turn of the 21st century, and in 
particular transiting exoplanets, created a resurgence of development 
in eclipse-modeling tools.  While assumptions of sphericity greatly 
simplify the problem, the extreme radius ratio of the bodies requires 
very high spatial resolution on the star, becoming very computationally 
expensive for methods that ``tile'' the bodies then numerically 
integrate the surface elements (``pixels''). In addition, the quality of 
transit data, especially {\it Kepler} observations, 
required very high-fidelity modeling. To overcome these challenges, 
analytic methods for computing transits were developed that were 
vastly faster than summing surface tiles. Most notably, the 
method presented in \citet{Mandel2002} - hereafter MA2002 -  
has seen widespread use and been employed in a great many investigations. 
The method presents analytic solutions, one for the four-parameter 
``nonlinear'' limb darkening law of \citet{Claret2000}, and another for the quadratic
limb darkening law \citep{Kopal1950}. Eleven different configurations of  planet-over-star are given and analytic expressions are provided for each case. The nonlinear limb darkening requires evaluations of hypergeometric functions while the quadratic law requires elliptic integrals (which are 
much faster to evaluate). \\

A comlimentary approach to computing transit light curves was presented 
by \citet{Gimenez2006a}. Inspired by Kopal's work, the method is based on the 
cross-correlation of two optical apertures. By casting the problem in 
the language of physical optics, the wealth of mathematics developed for 
that field could be employed. Specifically, the light loss during an 
eclipse is expressed as a Hankel (or Fourier-Bessel) transform
which can be evaluated via the sum of Jacobi polynomials.
The Gim\'{e}nez method is quite general, since the geometric factors are decoupled 
from the light curve factors. In this method the limb darkening is expressed as a power law 
expansion \citep{Kopal1950}. The method is valid for all manner of spherical two-body transits, occultations, and eclipses. The accuracy is set by the number of terms used in the summation of Jacobi polynomials. \citet{Gimenez2006b} also used the same approach to allow the fast computation 
of the Rossiter-McLaughlin (R-M) effect, which is a distortion in the radial velocity curve
observed during a partial eclipse or transit -- \citep{Rossiter1924,McLaughlin1924}.
 \\

Another method was developed by \citet{Kjurkchieva2013} that uses the
fact that while the stellar intensity decreases from the center to the 
limb (according to some limb darkening law), the intensity is uniform 
along any given concentric circle (i.e.\ the intensity is function of 
radius only). The intersection of the planet's boundary and these 
stellar concentric circles define arcs on the star that are 
``behind'' the planet (i.e.\ interior to the planet's disk).
The sum of these arcs is the light of the star blocked by the planet.
This allows the usual stellar surface area integral to be replaced 
with a single radial integral, computed numerically. 
A similar approach was taken by \citet{Kreidberg2015} 
who devised a fast method that made full use of the 
azimuthal symmetry of the stellar intensity, greatly simplifying the 
problem. Any radially symmetric limb darkening law is supported, and 
despite solving the area integral numerically, Kreidberg shows that 
this method is an order of magnitude faster than the MA2002
 analytic solution for the nonlinear limb darkening law. This is due to the poor
convergence properties of the evaluation of the Appell hypergeometric function.
We emphasize this important point -- while a closed-form analytic solution might appear 
to always be preferred, ultimately all such functions are computed numerically, 
and in some cases the cost of evaluation of special functions exceeds 
that of a straightforward numerical integral. \\

With the advent of space-based transit surveys, 
particularly the {\it Kepler} Mission (\citealt{Borucki2016} and 
references therein), light curves with continuous duration and precision 
orders of magnitude better than ground-based observations became 
available. This enabled the detection of exotic eclipses, 
a result of systems with more than two eclipsing bodies. 
Such systems include triply-eclipsing systems, circumbinary planets,
and overlapping transits in systems with several planets.
The {\it Kepler} Eclipsing Binary Star catalog (\citealt{Kirk2016}; 
see also \citealt{Orosz2015}) lists 14 cases of triply eclipsing systems 
with two particularly noteworthy examples:
KOI-126 \citet{Carter2011}, and KIC 2856960, \citet{Marsh2014}.
Circumbinary planets (e.g.\ \citealt{Doyle2011}) and systems 
with multiple transiting planets (e.g.\ KOI-94, \citealt{Hirano2012})
can exhibit complex syzygy\footnote{In this work we use the term syzygy to 
mean an alignment of three or more bodies along the observer's line of sight}
transits. Obviously such eclipses/transits  cannot be modeled by simply 
repeatedly applying the previous algorithms.  
A way to model these complex eclipses/transits was needed. \\

Spurred by necessity and fueled by ingenuity, the solution appeared 
remarkably quickly. \citet{Kipping2011} extended the work of Mandel \& Agol 
to be able to handle the case of a small moon of a transiting exoplanet. 
Soon after, \citet{Pal2012} presented a general method for handling transits 
of an arbitrary number of bodies of any size. While still treating 
the stars as spheres (i.e.\ disks on the plane of the sky), the method differs from 
the Mandel \& Agol and the Gim\'{e}nez/Kopal methods. In essence, 
by using Green's theorem, the integral over the visible surface areas of 
the eclipsed star is replaced with line integrals over the bounding 
arcs created by the circular edges of the transiting bodies. The exact 
analytic expressions for the linear limb darkening law, using incomplete 
elliptic integrals, is given by P\'{a}l. The accompanying code provided can 
handle linear and quadratic limb darkening laws for an arbitrary number of bodies.
More recently, \citet{Luger2017} significantly extended the work 
of \citet{Kreidberg2015} for any number of overlapping spheres 
(e.g.\ planet-planet eclipses). The semi-analytical method computes the 
eclipsed light as the sum of one-dimensional integrals of intensity-weighted elliptical 
segments on the occulted object. Notably, the method also allows for 
tabulated intensities in place of a parameterized limb-darkening law, 
irradiation of the planet by the host star (i.e.\ the planet's 
``phase curve''), and a planet ``hotspot'' offset from the sub-stellar 
point to simulate advection of the incident stellar flux by winds --
provided these effects are all radially symmetric. \\

In this paper, we present a modification and extension of the method 
pioneered by \citet{Pal2012} for simultaneous eclipses involving spherical bodies.
The method employs Green's theorem, but does not rely
on analytic closed-form expressions for the integrals, which can be 
difficult to derive. Rather, the integrals are evaluated numerically using highly 
accurate and efficient Gaussian quadrature. This provides flexibility 
that enables the method to easily handle any number of bodies of any 
size and all simple radial limb darkening prescriptions. An example of that
flexibility is the natural ease with which the R-M effect can be 
computed for multiple overlapping eclipses, including differential 
rotation of the stars. \\

In \S 2 we give a brief description of the general mathematical approach, 
and in \S 3 we describe in detail the method of construction of the arcs that define
the visible boundaries of the eclipsed star.
The ``1-forms'' required for the eclipse and R-M effect integrals 
are developed and tabulated in \S 4, where differential rotation is also discussed.
In \S 5 we compare our method with that of MA2002 and 
\citet{Gimenez2006a} for the simple two-body case, and with that of \citet{Pal2012}. 
In \S6 we present a worked example to illustrate the method, using KOI-126.
Several appendices are included in which we  provide detail on the derivation of 
the exterior derivatives, as well as on the R-M effect with differential rotation. Working
codes (in {\sc Matlab} and {\sc FORTRAN}) that can be used to compute 
the light curves and the R-M effect are made publicly available\footnote{The full {\sc Matlab} code can be downloaded from https://doi.org/10.5281/zenodo.1438555, and the {\sc FORTRAN} code can be downloaded from https://doi.org/10.5281/zenodo.1432722}. \\

\section{Method Overview}\label{sec:method_overview}

In the process of computing observables for eclipsing binary star systems one often encounters integrals of the form
\begin{equation}\label{eq:gensurf}
\iint\displaylimits_{S_{\rm{vis}}} f(x,y,z)\cos(\gamma)dS 
\end{equation}
where $f=f(x,y,z)$ is a quantity such as intensity or rotational radial velocity, $\cos(\gamma)$ is the foreshortening ($\gamma$ is the angle between the surface normal and the direction to the observer), $dS$ is the surface area element, and the integral is evaluated over the visible surface $S$. Throughout this work, $(x,y,z)$ is a right hand co-ordinate system with $x$ and $y$ defining the plane of the sky (POS) and $z$ points towards the observer. \\

The goal of this paper is to present a fast, robust method for evaluating such integrals. If we project opaque spheres onto the POS, we have 
\begin{equation}\label{eq:gensurfproj}
\iint\displaylimits_{S_{\rm{vis}}} f(x,y,z)\cos(\gamma)dS = \iint\displaylimits_{\rm{D} _{\rm vis}} F(x,y) dA
\end{equation}
where the surface area element is related to the disk area element by $dS = \sec(\gamma)dA$, and the integral is evaluated over the visible part of the disk, $D_{\rm vis}$. \\

In the language of the exterior calculus, the integrand $F(x,y)dA$ is a closed 2-form on the unit disk (scaled by the disk radius).
By the Poincar\'{e} Lemma, there exists a 1-form $P(x,y)dx+Q(x,y)dy$, the exterior derivative of which is $F$:
\begin{equation}\label{DPQdefine}
d\wedge\big[P,Q\big] \stackrel{\text{def}}{=} \frac{\partial Q}{\partial x} - \frac{\partial P}{\partial y} = F(x,y)
\end{equation}
where d is the exterior derivative operator. Applying Green's theorem we obtain:
\begin{equation}\label{eq:greenapply}
\iint\displaylimits_{S_{\rm{vis}}}f(x,y,z)\cos(\gamma)dS = 
\iint\displaylimits_{D_{\rm{vis}}}F(x,y)dA =
\oint\displaylimits_{\partial D_{\rm{vis}}}\big[P,Q\big] \boldsymbol{\cdot} \big[x', y'\big] d\varphi
\end{equation}
where $x=x(\varphi)$ and $y=y(\varphi)$ is a right-hand oriented parametric curve describing the boundary of the visible region, $x'$ and $y'$ are the derivatives of $x$ and $y$ with respect to $\varphi$, and the integral is evaluated over the boundary of the visible disk ($\partial D_{\rm vis}$). Thus, the evaluation of the visible surface integral reduces to finding the description of the oriented parametric curve and to expressing the 1-form $Pdx+Qdy$ in terms of simple functions. \\

\section{Development of the Bounding Curve: Construction from Arcs}\label{sec:boundingcurve}

During a multibody eclipse involving $\cal{N}$ bodies, we seek to compute the boundary of the visible portion of each body at a given time, $t_0$.  It is assumed that the bodies are spherical and given in the front-to-back (i.e.\ nearest to furthest) order from the observer. It is also assumed that the light travel time effect (LTTE) has been accounted for in the positioning of the bodies' centers at time $t_0$ (i.e.\ in the local frame, not the observer's frame, in which we are observing the LTTE corrected two-dimensional view of the system on the POS). Because the projection of these spherical bodies on the POS are circles, our bounding curves will be composed of segments of circular arcs. \\

Consider a body, denoted by $N$, with radius $R_N$ and central POS co-ordinates $(\rm \xi_{N,x}, \xi_{N,y})$. The visible portion of Body $N$ is bounded by arcs that are either on the circumference of Body $N$ itself or on the circumference of one or more additional bodies obstructing it. Consider one of these bodies, Body $M$, with radius $R_M$ and POS co-ordinates $(\rm \xi_{M,x}, \xi_{M,y})$. Given these POS center positions and radii, the parametric equations of a bounding arc within the disk of body $N$ may be written in terms of co-ordinates with origin at the center of Body $N$, normalized by body $N$'s radius:
\begin{align}
\begin{aligned} \label{eq:xiyi_coord}
x &= \big[R_M\cos(\varphi)+\xi_{M,x}-\xi_{N,x}\big]/R_N \\ 
y &= \big[R_M\sin(\varphi)+\xi_{M,y}-\xi_{N,y}\big]/R_N\\
\end{aligned}
\end{align}
The starting and ending angles are measured from the positive x-axis in the POS co-ordinates of body $M$. In all cases the ending angle will be greater than the starting angle to preserve the right handed orientation of the arc. To do that, the domain of an arc is restricted\footnote{Consider the same arc written 3 different ways: $(-10^{\circ},20^{\circ})$,$(350^{\circ},20^{\circ})$, and $(350^{\circ},380^{\circ})$. 
Only in the last pair are both angles positive with the ending angle greater than the starting angle. Thus, we consider arcs in the range of $[0,4\pi)$.} to $[0,4\pi)$. 
To denote these various arcs that are described by Equation (\ref{eq:xiyi_coord}) we use the following notation:
\begin{align}\label{eq:Arc_NM}
Arc_{NM}=\bigg[\text{Back Body $N$, Arc Centered on Body $M$, Starting Angle, Total Angle, Ending Angle}\bigg] 
\end{align}
$Arc_{NM}$ can be thought of as ``an arc on the back body ($N$) centered on a front body ($M$) with the given starting and ending angles''. Note that $N \geq M$, and the Total Angle is used in some of the integration formulae and is otherwise added for computational convenience and debugging. \\

In general, the visible portion of Body $N$ is formed in part by the ``shadows'' cast by the bodies in front of it. Therefore, constructing the path that bounds this visible potion requires the use of the boundary of the shadowing bodies. We call the former boundary the ``visible path'' and the latter boundary the ``shadow path''.  We initialize the process with a front body which is entirely visible, hence the visible path is the boundary of the body itself. Likewise, since there are no other bodies, the shadow path is the boundary of the body itself. We use that shadow path to compute the visible path of the second body, which is behind the first one. The two bodies together form a shadow path, which is then used to compute the visible path on the next body. The process continues until the visible path of the body which is furthest from the observer is computed. In Appendix A, we provide a thorough discussion and worked examples on how to go about the process of arc construction. \\

\section{Limb Darkening and Rossiter-McLaughlin effect 1-forms}\label{sec:LimbRM1forms}

After developing the right hand oriented parametric curve describing the boundary of the visible region in \S\ref{sec:boundingcurve}, we now turn to expressing the 1-form $Pdx + Qdy$ in terms of simple functions in the cases of intensity (limb darkening) and radial velocity (the R-M effect).  This will complete the development of the line integral over the visible boundary (Equation \ref{eq:greenapply}). Note that in cases where we have data in different bandpasses, we do not need to redo the generation of the bounding arcs, but rather use those arcs with the appropriate limb darkening to compute the intensity in each bandpass. \\

\subsection{Limb Darkening}\label{sec:limbdark}

First consider these various parametrized limb darkening laws: 
\begin{align}
&I(\mu)/I_0 = 1 - c_1(1-\mu) &  &\text{Linear} \label{eq:linear_law} \\
&I(\mu)/I_0 = 1 - c_1(1-\mu)-c_2(1-\mu)^2&  &\text{Quadratic} \label{eq:quad_law} \\
&I(\mu)/I_0 = 1 - a(1-\mu)-b(1-\sqrt\mu)& &\text{Square Root} \label{eq:sqrt_law} \\
&I(\mu)/I_0 = 1 - \sum_{j=1}^4 c_j(1-\mu^{\frac{j}{2}})& & \text{Claret (4 parameter nonlinear)} &\label{eq:Claret4param} \\
&I(\mu)/I_0 = 1 - l_1(1-\mu)-l_2\mu\log(\mu)& &\text{Logarithmic} & \label{eq:log_law}
\end{align}
Note the following: (i)  $I_0$ is the intensity at the center of the disk. (ii) $r^2=x^2+y^2 \text{~~and~~} \mu=\sqrt{1-r^2}$. (iii) When modeling the data, \cite{Kipping2013} recommends re-parametrization of the limb darkening laws that allow uniform sampling of the physical parameter space. For example, the re-parametrized Quadratic Law has coefficients $q_1=(c_1+c_2)^2$ and $q_2=0.5c_1(c_1+c_2)^{-1}$. (iv) Kipping's log term differs from the one given here, in Equation~\ref{eq:log_law} \citep[see][]{Espinoza2016} (v) The Exponential Law is not physical, so it is omitted \citep[see][]{Espinoza2016}. (vi) Kipping (2016) proposed a 3-parameter limb darkening law that is a subset of the Claret 4-parameter law (the $c_1$ term is set to zero). (vii) Finally, it is understood that the limb darkening coefficients are wavelength or bandpass dependent. (viii) Equations \ref{eq:RMgfunc_summary} explicitly give the 1-forms $[P, Q]$ for any limb darkening given by: $I(\mu)/I_0=f(r^2)=f(1-\mu^2)$ (where $f$ is a continuous function on the disk $D$), and for the R-M effect based on that limb darkening. The specific form of the integral $G$ (Equation \ref{eq:RMgfunc_integral}) will determine if $[P,Q]$ can be expressed in closed form, by a special function, or will require numerical evaluation. \\

As can be seen in Equations \ref{eq:linear_law}-\ref{eq:log_law}, all of the laws are a linear combination of a few common terms, namely, $\bm{1},~\mu^{\frac{1}{2}},~\mu,~\mu^{\frac{3}{2}},\mu^2$ and $\mu\log(\mu)$, where $\bm{1}$ is the identically one function. If we form a vector $\bm{C}$ of the coefficients and a vector $\bm{\Psi}$ of the simple functional expressions, each limb darkening law may be written as a dot product of $\bm{C}$ and $\bm{\Psi}$.
\begin{equation} \label{eq:CdotPsi}
I(\mu)/I_0 = \bm{C} \boldsymbol{\cdot} \bm{\Psi}
\end{equation}
Our goal is to separate the integration over the boundary of the visible region from the application of the specific limb darkening coefficients. This will allow the computation of intensity for varying bands without redoing the boundary integration. For example, expanding the terms of the Quadratic law gives the equation of the linear combination as:
\begin{align} \nonumber
\begin{aligned}
& I(\mu)/I_0 =\big[1-(c_1+c_2)\big] \bm{1}+(c_1+2c_2)\mu-c_2\mu^2 \\ 
\Longrightarrow~~~
& \bm{C}=\big[\big(1-(c_1+c_2)\big),(c_1+2c_2),-c_2\big] \text{~~and~~} \bm{\Psi}=\big[\bm{1},\mu,\mu^2\big] 
\end{aligned}
\end{align}
The dot product forms of Equations  \ref{eq:linear_law}-\ref{eq:log_law} are then given by:
\begin{align}
&\bm{C}=\big[(1-c_1),c_1\big] \text{~~and~~} \bm{\Psi}=\big[\bm{1},\mu\big] &  &\text{Linear} \label{eq:linear_law_dot} \\
&\bm{C}=\big[\big(1-(c_1+c_2)\big),(c_1+2c_2),-c_2\big] \text{~~and~~} \bm{\Psi}=\big[\bm{1},\mu,\mu^2\big] &  &\text{Quadratic} \label{eq:quad_law_dot} \\
&\bm{C}=\big[\big(1-(a+b)\big),b,a\big] \text{~~and~~} \bm{\Psi}=[\bm{1},\mu^{\frac{1}{2}},\mu\big]& &\text{Square Root} \label{eq:sqrt_law_dot} \\
&\bm{C}=\big[\big(1-(c_1+c_2+c_3+c_4)\big),c_1,c_2,c_3,c_4\big] \text{~~and~~} 
\bm{\Psi}=\big[\bm{1},\mu^{\frac{1}{2}},\mu,\mu^{\frac{3}{2}},\mu^2\big] & & \text{Claret (4 parameter nonlinear)} &\label{eq:Claret4param_dot} \\
&\bm{C}=\big[(1-l_1),l_1,-l_2\big]  \text{~~and~~} \bm{\Psi}=\big[\bm{1},\mu,\mu\log(\mu)\big] & &\text{Logarithmic} & \label{eq:log_law_dot}
\end{align}
Table \ref{tab:LimbRM_1Forms} provides the 1-forms $\big[P,Q\big]$ which are the exterior anti-derivatives of the $\Psi$ vector component functions. See Appendix \ref{sec:appendix_exderiv} for the derivations. \\

Using this form of the limb darkening laws, we now update Equation (\ref{eq:greenapply}):
\begin{equation}\label{eq:greenapply_updated}
I_{\rm vis}=\iint\displaylimits_{D_{\rm vis}}\big(\bm{C} \boldsymbol{\cdot} \bm{\Psi}\big)dA
\end{equation}
The fractional flux can be defined as the quotient
\begin{equation}\label{eq:f_vis_def}
{\cal F}_{\rm vis}=I_{\rm vis}/I_{\rm total}
\end{equation}
The computation of the fractional flux requires two evaluations of the $I_{\rm vis}$ integral, so we seek a computational form of this integral. By linearity of integration we obtain:
\begin{align} \label{eq:intensity_integral}
\begin{aligned}
I_{\rm vis}=\bm{C} \boldsymbol{\cdot} \bigg[\iint\displaylimits_{D_{\rm vis}}\Psi_1dA,\,\,\boldsymbol{\ldots}\,\,,\iint\displaylimits_{D_{\rm vis}}\Psi_ndA\bigg]
\end{aligned}
\end{align}
and by Green's Theorem, we obtain the following form of $I_{\rm vis}$:%
\begin{align} \label{eq:intensity_integral2}
\begin{aligned}
I_{\rm vis}=\bm{C} \boldsymbol{\cdot} \bigg[
\oint\displaylimits_{\partial D_{\rm vis}}\big[P_1,Q_1\big] \boldsymbol{\cdot} \big[x',y']d\varphi,\,\,\boldsymbol{\ldots}\,\,,
\oint\displaylimits_{\partial D_{\rm vis}}\big[P_n,Q_n\big] \boldsymbol{\cdot} \big[x',y']d\varphi\bigg]
\end{aligned}
\end{align}
From $\S\ref{sec:boundingcurve}$, the boundary curve of the visible portion of star $N$ from the observer is a set of circular arcs given by the $VisiblePath(BodyN)$ stack, where each arc is represented in the form of Equation (\ref{eq:Arc_NM}):
\begin{equation}\label{eq:arc_ij}
Arc_{NM} = \big[N,M,\varphi_{0},\Delta\varphi,\varphi_{1}\big]
\end{equation}
In terms of co-ordinates with origin the center of body $N$ scaled by body $N$'s radius $R_N$, we have the following parametric description of this arc within the disk of body $N$:
\begin{align}
\begin{aligned} 
x &= \big[R_M\cos(\varphi)+\xi_{M,x}-\xi_{N,x}\big]/R_N \nonumber \\ 
y &= \big[R_M\sin(\varphi)+\xi_{M,y}-\xi_{N,y}\big]/R_N\\
\end{aligned}
\end{align}
where $R_N$ is the radius of Body N, $R_M$ is the radius of Body $M$ (with its center the point in the POS having co-ordinates $\xi$), and $\varphi_{0} \leq \varphi \leq \varphi_{1}$, with derivatives,
\begin{align}\label{eq:xiyi_deriv}
\begin{aligned}
x'&= \frac{-R_M}{R_N} \sin(\varphi)~~~~~~~
y' &= \frac{R_M}{R_N}\cos(\varphi)
\end{aligned}
\end{align}
In addition, $x$ and $y$ simplify when $M=N$:
\begin{equation}\label{eq:xiyi_simplify}
x = \cos\varphi~, ~~ y=\sin\varphi
\end{equation}
Thus,
\begin{equation}\label{eq:limbdarkintegral_new}
\oint\displaylimits_{\partial D_{\rm{vis}}}\big[P_j,Q_j\big] \boldsymbol{\cdot} \big[x', y'\big] d\varphi  = \sum_{i=1}^{\rm{\#~of~Arcs}} (\pm1) 
\int\displaylimits_{\varphi_{j_0}}^{\varphi_{j_1}} \big[P_j,Q_j\big] \boldsymbol{\cdot} \big[x'_i, y'_i\big] d\varphi
\end{equation}
where $\pm1$ is the orientation. The visible region is outside of all bodies $M < N$, thus the orientation is $-1$. When $M=N$, the visible region lies in body $N$, thus the orientation is $+1$. In the case $N=M$, implying $r=1$ and $\mu=0$, the 1-forms $\big[P_j,Q_j\big]$ greatly simplify, with the integrand $\big[P_j,Q_j\big] \cdotp \big[x', y'\big]$ now easily integrated ($P$ and $Q$ are given in Table \ref{tab:LimbRM_1Forms}, are derived in Appendix B and given in closed form in Appendix \ref{sec:appendix_closedform}). \\

\subsection{The Rossiter McLaughlin Effect}\label{sec:RMeffect}

Now consider the R-M effect, which is a perterbation of the eclipsed star's apparent radial velocity due to obscuring a portion of the star's rotationally radial velocity field during an eclipse event. \citet{Gimenez2006b} developed a computational method based on \citet{Kopal1979}. He begins with the following equation (Equation 1, \citealt{Gimenez2006b}):
\begin{equation}\label{eq:deltaRV}
\delta RV = \frac{ \mathlarger{\iint}\displaylimits_{S_{\rm{vis}}}vI\cos(\gamma)dS }{\mathlarger{\iint}\displaylimits_{S_{\rm{vis}}}I\cos(\gamma)~dS}                   
 \end{equation}
where $\delta RV$ is the radial velocity RV perturbation, $v$ is the rotationally induced radial velocity of the star, $I=I(\mu)$ is any limb darkening law in its dot product form (Equations~\ref{eq:linear_law_dot}-\ref{eq:log_law_dot}), and $\gamma$ and $dS$ are as before.
Then, following Equation (\ref{eq:greenapply}) for both the numerator (rotational radial velocity) and the denominator (intensity), we get
\begin{equation}\label{eq:deltaRV2}
\delta RV = \frac{\mathlarger{\iint}\displaylimits_{D_{\rm{vis}}}v(x,y)IdA }{\mathlarger{\iint}\displaylimits_{D_{\rm{vis}}}IdA} =
\frac{\mathlarger{\oint}\displaylimits_{\partial D_{\rm{vis}}}\big[P_{\rm RM},Q_{\rm RM}\big] \boldsymbol{\cdot} \big[x', y'\big] d\varphi}{\mathlarger{\oint}\displaylimits_{\partial D_{\rm{vis}}}\big[P_{\rm FF},Q_{\rm FF}\big] \boldsymbol{\cdot} \big[x', y'\big] d\varphi}                
 \end{equation}
where we use the abbreviation $FF$ to denote flux fraction. Note that the denominator is the visible flux. \\

Define the rotation axis of the star in the dynamic co-ordinate system  $(x,y,z)$, with the observer on the positive $z$ axis, as defined by Equation (\ref{eq:gensurf}). Note that this differs from the system defined first by \citet{Hosokawa1953} and then used again by \citet{Gimenez2006a}.  Since their emphasis was on simple binary systems and planetary transits, their $y$ axis was just the projection of the  orbital pole on the POS.  In this paper, we do not assume a simple 2 body system but rather a multi-body system.   
\begin{align}\label{eq:ThetaPhi_define}
\begin{aligned}
\Theta_{\rm{rot}} &\stackrel{\text{def}}{=} \text{angle in the $(x,y)$ plane from the y-axis} \\
\Phi_{\rm{rot}} &\stackrel{\text{def}}{=} \text{angle from the $z$ axis (co-latitude)}
\end{aligned}
\end{align}
If the axis of rotation of the star is the $z$-axis, then the righthand surface velocity field on the unit sphere is given by
\begin{equation}\label{eq:vfield}
\vec{v}(x,y,z)=\omega (-y,x,0)
 \end{equation}
where $\omega$ is the angular velocity in radians per day. Note that $\|(-y,x,0)\|$ is the distance to the axis of rotation (the $z$ axis). \\

Using Rotation Transformations (orthonormal matrices with determinant equals to $1$) move the $z$ axis to the rotation axis described by $\big(\Theta_{\rm{rot}},\Phi_{\rm{rot}}\big)$. Since the transformations are length and orientation preserving, the transformation
of the velocity field is also preserved as the velocity field generated by a righthand rotation about the axis $\big(\Theta_{\rm{rot}},\Phi_{\rm{rot}}\big)$. The radial velocity function is simply the $z$-component of this velocity field. Namely,
\begin{equation}\label{eq:v_of_xy}
v(x,y)=\omega \big[\sin(\Phi_{\rm rot})\cos(\Theta_{\rm rot})x+\sin(\Phi_{\rm rot})\sin(\Theta_{\rm rot})y\big]
 \end{equation}
for $(x,y)$ any point on the POS disk normalized to the unit disk. Note that the observed rotational velocity is $v_{\rm rot}\sin i=R_\ast \omega \sin(\Phi_{\rm{rot}})$, with $R_*$ the stellar radius. \\

Define
\begin{align}\label{eq:defAB}
\begin{aligned}
&A \stackrel{\text{def}}{=}\sin(\Phi_{\rm rot})\cos(\Theta_{\rm rot})\\
&B \stackrel{\text{def}}{=}\sin(\Phi_{\rm rot})\sin(\Theta_{\rm rot})\\
\end{aligned}
\end{align}
Thus
\begin{align}\label{eq:omega_AxBy}
\begin{aligned}
&v(x,y)=\omega \big(Ax+By\big)
\end{aligned}
\end{align}
and the R-M effect integrand is simply proportional to the product of a limb darkening law in dot product form (Equation \ref{eq:CdotPsi}) with $(Ax + By)$. Again, this forms a linear combination which may be expressed in the dot product form.
\begin{align}
\begin{aligned}
vI(\mu)=(\bm{C}_{\rm RM}\boldsymbol{\cdot} \bm{\Psi}_{\rm RM}\big)
\text{~~where~~} 
\bm{C}_{\rm RM}=\omega\big[A\bm{C},B\bm{C}\big] \text{~~~and~~~} \bm{\Psi}_{RM}=\big[x\Psi,y\Psi\big].
\end{aligned}
\end{align}
The lengths of $\bm{C}_{\rm RM}$ and $\bm{\Psi}_{\rm RM}$ are twice the lengths of the corresponding terms used for the flux fraction computation. The exterior anti-derivative 1-forms for the expressions in $\bm{\Psi}_{\rm RM}$ are derived in Appendix \ref{sec:appendix_exderiv} and can be found in the second section of Table \ref{tab:LimbRM_1Forms}. The R-M effect numerator is, therefore, evaluated in exactly the same manner as the limb darkening and the denominator is simply the visible intensity. It is important to note that the domain of Equation (\ref{eq:omega_AxBy}) is the unit disk, and the units are those of $\omega$, i.e.\ ${\rm radians~sec^{-1}}$.  Multiplying  Equation (\ref{eq:omega_AxBy}) by the stellar radius, $R_\ast$, in meters, rescales the problem:
\begin{equation}\label{eq:rescale_vxy}
R_\ast v(x, y) = R_\ast \omega(Ax + By)=\omega( AxR_\ast + ByR_\ast) 
\end{equation}
in units of ${\rm m~s^{-1}}$.  Note also that $(xR_\ast, yR_\ast)$ are the POS co-ordinates in meters, and 
$\bm{C}_{RM}$ in ${\rm m~s^{-1}}$  is $R_\ast \bm{C}_{RM}$. \\

If the limb darkening law is a continuous function of $r^2$ on the unit disk, then Equations (\ref{eq:RMgfunc_integral}) and (\ref{eq:RMgfunc_summary}) in appendix \ref{sec:appendix_exderiv}  provide quadrature formulae for the R-M effect 1-forms.  We can generalize the computation of the R-M effect to account for differential rotation, in the case of linear and quadratic law limb darkening. As one might expect, the derivation of the 1-forms for differential rotation is much more involved than that of the 1-forms of a rigid-body rotation. Depending on what limb darkening one assumes (Linear vs Quad law), there can be up to 42 terms, so some attention to detail is required (see Appendix \ref{sec:appendix_rm}). It may be possible to extend the analysis of differential rotation to other limb darkening laws, although that's beyond the scope of this work.\\

\begin{figure}[ht!]\
\plotone{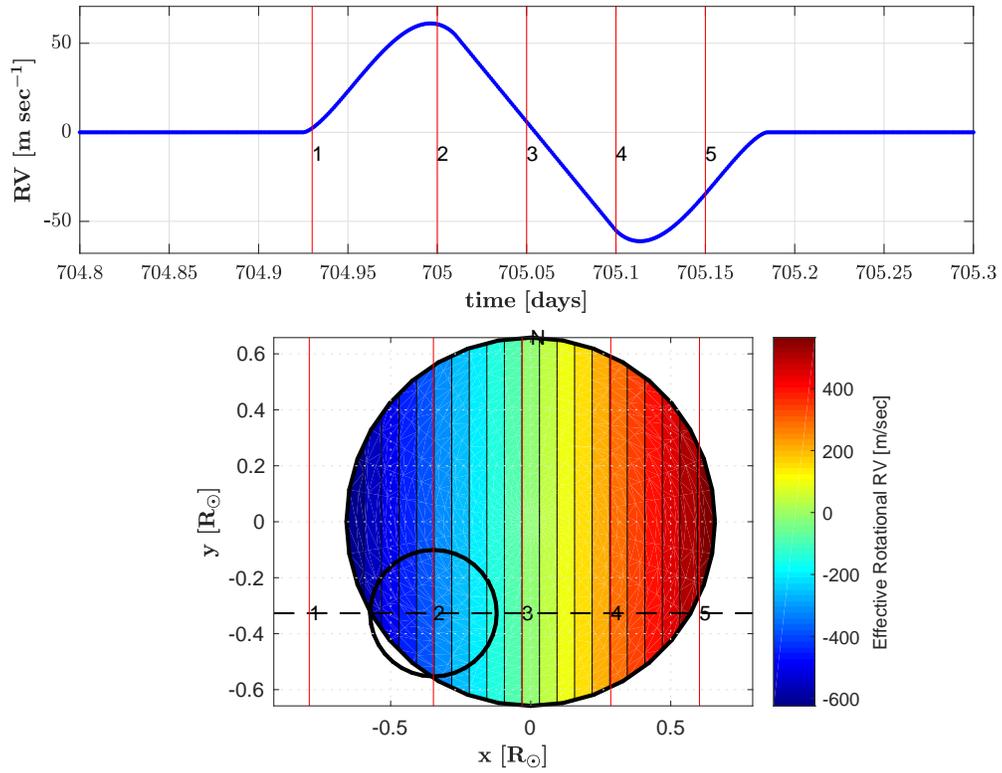}
\caption{Bottom: A smaller body, denoted by the bold circle, moves across a larger star along the path given by the thick dashed line. The effective rotational radial velocity, ${\cal V}$, of the larger star (rotational period $P_1=35.1 {\rm ~days}$, $R_1=0.658~R_\odot$) is shown using the colors, with blue denoting large negative numbers and red denoting large positive numbers. The angles, defined in Equation (\ref{eq:ThetaPhi_define}), are $\Phi_{\rm rot}= 90^\circ$, $\Theta_{\rm rot }=0^\circ$. The ``North pole'' of the star is marked with the letter ``N''. There is no limb darkening  and no differential rotation. Top: The R-M signal in ${\rm m~s^{-1}}$  is shown as a function of time. The numbers $1-5$ in the bottom panel are positions of the center of the smaller body, and correspond to those times at the top panel.}
\label{fig:Mock16_0_90_0_noLD}
\end{figure}

\begin{figure}[ht!]\
\plotone{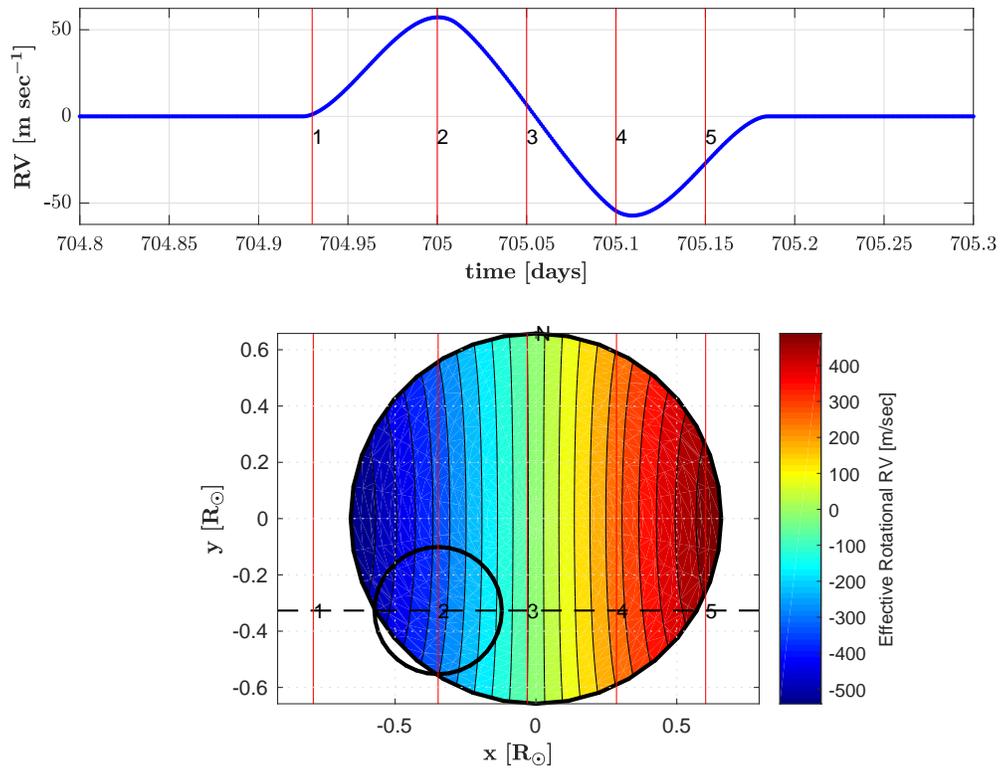}
\caption{Similar to Figure \ref{fig:Mock16_0_90_0_noLD}, with no differential rotation, but with quadratic limb darkening ($c_1=0.77113,~~ c_2= -0.050753$).}
\label{fig:Mock16_0_90_0_LD}
\end{figure}

\begin{figure}[ht!]\
\plotone{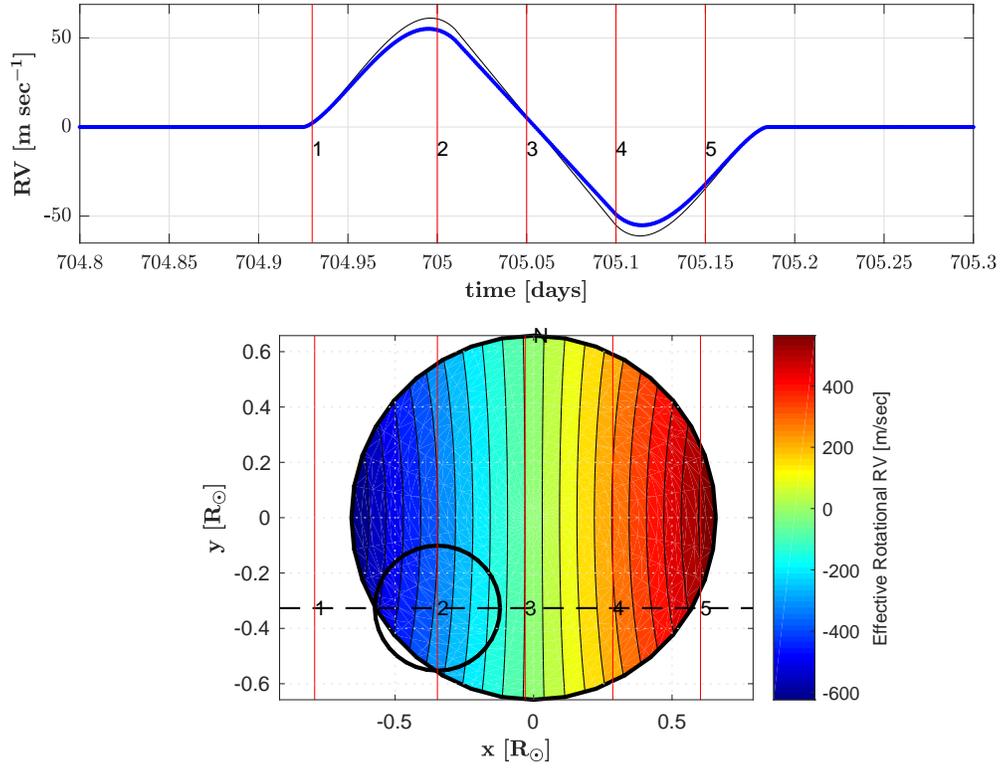}
\caption{Similar to Figure \ref{fig:Mock16_0_90_0_noLD}, with no limb darkening, but with differential rotation ($\epsilon=0.4$, see Equation (\ref{eq:Fover_unitdisk}) for its definition). For comparison, the light black line in the top panel shows the R-M signal from Figure \ref{fig:Mock16_0_90_0_noLD}, a case with neither differential rotation nor limb darkening.}
\label{fig:Mock16_0_90_0p4_noLD}
\end{figure}

\begin{figure}[ht!]\
\plotone{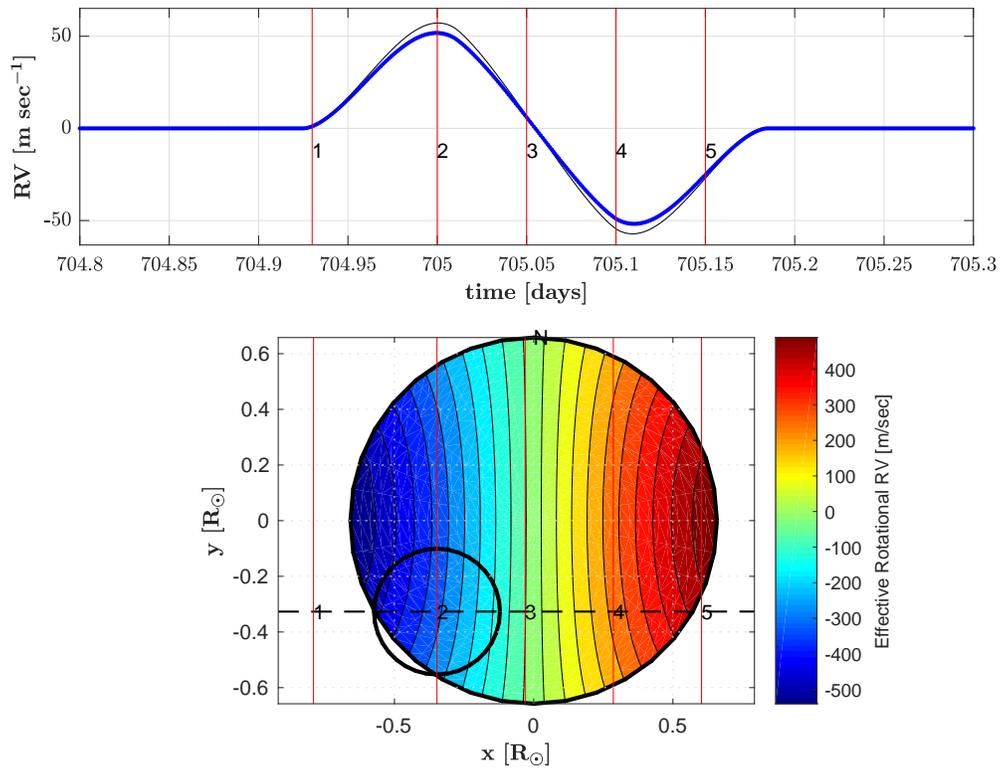}
\caption{Similar to Figure \ref{fig:Mock16_0_90_0_noLD}, but  with both quadratic limb darkening ($c_1=0.77113,~~ c_2= -0.050753$) and differential rotation ($\epsilon=0.4$). For comparison, the light black line in the top panel shows the R-M signal appearing in Figure \ref{fig:Mock16_0_90_0_LD}, a case with no differential rotation but with quadratic limb darkening.}.
\label{fig:Mock16_0_90_0p4_LD}
\end{figure}

\begin{figure}[ht!]\
\plotone{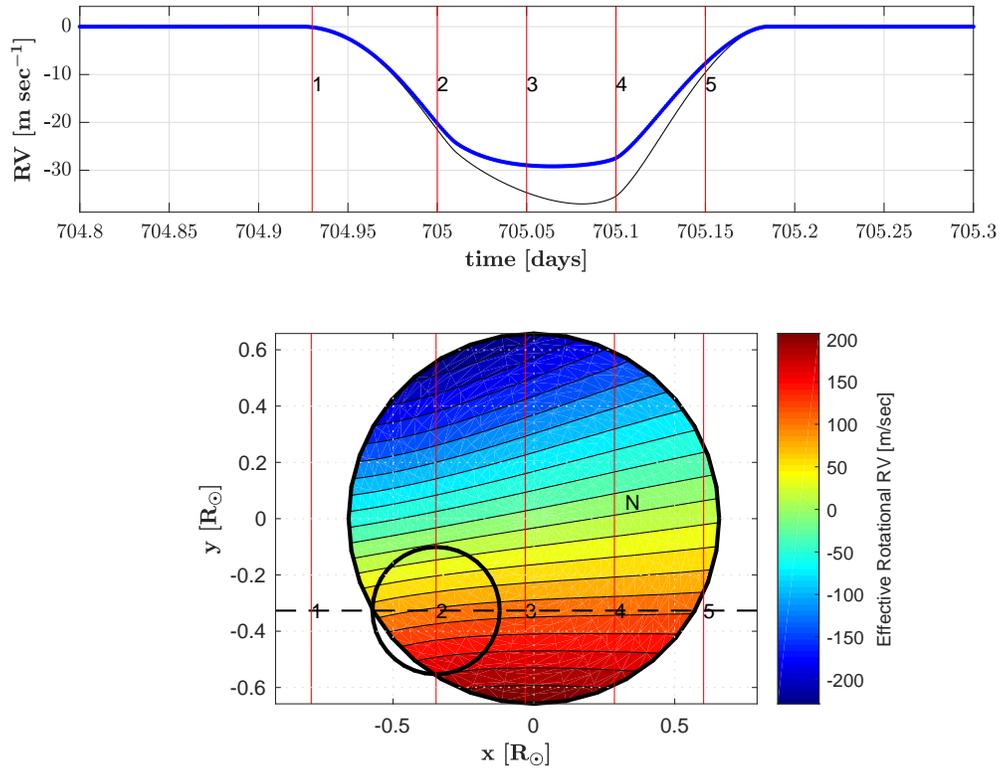}
\caption{Similar to Figure \ref{fig:Mock16_0_90_0p4_LD}, but with a misaligned rotation axis ($\Phi_{\rm rot}= 30^\circ$, $\Theta_{\rm rot }=-80^\circ$). For comparison, the light black line in the top panel shows the R-M signal for a star with the same axial orientation but rotating as a rigid body (i.e.\ $\epsilon=0$).}
\label{fig:Mock16_m80_30_0p4_LD}
\end{figure}

For convenience we will refer to the quantity $v(x,y)I(\mu)\cos(\gamma)$ as the ``effective rotational radial velocity'', ${\cal V}(x,y)$.  To visualize how the R-M effect can change with limb darkening, with the change in orientation of the rotational axis, and with differential rotation, we computed ${\cal V}$ in a mock system in which a small star transits a much larger star (Figures \ref{fig:Mock16_0_90_0_noLD}-\ref{fig:Mock16_m80_30_0p4_LD}). We start with the simple case (Figure \ref{fig:Mock16_0_90_0_noLD}) of no limb darkening, an aligned spin axis ($\Phi_{\rm rot}=90$, $\Theta_{\rm rot}=0$, meaning the star's rotation axis is parallel to the y axis), and no differential rotation. In this case, the contours of ${\cal V}$ are vertical lines, hence the value of ${\cal V}$ at each point represents the true Doppler shift at that point. The R-M signal in this case is a symmetric ``Z-wave''. We now add limb darkening (Figure \ref{fig:Mock16_0_90_0_LD}). The contours of ${\cal V}$ now become symmetrically curved towards the rotation axis, hence, ${\cal V}$ is not a true Doppler shift any longer. Since $|{\cal V}|$ is symmetric about the rotation axis, the R-M signal is, likewise, symmetric. The third case (Figure \ref{fig:Mock16_0_90_0p4_noLD}) shows an aligned system with no limb darkening and with differential rotation. The contours of ${\cal V}$ are symmetrically curved in a manner similar to that of the second case. As before, the R-M signal is symmetric. When compared to the first case, we note that the amplitude of the R-M signal is smaller when differential rotation is included. The fourth case (Figure \ref{fig:Mock16_0_90_0p4_LD}) shows an aligned system with both limb darkening and differential rotation. The ${\cal V}$ contours become curved to a greater extent than in the previous cases. However, since the symmetry of the vector field about the rotational axis remains, the R-M signal is, likewise, symmetric. When compared to the second case, the amplitude of the R-M signal is reduced. Finally (Figure \ref{fig:Mock16_m80_30_0p4_LD}), we show an example where the rotation axis of the star is misaligned with the orbit of the transiting body ($\Phi_{\rm rot}=30$, $\Theta_{\rm rot}=-80$). The ${\cal V}$ contours show no symmetry at all, and therefore the R-M effect exhibits no symmetry. When compared to a similar (misaligned) system with rigid body rotation, the difference between the R-M signals is quite noticeable. 

\subsection{The Computation and Accuracy of the Method}\label{sec:method_conversion}

 The computation of the light loss and the R-M effect through a transit
involves the sum of integrals of simple functions (Equation
\ref{eq:limbdarkintegral_new}).  For the flux fraction, each limb
darkening law has a vector of simple functions with components
involving $\mu$ (e.g.\ $\mu^0$, $\mu^{1/2}$, $\mu^1$, etc.,
see Equations \ref{eq:linear_law_dot} through \ref{eq:log_law_dot}),
and each of these functions of $\mu$ has a 1-form associated
with it (Table 1).  For the R-M effect, these simple functions
involve $\mu$, $x$, and $y$
(e.g.\ $\mu^2 x$, $\mu y$, etc.), and the associated 1-forms are
also given in Table 1.  Each 1-form used eventually leads to a definite
integral that one would need to evaluate.  When the integration arc
is on the boundary of the  back body (that is when $M=N$), $\mu=0$ and the 1-forms
simplify greatly, leading to
closed-form expressions for all of
the definite integrals.  When the integration arc
is the boundary of the front body projected onto the back body (that is when $M<N$), the definite integrals
associated with the $\mu^0$ and $\mu^2$ terms for the flux fraction
and the $x$, $y$, $\mu^2x$, and $\mu^2y$ 
terms for the R-M effect (with no
differential rotation) lead to definite integrals that 
can be  evaluated in terms of a sum of simple
functions.   We give all of these
closed-form expressions in Appendix E.  \\

Apart from the specific cases given in Appendix E, 
the computation of the flux fraction and the R-M effect will 
involve definite integrals that need to be evaluated
numerically, and we use Gaussian quadrature for this purpose.We can show, using standard techniques, that we can compute the definite integrals to any degree of precision desired.
To do this,  each integration arc is divided up
into $K$ ``panels'' according to
\begin{align}\label{eq:Mpanels}
\begin{aligned}
K = {\rm NINT}\big[2p(\varphi_1 - 
\varphi_0)+1\big]T_{\rm pan}
\end{aligned}
\end{align}
where NINT is the ``nearest integer'' function, 
$T_{\rm pan}$ is an integer ``tolerance'' parameter,
$\varphi_{0}$ and $\varphi_{1}$ are the starting and ending angles (in radians) 
of the arc, respectively, and where
$p=R_M/R_N$ is the ratio
of the front body's radius to the back body's radius. 
The code then employs  Gaussian 4-point
quadrature as the integration method for each panel. 
The error
estimate for the computation of the definite integrals that
give the fractional flux (Equation
\ref{eq:f_vis_def}) is based on the asymptotic error formula for
Composite Gaussian Integration and on the Aitken extrapolation method
for linearly convergent sequences \citep{Atkinson1989}.  The
asymptotic error formula for Gaussian 4-point quadrature depends on
the differentiation order of the integrand.  Equation \ref{eq:dGdr} in appendix
\ref{sec:appendix_numproperties} implies that the order will always be at least 1.  
In this case the asymptotic error formula for Composite Gaussian Integration is
$O(h^2)$ where $h$ is the panel width.  If we now define a sequence of
fractional fluxes with the panel length reduced by half ($T_{\rm pan}$ doubled)
for successive members, this sequence will converge linearly with the
ratio of successive differences being 0.25 or less.  We can then apply the Aitken extrapolation
to obtain a much better approximation to the sequence limit.  The
difference between the sequence and the extrapolated values provides
the error estimate.
To illustrate this, we apply the process to the example containing four
bodies  which we discuss in Appendix \ref{sec:appendix_workedexample}. The values for each step in
the error estimation process are shown in Table
\ref{tab:example_4body_error}.  The first panel in Table
\ref{tab:example_4body_error} lists the flux fractions for all but the
first (unobscured) body for 7 different values of the tolerance $T_{\rm pan}$.
Following the Aitken extrapolation algorithm, we compute successive
flux differences for each of the bodies.  We then compute the ratio of
the differences. Note that these ratios for each body are almost
constant (property of linear convergence) and they are all less than
0.25.  Finally, we apply the Aitken formula resulting in
the extrapolated flux.  Comparing the flux to the extrapolated flux
provides the error estimate. In this case every doubling of the
subintervals results in an error reduction of a factor of $\sim$ 0.18, the ratio of
the differences.  When the tolerance is $T_{\rm pan}=128$ the
error in the fractional flux is $\approx 10^{-12}$ or better.
In the discussion that follows we will assume that light curves computed
using $T_{\rm pan}=128$ are ``exact''. \\

The quadrature error one gets for a given number of panels
$K$ depends on the rate of convergence of the process which, in
turn, depends on the bounds of higher-order derivatives of
the integrand.  When the bounds are not
available (for example, owing to singularities), higher-order
convergence might not occur.  An instructive example is the
simple integral
\begin{equation}
I = \int_a^1\sqrt{x}dx = {\frac{2}{3}}\left(1-a^{3/2}\right)
\quad\rm{for}\,\,0\le a<1 \nonumber
\end{equation}
The first derivative of the integrand
has a singularity at
$x=0$.
When $a=0.1$, 6 panels are needed to give a quadrature
error of less than $10^{-8}$, using 4-point Gaussian quadrature.
As $a\rightarrow 0$, the number of panels required to keep the same
accuracy increases.  For example, 
when $a=0.05$, 11 panels are needed to get an error less than
$10^{-8}$, and when $a=0.01$, the number of panels required is
$K=34$. When our simple integral becomes
\begin{equation}
I = \int_a^1{x}^{3/2}dx = {\frac{2}{5}}\left(1-a^{5/2}\right)
\quad\rm{for}\,\,0\le a<1 \nonumber
\end{equation}
the singularity at $x=0$ does not occur until the second derivative
of the integrand.
In this case,
the number of panels required to get a quadrature
error of $10^{-8}$ becomes $K=4$, 6, and 12 when
$a=0.1$, 0.05, and 0.01, respectively.  This exercise will be useful
in the following discussion. \\

We have a code that can compute the flux fractions for any number
of mutually overlapping bodies to any degree of desired precision
by using a large number of quadrature panels for each arc on the visible
boundaries.  In most practical applications,  a fractional precision in the models
down to $\approx 10^{-15}$ is unnecessary.  
Depending on the available observational data,
one might only need model light curves with fractional precisions
of $\approx 10^{-6}$ to $10^{-8}$.
In addition, in most cases only two overlapping bodies need
to be considered at any given time.  Given this fact, we can make some modifications
in the algorithm to increase the performance while giving reasonably small
quadrature errors.  There are four types of changes we can make, and we
discuss these below. \\

\begin{figure}[t]
\includegraphics[angle=-90,scale=0.7]{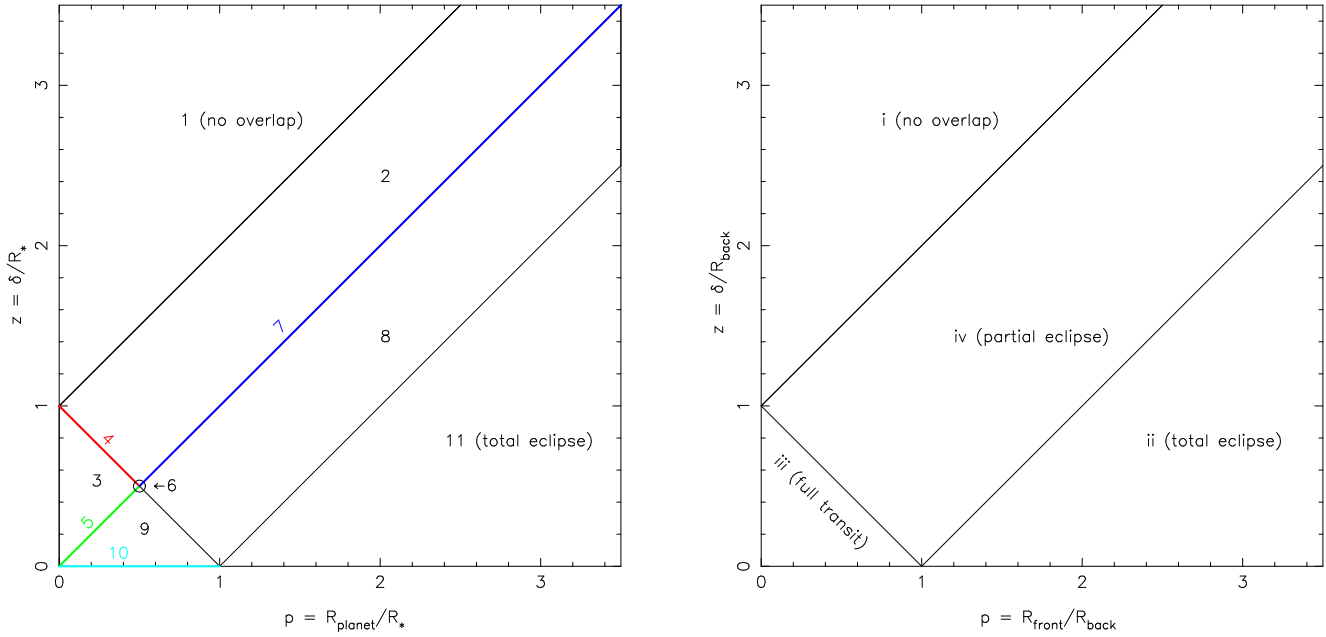}
\caption{Left:
The $(p,z)$ plane showing the various cases used to compute eclipse
events in the Mandel \& Agol algorithm, where $p$ is the ratio of
the radii and $z$ is the distance between the centers, normalized
to the radius of the body in back.
Regions 4, 5, 7, and 10 are lines and region 6 is a single point.
Different functions may need to be evaluated for different regions.
In the MA2002 notation, the ``planet'' is always the body
in front and the ``star'' is always the body in back.
Right:  
The $(p,z)$ plane showing the four cases used to compute eclipse
events in our algorithm.  The only distinction between case iii and iv
is the number of function evaluations used.
For clarity we have used $R_{\rm front}$  as the radius of the body closer to the observer 
(called $R_M$ in the text), and $R_{\rm back}$ to denote the radius of the more distant star 
(called  $R_N$ in the text).}
\label{fig:drawMAregions}
\end{figure}

When working with only two bodies, the two important parameters 
that determine the flux fraction are
the ratio of the radii of the two
objects and the separation of the centers in the POS.
In the notation of  MA2002  the parameter
$p$ is given as  $p\equiv R_{\rm planet}/R_*$,
where $R_{\rm planet}$ is always the radius of the front
body and $R_{*}$ is the radius of the back body.  In addition,
MA2002 define a parameter $z\equiv \delta/R_*$,
where $\delta$ is the separation of the centers in the POS.
In our notation these quantities are
$p\equiv R_{M}/R_{N}$ and 
$z\equiv \delta/R_{N}$.
MA2002 partition the ($p,z$) plane into
11 regions (see Figure \ref{fig:drawMAregions}), 
and these will be discussed in
greater detail below.  For our discussion, we have four regions
as shown in Figure \ref{fig:drawMAregions},
two of which are trivial:  (i) when $z>p+1$ there is
no overlap (this is Region 1 in MA2002);
(ii) when $z<p-1$ we have a total eclipse
(this is Region 11 in MA2002);
(iii) when $z \le 1-p$ we have a full transit
(this area contains 
Regions 3, 4, 5, 6, 9, and 10 in MA2002);
and (iv) when $z> 1-p$ and $p-1 \le z \le p+1$
we have a partial eclipse (this area contains
Regions 2, 7, and 8 in MA2002).     \\

\begin{figure}[t]
\includegraphics[angle=-90,scale=0.7]{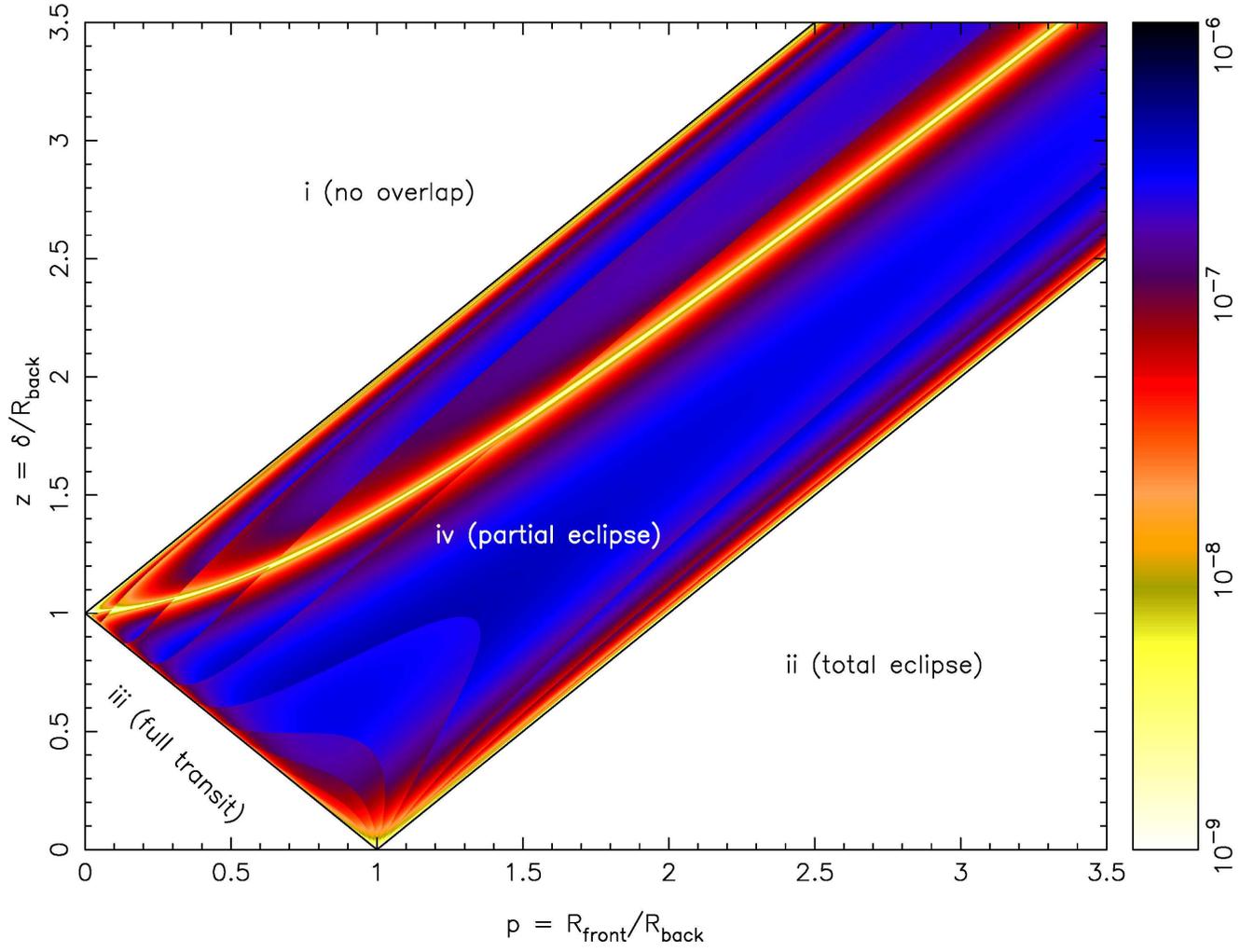}
\caption{The quadrature error for the definite integral associated
with the linear term in the limb darkening law over the
 $(p,z)$ plane.  The darker colors represent
errors on the order of a few parts per million.  Errors
below $10^{-9}$ are given as white.
}
\label{fig:compressploterror_tol2}
\end{figure}

We begin the algorithm refinement
by first getting some idea of the quadrature errors when the
number of quadrature panels is small.
We use the quadratic limb darkening law with coefficients
of 0.6 and 0.2. We divide up the ($p,z$) plane
into a $2880\times 2880$ grid between 0.0 and 3.5 along each
axis.  At each point in the grid we compute the flux fraction
using a tolerance of $T_{\rm pan}=2$ and $T_{\rm pan}=128$ and 4-point
integration, and find the absolute value of the difference between the two flux fractions.
We adopt this difference as a measurement of the quadrature error.  
Figure  \ref{fig:compressploterror_tol2} shows a color map of the results.  
With two notable exceptions, the quadrature errors are on the order of a few times $10^{-7}$, where
the out-of-eclipse flux is normalized to unity.
The first exception is the triangle in the lower left of the diagram where the errors 
are less than $\approx 10^{-10}$.  This is case (iii),
discussed above where there is a full transit. Here the front body
does not touch the limb of the back body (as seen on the POS), and
the quantity $\mu$ is always greater than zero 
along the integration path.  Higher-order derivatives
of the integrands for the $\mu^{1/2}$, $\mu^1$ and $\mu^{3/2}$
terms will have factors of $\mu$ to various powers in the
denominators, but since $\mu>0$ always, the 
higher-order derivatives will not suffer from singularities and 
hence the rate of convergence of the integration will be very high.
On the other hand, when we consider case (iv) discussed above
(e.g.\ a partial eclipse), the shadow boundary of the 
body in front will intersect the limb of the body in back.  
Thus, generally speaking, there will be some of the higher-order
derivatives that will have a singularity when $\mu=0$, thereby
slowing convergence.  
The second exception where the errors are smaller
occurs along the curve described by
$z=\sqrt{p^2+1}$.
As we can see from Figure \ref{fig:compressploterror_tol2},
the errors are $\approx 10^{-9}$ along this 
curve.   One can show that along this
curve, the integrand is effectively multiplied by a factor of $\mu^2$,
which gives us two more higher-order derivatives before
singularities occur.  As a result, 
along this curve the convergence is faster and
the errors are much smaller.  \\

Now that we have some baseline-level estimate of the number of function evaluations
that are required to get a reasonably small quadrature error, 
our first improvement to the algorithm
involves using higher-order Gaussian quadrature.
The number of function evaluations required to numerically compute
the definite integrals would be $N_{\rm op}=4K$ for 4-point Gaussian quadrature
and $K$ panels.  Alternatively, one could use $(4K)$-point
Gaussian quadrature and one panel to evaluate the definite integrals
using the same number of function evaluations.  The application of
the Aitken extrapolation technique to evaluate the quadrature error
for these high order quadrature cases is less straightforward since the order changes.  Fortunately we can
always use the difference between a model with a given tolerance and the same model 
computed using $T_{\rm pan}=128$ (or higher), and 4-point integration to get a good 
estimate of the quadrature error since the latter
model can be shown to be very nearly exact.  
We have found from numerical experiments
that using a higher order Gaussian quadrature over one panel always produces smaller errors
than using 4-point Gaussian quadrature and $K$ panels, when the number of
function evaluations is the same for both.  This result is a consequence of the
fact that the error formulae for Gaussian integration involve higher-order derivatives
of the integrand and various constant coefficients, and
while the derivatives of the integrands may suffer from singularities past the
first or second order (thereby slowing convergence), 
the values of the constant terms do decrease with
increasing order.  \\

Thus, for our first modification to the algorithm, we set the number of
{\em operations} as  
\begin{align}
\begin{aligned}\label{eq:Mopt}
N_{\rm op} = {\rm NINT}\bigg\{\big[3.7p(\varphi_1 - 
\varphi_0)+1\big]T_{\rm op}\bigg\}
\end{aligned}
\end{align}
where the meaning of the variables are the same as in Equation
\ref{eq:Mpanels}. The factor of 3.7 was arrived at through numerical experimentation.
Also note that in this case, the tolerance parameter $T_{\rm op}$ is not necessarily
an integer.
Once we have the number of operations set, $N_{\rm op}$-point Gaussian quadrature
is used for a single  panel, up to order 64.  If the number of desired operations exceeds 64, then
the integration arc is subdivided with the requirement that the
quadrature order is the same for each subarc and is as large as
possible without exceeding 64. In addition, we set the minimum
value of $N_{\rm op}$ to 8. \\

Our second modification to speed up the routine targets values
of $p$ and $z$ where there is a full transit 
(that is, when $z\le 1-p$).  As discussed above,
the quadrature errors there are very small owing to the 
properties of the higher-order derivatives of the integrand.  Since
the convergence in this region is relatively fast, we can get
by with fewer overall function evaluations, $N_{\rm op}$, and we start with
$N_{\rm op}=8$.  
Generally speaking,
the quadrature errors are the smallest when both $p$ and $z$ are small, and
increase as one moves up diagonally in the region.  Consequently,
we define triangles with vertices given by $p=0$ and $z=1$,
$p=1$ and $z=0$, and $p=p_0$ and $z=z_0$ (where 
$p_0=z_0$).  For a given point $(p,z)$,
$N_{\rm op}$ is increased by 1 when the following conditions are met:
if it is inside the triangle with a vertex
of (0.25,0.25), if that point is inside the triangle with vertex 
(0.35,0.35),  if it's inside the triangle
with vertex at (0.40,0.40) and, finally,
if the point is between the lines 
given by $z=0.997-p$ and $z=1-p$.  \\

The third modification we can make to increase the algorithm
performance is for the case where there are 
only two bodies and only the flux fraction is required.
The co-ordinate system in the POS can be rotated in such a way
so that the center of the front body lies 
on the $x$-axis.
In that case, the starting and ending
angles of the arc will be symmetric about the
$x$-axis. For example $\varphi_0=330^{\circ}$
and $\varphi_1=390^{\circ}$, or   
$\varphi_0=170^{\circ}$
and $\varphi_1=190^{\circ}$.  In cases like these, we can integrate
over half of the arc (thereby using half the number of function
evaluations) and multiply the quadrature sum by 2 to get the
final flux fraction.  This modification won't work when computing
the R-M effect, as both the $x$ and $y$ co-ordinates
of the center of the body in front are important, which makes
the axis rotation impossible. \\

Similarly to the third modification, the final one applies to the case of having only two bodies.
Many of the procedures described in Appendix \ref{sec:appendix_workedexample} 
to construct the arcs in the general routine for $\cal{N}$ bodies are not necessary when there are
only two bodies. For example checking for arc intersections or computing the
shadow boundary. In addition, having to store only the angles greatly simplifies the code. 
This streamlined version of the routine is about 10\% faster than
the general routine called when ${\cal N}=2$.  It is
straightforward to decide whether to call the general routine
or the two-body routine based on the sizes of the bodies
and the distances between them. \\

\begin{figure}[t]
\includegraphics[angle=-90,scale=0.75]{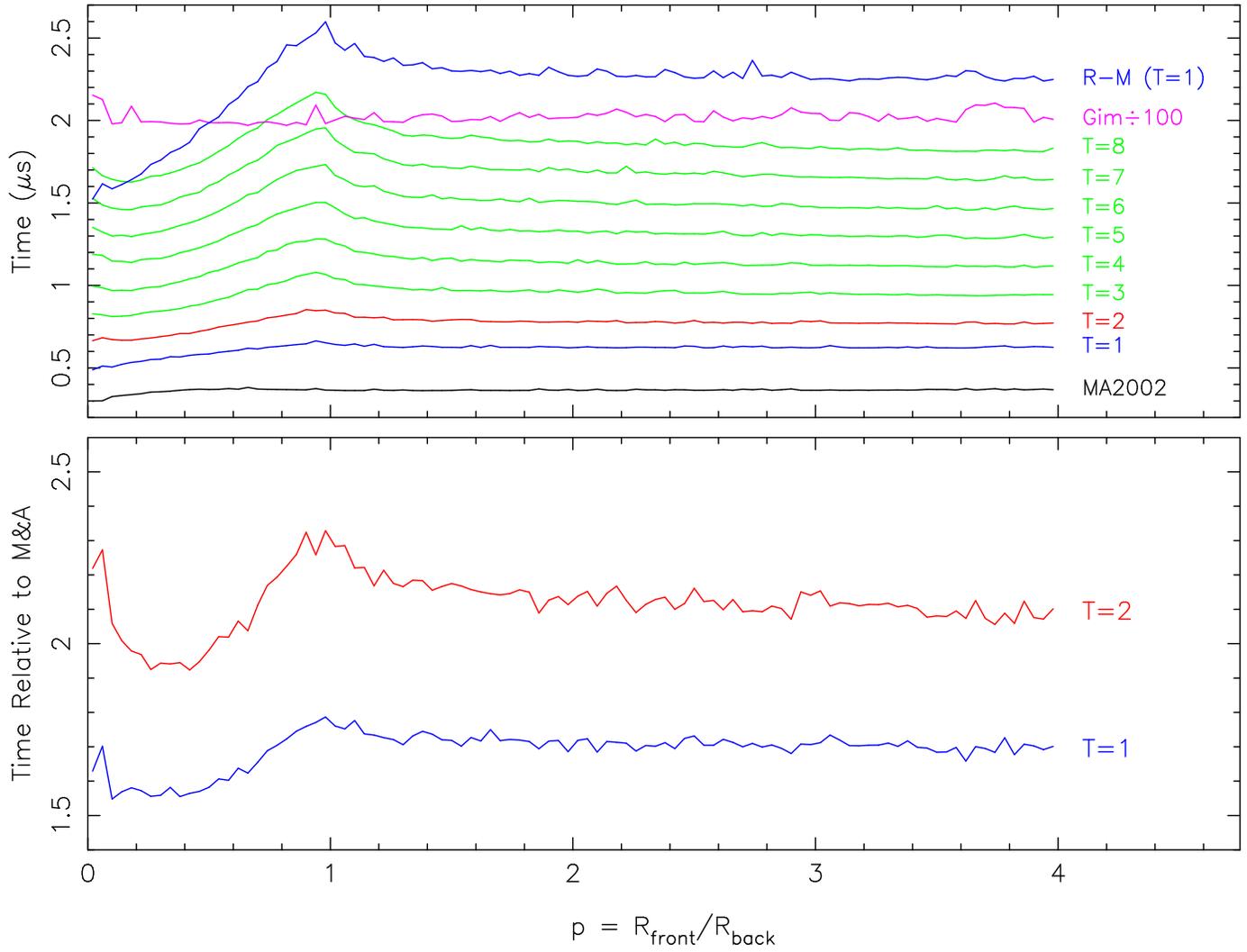}
\caption{Top:  The time in microseconds needed to compute
a flux fraction as a function of the radius ratio $p$.
The black line is for the Mandel \& Agol {\tt occultquad}
routine, the bottom curves are for $T_{\rm op}=1$ to $T_{\rm op}=8$ in steps of 1, 
the magenta curve is for the Gim\'{e}nez routine (which includes
the R-M effect) where the times were divided by 100, and
the top blue curve is for the R-M effect and flux fraction for $T_{\rm op}=1$.
Bottom:  The ratio of the $T_{\rm op}=1$ and $T_{\rm op}=2$ times relative 
to the MA2002 {\tt occultquad} routine.  
}
\label{fig:plottiming}
\end{figure}

\begin{figure}[t]
\plotone{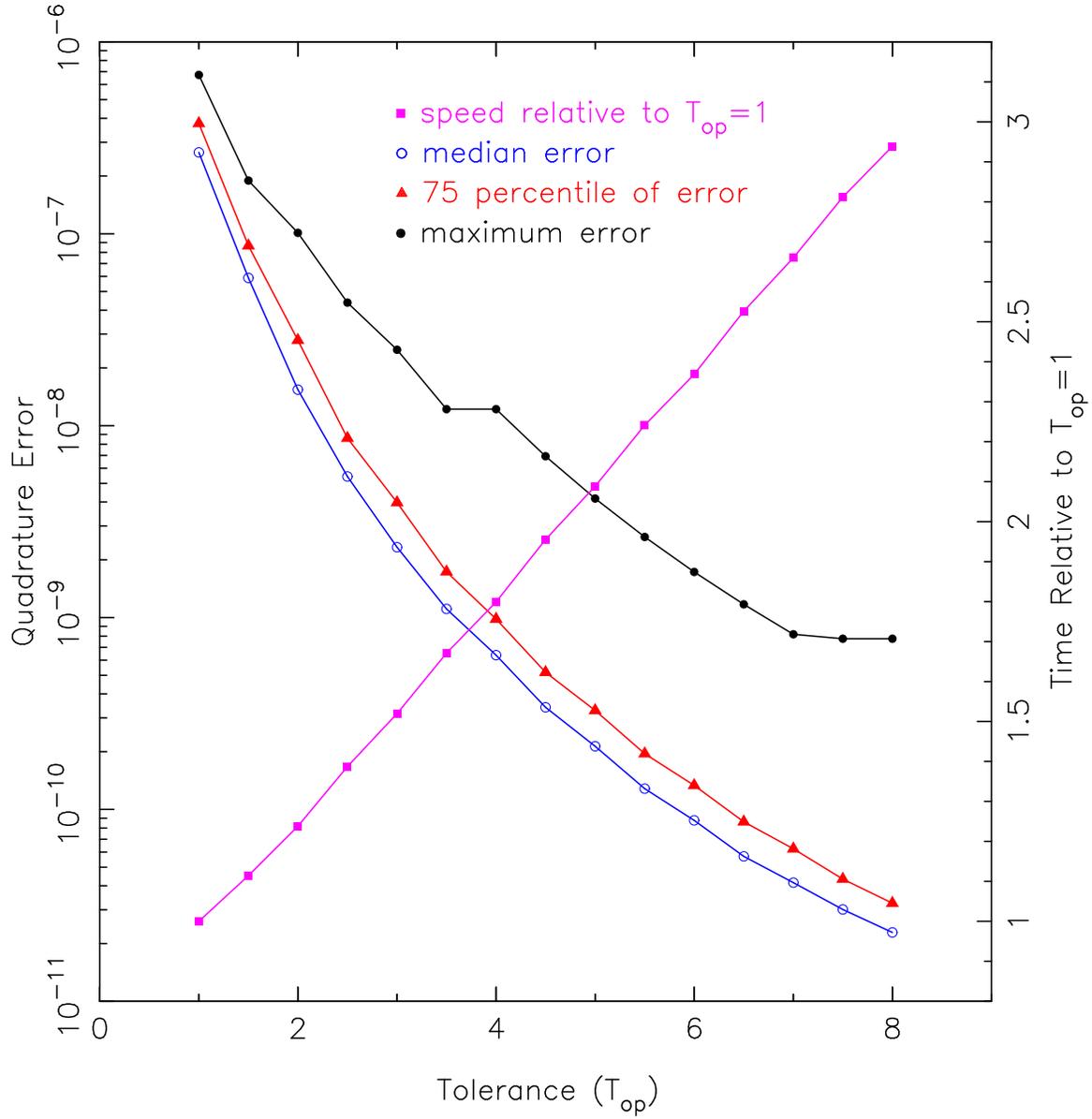}
\caption{The quadrature errors (left-hand scale)
and relative algorithm speed (right-hand scale) for
the computation of the flux fraction as a function of the 
tolerance $T_{\rm op}$.  The various curves show the maximum
error (black filled circles), the 75 percentile error (red triangles),
and the median error (blue open circles).  
The magenta filled squares give the speed relative to the $T_{\rm op}=1$ models.
}
\label{fig:plotspeedvsaccuracy}
\end{figure}

We have performed extensive tests to measure the speed of our routines.  
For a given value of $p$, the value of $z$ was varied
from $z=p-1$ to $z=p+1$ in steps of $10^{-6}$ (when
$p<1$ the lower bound on $z$ is zero) and a flux fraction was
computed for each $z$ using the quadratic law.  The time required to compute these flux fractions
was measured using the {\sc FORTRAN} function {\sc cpu\_time()}.
These tests were done on a workstation with an Intel Xeon
W-2155 CPU running at 3.30 GHz using the 
Portland Group {\sc FORTRAN}
compiler version 18.4 with the {\sc -fast} optimization flag.
Figure \ref{fig:plottiming} shows the results.   For a given
tolerance,  the time required to compute a flux fraction
rises from small values of $p$, reaches a peak near
$p=1$, then drop off modestly thereafter.  Past
$p\approx 3$, the curves are flat.
When $T_{\rm op}=4$, the time to compute a single flux fraction is
about one microsecond.
Figure \ref{fig:plotspeedvsaccuracy} shows the time required
relative to $T_{\rm op}=1$ for values of $T_{\rm op}$ ranging from 1 to 8 in steps
of 0.5.   The time vs.\ tolerance curve is nearly linear,
and the slope is modest since
models with $T_{\rm op}=8$ take about three times longer
than models with $T_{\rm op}=1$.
For other limb darkening laws, the times required
relative to the quadratic law are 0.93 for the linear law, 1.55 for the 
logarithmic law, and 1.89 for the square root law.  \\

When computing the R-M effect, some numerical experimentation
has shown that more function evaluations are required to ensure
the quadrature error is less than $\approx 0.1 {\rm m~s^{-1}}$.
The multiplier in Equation \ref{eq:Mopt} is 12.0 instead of 3.7.
Also, 
the convergence of the Gaussian
quadrature for $p>1$ near the onset of
total eclipses is relatively slow, so the value of
$N_{\rm op}$ is increased by 4 when $p>1$ and $z<\sqrt{p(p+1)}$
(these conditions were arrived at after the inspection of
many error maps).  When $T_{\rm op}=1$, it takes about 3.6 times longer
to compute a flux fraction along with the R-M correction compared to
computing the flux fraction alone. \\

\begin{figure}[t]
\includegraphics[angle=-90,scale=0.75]{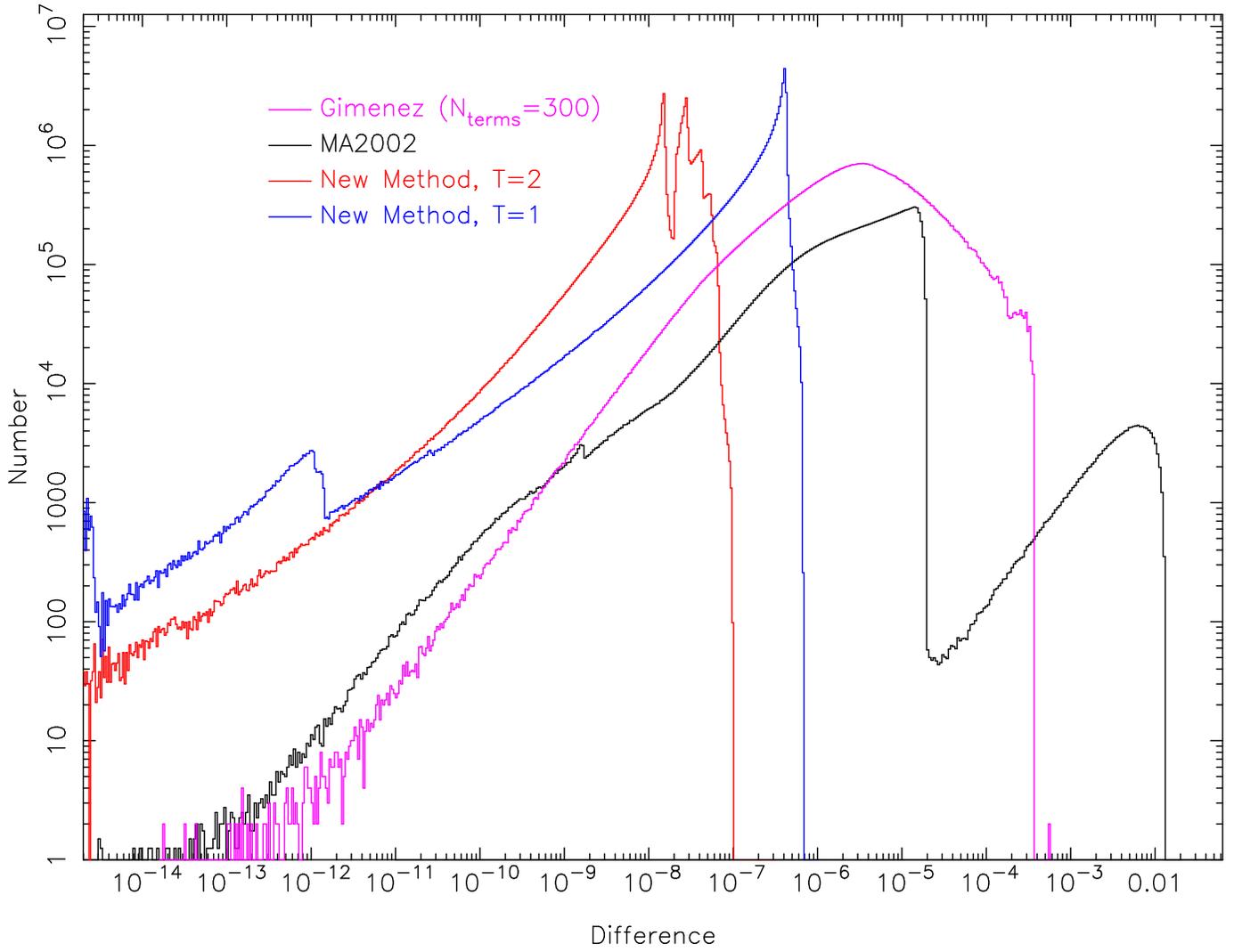}
\caption{The frequency distributions for the quadrature errors
associated with the flux fraction computation for the quadratic 
limb darkening law for values of the radius ratio $p$ between 0 and
100.  The curves are for the {\tt occultquad} routine of
MA2002 (black), the \citet{Gimenez2006a} routine (magenta),
our new method with $T_{\rm op}=1$ (blue) and our new method with $T_{\rm op}=2$
(red).
}
\label{fig:plothist}
\end{figure}

To fully characterize the quadrature
errors, we computed flux fractions
for $5 \times 10^7$ randomly selected
points in the $(p,z)$ plane for $0<p<100$ and 
$p-1<z<p+1$ (when $p<1$ the lower bound on $z$ is of course zero).
The flux fractions were computed using the quadratic
law with coefficients of 0.6 and 0.2, for values of $T_{\rm op}$ ranging from 1 to 8
in steps of 0.5, and
for $T_{\rm pan}=128$.  We also computed models using $T_{\rm op}=2$ for each of 
the linear, square root, and
logarithmic limb darkening laws.
Using the $T_{\rm pan}=128$ model as the reference, we computed
the quadrature errors for the other models.  Figure \ref{fig:plothist}
shows the frequency distributions for the $T_{\rm op}=1$ and $T_{\rm op}=2$ quadratic
law cases. When $T_{\rm op}=1$, the maximum quadrature error is $6.72\times 10^{-7}$, and
when $T_{\rm op}=2$, the maximum quadrature error is $1.00\times 10^{-7}$.
The histograms for the  $T_{\rm op}=2$ models with the other
limb darkening laws look very similar to the histogram
for the quadratic law.  The maximum errors (for $T_{\rm op}=2$) are
$1.13\times 10^{-7}$ for the square root law,
$1.00\times 10^{-7}$ for the logarithmic law,
and $5.82\times 10^{-8}$ for the linear law.\\

Figure \ref{fig:ploterrortrend1} shows how the quadrature errors depend
on the ratio of radii $p$.  To generate this plot, we took 3000 values of
$\log p$ ranging between $-3.0$ and 2.0 in equal steps. 
For a given bin centered on a particular value of
$p$, 100,000 values of $z$ were randomly chosen
in the appropriate range, and the quadrature errors
were found for all of the cases. Then
the maximum quadrature error
and the 99 percentile  error in that bin
were found.  For planet transits (i.e.\ small $p$),
the maximum quadrature error is $10^{-8}$ or smaller.  When 
the value of $p$ is around $p=1$, the quadrature errors 
flatten out, and they never exceed the thresholds mentioned previously.
Finally, Figure \ref{fig:plotspeedvsaccuracy}
shows how the errors change as the tolerance parameter $T_{\rm op}$ changes.
As one might expect, the maximum quadrature error decreases as
$T_{\rm op}$ increases.  Furthermore, the median of the error distribution is
usually smaller than the maximum error 
by an order of magnitude or more. \\

\begin{figure}[t]
\includegraphics[angle=-90,scale=0.75]{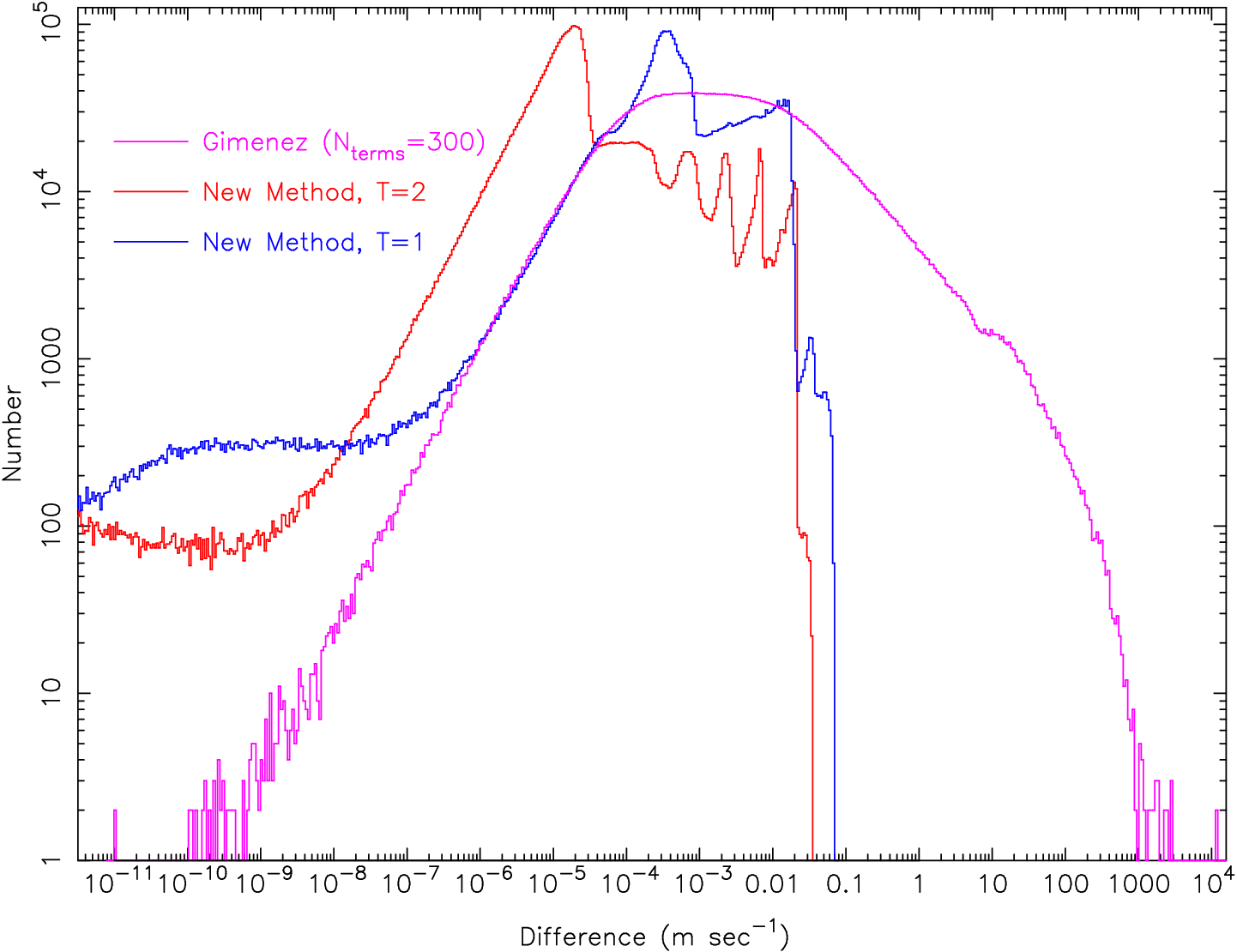}
\caption{The frequency distributions for the quadrature errors
in ${\rm m~s^{-1}}$ associated with the R-M computation for the quadratic 
limb darkening law for values of the radius ratio $p$ between 0 and
100.  The curves are for  the \citet{Gimenez2006a} routine (magenta),
our new method with $T_{\rm op}=1$ (blue) and our new method with $T_{\rm op}=2$
(red).
}
\label{fig:plotRVhist}
\end{figure}

\begin{figure}[t]
\plotone{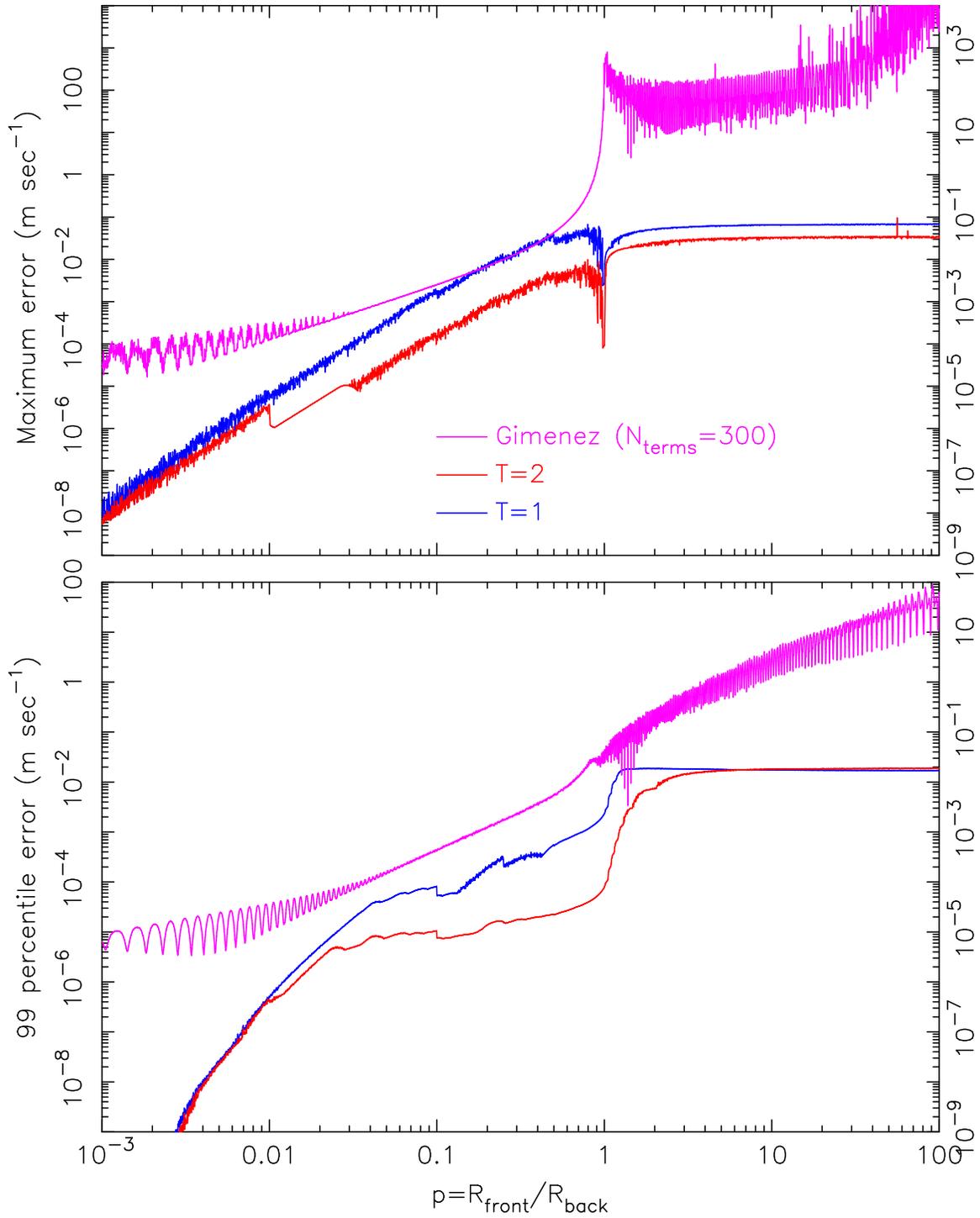}
\caption{The maximum quadrature error
(top) and the 99 percentile error (bottom) for the 
computation of the 
R-M effect using the quadratic limb darkening
law as a function of the radius ratio $p$.   
The curves are for  
the \citet{Gimenez2006b} routine (magenta),
our new method with $T_{\rm op}=1$ (blue) and our new method with $T_{\rm op}=2$
(red).}
\label{fig:plotRVerrortrend}
\end{figure}

For the computation of the R-M effect, we have at least three parameters.
At a minimum, we need the ratio of radii $p$ and the relative $x$ and
$y$ co-ordinates of the two bodies on the POS.  For simplicity
we consider cases when $\Phi_{\rm rot}=90^{\circ}$ and
$\Theta_{\rm rot}=0$.  We computed the R-M effect for 
$5 \times 10^7$ cases with $0<p<100$ and relative $x$ and $y$ co-ordinates
appropriate for each $p$ for $T_{\rm op}=1$ and $T_{\rm op}=2$.  We used the quadratic
limb darkening law with coefficients of 0.6 and 0.2, and  assumed
a rotation period of 10 days and a radius of $1\,R_{\odot}$ for
the back body.
Figure \ref{fig:plotRVhist}
shows the distribution of the quadrature errors (in ${\rm m~s^{-1}}$).
When $T_{\rm op}=1$, the maximum quadrature error is $\sim 0.067~{\rm mm~s^{-1}}$.  
Figure \ref{fig:plotRVerrortrend} shows the maximum quadrature
error and the 99 percentile error as a function of $p$,
where the same procedure used to make Figure \ref{fig:ploterrortrend1}
was used.  
For planet transits
where $p\lesssim0.1$ the quadrature errors are less than about
$10^{-3}~{\rm m~s^{-1}} $.
When $p>1$, the maximum quadrature errors are around $10^{-2} {\rm m~s^{-1}}$. \\

When working with three or more bodies, we no longer have a simple
plane in which we can map out the uncertainties.  
However, all of the arcs in multibody cases are subarcs
of all pairwise two-body cases.  Thus, all of the integrands are the same,
although the integration intervals might be smaller and/or broken into
pieces.  Therefore, the behavior of the
quadrature should be similar.  In many of the multi-body cases
the integrand does not change sign from the two-body cases, 
and this would result in the quadrature
error of a subarc being roughly the same as the
two-body quadrature error. \\

\begin{figure}[t]
\plotone{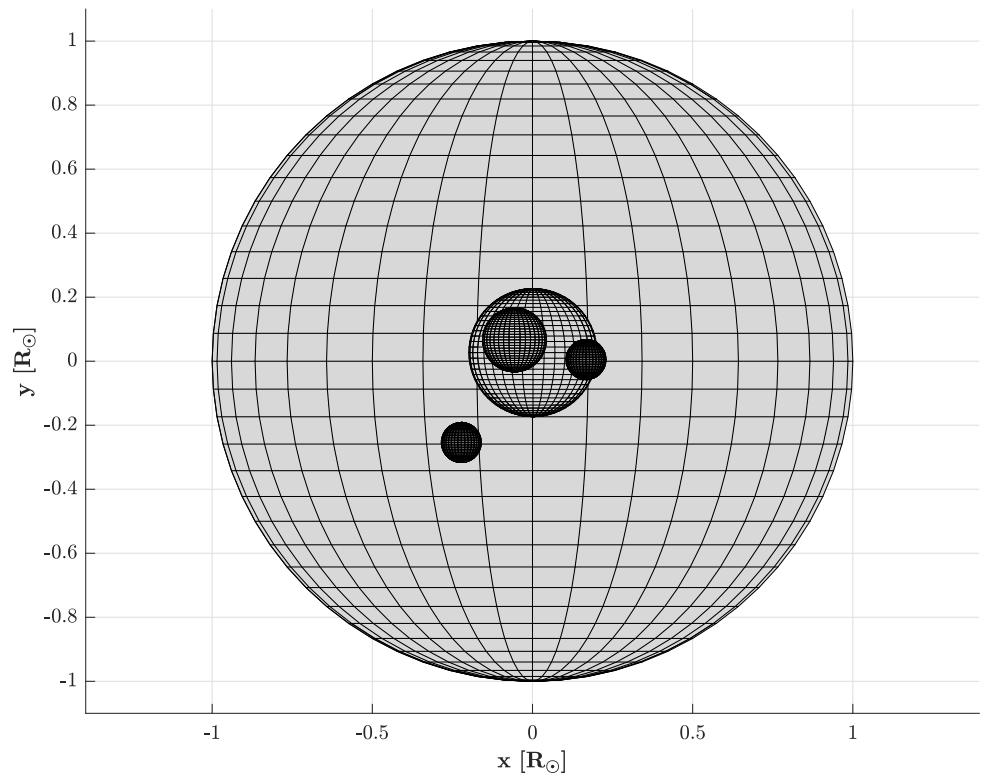}
\caption{The spatial configuration of a mock 5-body
system discussed in \S4.3, shown near the time of the transit of
four smaller bodies across the largest body.}
\label{fig:Jerrys5BodyConfig}
\end{figure}

\begin{figure}[t]
\plotone{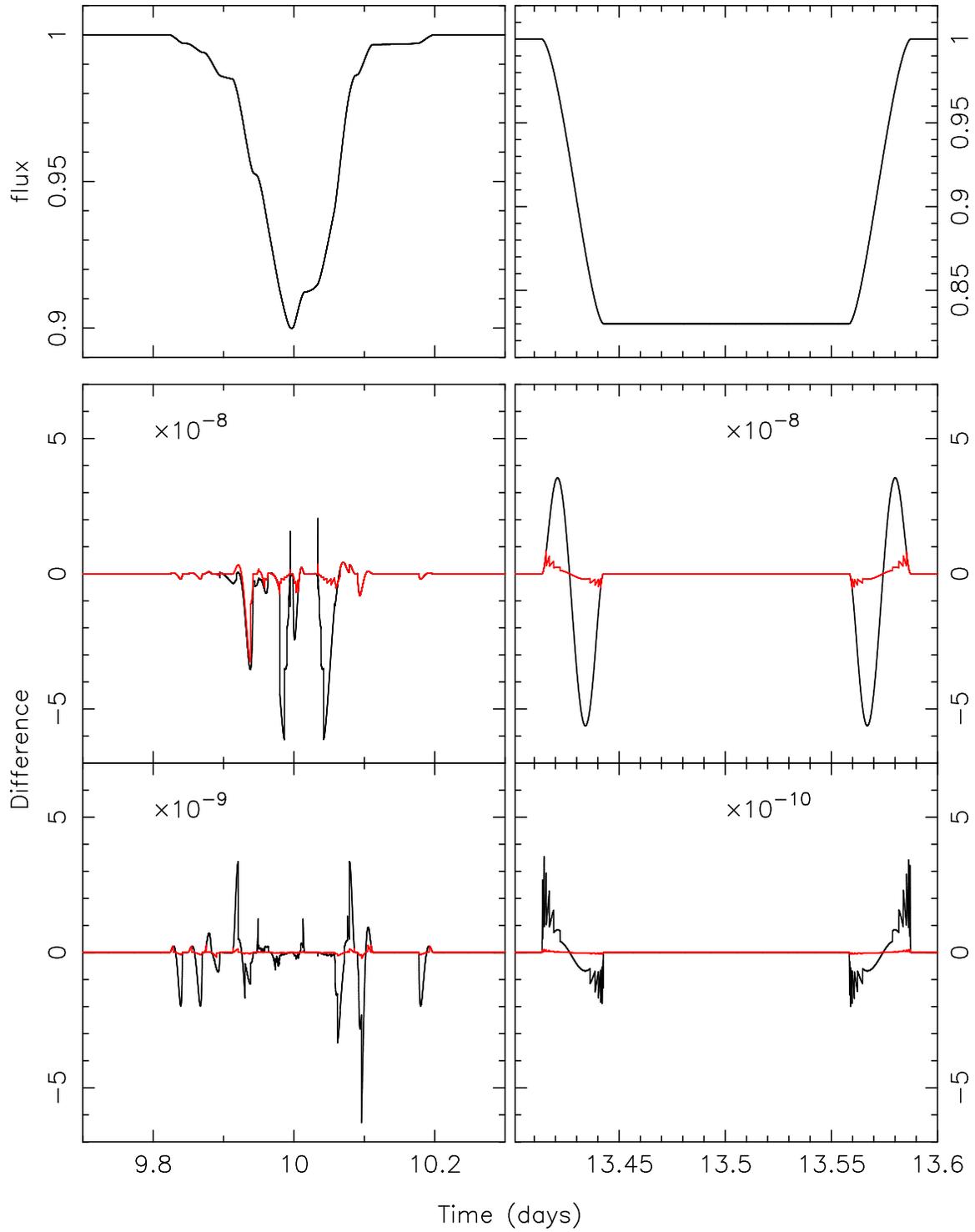}
\caption{Top:
The light curve of a mock five-body system during
the transit of the four smaller bodies across the largest
body (left), and during a total eclipse of the second
body by the first (right).  Middle:
The difference between the nominal light curve made
with $T_{\rm pan}=128$ and the light curves made with $T_{\rm op}=1$ (black), 
and $T_{\rm op}=2$ (red).
Bottom:
The difference between the nominal light curves with
$T_{\rm pan}=128$ and the light curves made with $T_{\rm op}=4$ (black), 
and $T_{\rm op}=8$ (red)}
\label{fig:convergefig3}
\end{figure}

\begin{figure}[t]
\plotone{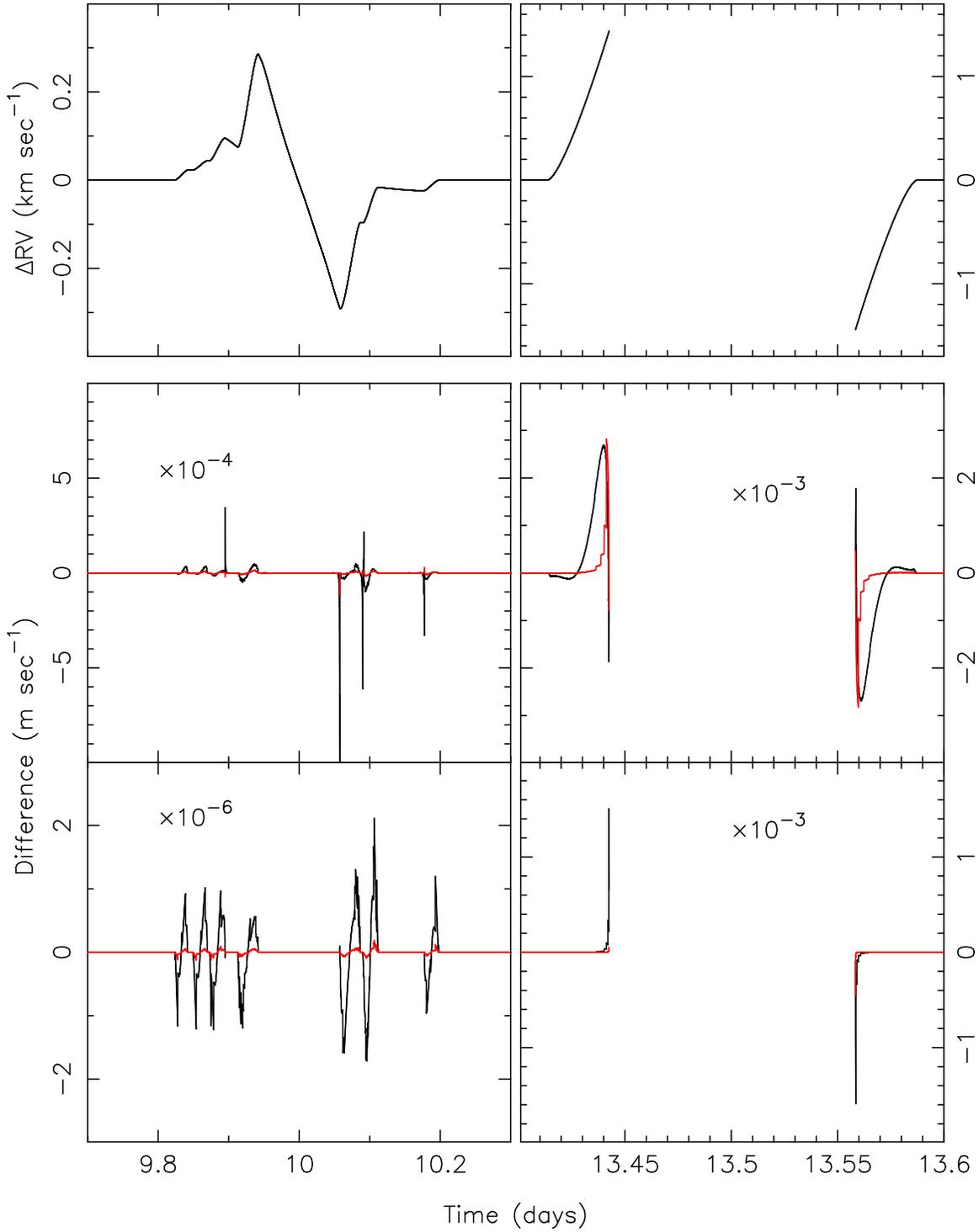}
\caption{Top: The R-M effect of a mock five-body
system during the transit of the four smaller bodies across
the largest body (left), and during a secondary eclipse
(right).
The orbital motion of the star has been removed.
Middle:  
The difference between the nominal R-M curve with $T_{\rm pan}=128$
and the R-M curve with $T_{\rm op}=1$ (red), and $T_{\rm op}=2$ (black) 
The units of the differences
are ${\rm m~s^{-1}}$, and the curves have been scaled
as indicated.  Bottom: 
The difference between the nominal R-M curve and 
the R-M curve with $T_{\rm op}=4$ (black), and $T_{\rm op}=8$ (red).}
\label{fig:convergeRVfig3}
\end{figure}

To illustrate the errors and how the errors decrease with
increasing tolerance, 
we computed light curves of a mock 5 body 
system (see Figure \ref{fig:Jerrys5BodyConfig})
with masses of 
$M_1=0.8\,M_{\odot}$,
$M_2=0.2\,M_{\odot}$,
$M_3=5\times 10^{-4}\,M_{\odot}$,
$M_4=4\times 10^{-4}\,M_{\odot}$, and
$M_5=2.5\times 10^{-4}\,M_{\odot}$, radii of
$R_1=1.0\,R_{\odot}$,
$R_2=0.2\,R_{\odot}$,
$R_3=0.1\,R_{\odot}$,
$R_4=0.0625\,R_{\odot}$, and
$R_5=0.0625\,R_{\odot}$, and fluxes of
$F_1=0.80$,
$F_2=0.17$,
$F_3=0.01$,
$F_4=0.01$, and
$F_5=0.10$.  The initial orbital parameters were circular but a full
integration of the laws of motion advanced the bodies (for a detailed
description of our integration see \citet{Welsh2015} and
\citealt{Hairer2002}). The initial periods were $P_1=7$ days, $P_2=70$
days, $P_3=210$ days, and $P_4=500$ days.  The orbital inclinations
were close to $90^{\circ}$, and the nodal angles were close to
$0^{\circ}$.  The initial positions of the bodies were such that the
four smaller bodies all transited the largest star near day 10.0.  
At this time we have a complex pattern of arcs describing the visible
boundary of the most distant star, and this pattern rapidly changes
before and after this time.
The model light curves were computed (assuming  quadratic limb darkening)
with tolerances of $T_{\rm op}=1,2,4,8$, and $T_{\rm pan}=128$.  Figure
\ref{fig:convergefig3} shows a close-up of the multi-body transit
event at day 10, as well as a close-up of a total eclipse of the second
body at day 13.5.  We plot the curves showing the differences
between the $T_{\rm pan}=128$ curve and the other curves.  For $T_{\rm op}=1$,
the errors in the computed fluxes are a
few parts in $10^7$. As the tolerance gets larger, 
the differences get     
smaller, as expected.  
Figure \ref{fig:convergeRVfig3} shows the
R-M curves for the multi-body event
at day 10,  and for a secondary eclipse at day 13.5,
assuming synchronous rotation, $\Phi_{\rm rot}=90^{\circ}$,
and
$\Theta_{\rm rot}=0^{\circ}$.  For the syzygy event
near day 10, the quadrature errors are smaller than $10^{-3}~{\rm m~s^{-1}}$
for $T_{\rm op}=1$, and considerably smaller than
that for $T_{\rm op}=2$, 4, and 8.  The quadrature
errors are somewhat larger just before the onset of the
total secondary eclipse owing to the lower differentiability
order near there, which slows the convergence of the quadrature
process.  Even so, the spikes there are at the level of
a few times $10^{-3}~{\rm m~s^{-1}}$, which seems adequate
for most practical applications.

\section{Comparison with Other Methods}\label{sec:methodscompare}

For comparison purposes,
we have  a direct implementation of
the MA2002 {\tt occultquad} routine for quadratic limb darkening. 
The accuracy of this routine is not adjustable. 
We also wrote our own routine to implement the \citet{Gimenez2006a}  
method for computing the flux fraction for
the quadratic limb darkening law and the \citet{Gimenez2006b}
method for computing the R-M effect, which also uses the quadratic
limb darkening law.
The accuracy of this routine is adjustable by  
specifying the parameter $N_{\rm terms}$,
which is the number of terms in the polynomial expansions
(namely Equation 11 from \citealt{Gimenez2006a} and Equation 11
from \citealt{Gimenez2006b}).
A larger number of polynomial terms provide more
accurate light curves but, obviously, at the expense of longer computation
times.  We typically use $N_{\rm terms}=300$.  \\

We performed speed tests of these two algorithms using the 
same procedure as above, and the results are shown in
Figure \ref{fig:plottiming}.  When $p\gtrsim 0.5$, {\tt
occultquad} requires about 0.36 microseconds to compute a flux
fraction, which is about 1.7 and 2.1
times faster compared to our method
with $T_{\rm op}=1$ and $T_{\rm op}=2$, respectively. 
The Gim\'{e}nez routine with $N_{\rm terms}=300$ 
is significantly slower as it
requires about 200
microseconds to compute a flux fraction along with the R-M effect, which is $\approx 80$
times slower than our R-M routine using $T_{\rm op}=1$. 
We note, however, that within the workings of the our photodynamical code, 
the time required to compute the flux fraction, with any of the methods,
is relatively small compared to the time required for integrating the Newtonian
equations of motion, and computing the fitness. For example, in the case
of Kepler-16, running the DEMCMC-based optimizer is only a factor of 2
slower than running the same code with {\tt occultquad} or our routine.
In other words, in the context of our full photodynamical modeling, a factor
of a few in the speed difference of the flux fraction routine is not really noticeable. \\

\begin{figure}[t]
\plotone{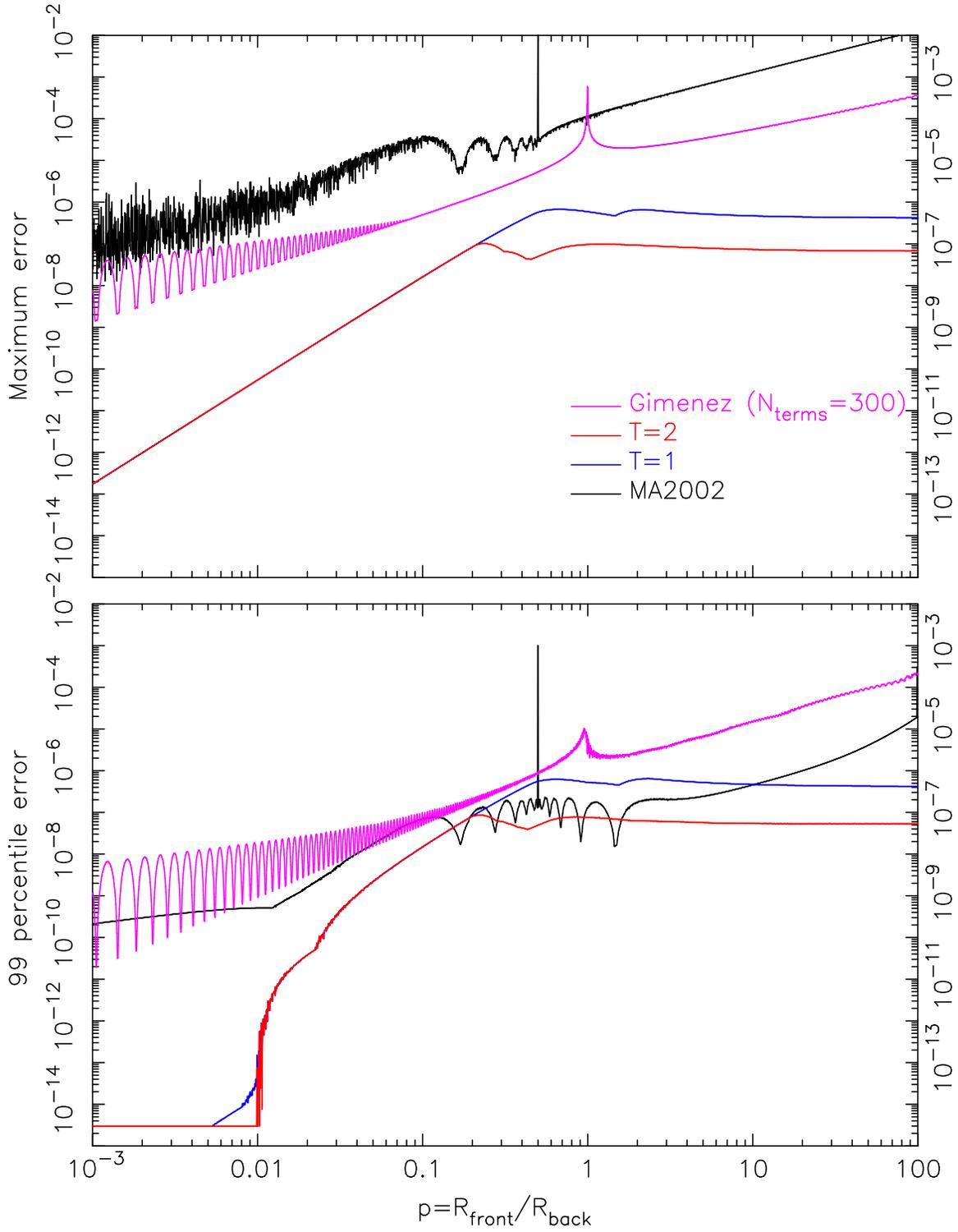}
\caption{The maximum quadrature error
(top) and the 99 percentile error (bottom) for the 
computation of the flux fraction using the quadratic limb darkening
law as a function of the radius ratio $p$.   
The curves are for the MA2002 {\tt occultquad} routine (black), 
the \citet{Gimenez2006a} routine (magenta),
our new method with $T_{\rm op}=1$ (blue) and our new method with $T_{\rm op}=2$
(red).
}
v\label{fig:ploterrortrend1}
\end{figure}

We also characterized the quadrature errors for these
two routines using the same procedures described above.
Figure \ref{fig:plothist}\ shows the results.  The quadrature errors
for the Gim\'{e}nez routine are smaller than a few times
$10^{-4}$.  The mode of the error distribution for the
{\tt occultquad} routine is around $10^{-5}$.  In addition, this
distribution has a small extension to errors of up
to about 0.01.
Figure \ref{fig:ploterrortrend1} shows how the quadrature errors depend
on the ration of radii $p$.  For planet transits (i.e.\ for small
$p$), the 99 percentile quadrature  errors for 
{\tt occultquad} and for the Gim\'{e}nez routine are smaller than 
$10^{-6}$, and in most cases considerably smaller.  However, the maximum error for {\tt occultquad}
is about $10^{-5}$ when $p=0.1$.  Furthermore, the maximum
error  goes up with increasing $p$ and reaches $\sim 0.01$ 
when $p=100$.  Also note a rather significant spike in the 
{\tt occultquad} curve
near $p=0.5$, where the 99 percentile error is about 0.001 and the
maximum error reacher $\sim 0.01$.  The quadrature errors for
the Gim\'{e}nez routine also grow with increasing $p$, and there is also
 a modest spike, although it occurs  at $p=1$ instead of
$p=0.5$. We note that we do have the option of increasing $N_{\rm terms}$
in order to reduce the quadrature errors. \\

We can see from Figure \ref{fig:ploterrortrend1} that for
a given
value of the ratio of radii $p$, the 99 percentile
quadrature error for the {\tt occultquad} routine
is usually two or three orders of magnitude smaller
than the maximum quadrature error.   This suggests that
there is a small but significant tail in the quadrature
error distribution that
extends to relatively large values.  
To see why that is, we refer again to Figure \ref{fig:drawMAregions},
which shows the $(p,z)$ plane, where 
$p\equiv R_{\rm planet}/R_{*}$ in the Mandel \& Agol notation
(in our notation $p\equiv R_{M}/R_{N}$),  
and where $z\equiv \delta/R_{\ast}$ in the Mandel \& Agol notation 
(in our notation $z\equiv \delta/R_{N}$)
with $\delta$ being the separations of the centers on the POS.
In the MA2002 {\tt occultquad} routine, 
the $(p,z)$ plane is divided up into 11 regions as indicated in the figure.
For example, region 1 is where no transit events occur. 
In region 2 the planet disk lies on the limb of the star but does not cover
the center of the stellar disk.  Regions 4, 5, 7, and 10 are lines in the plane, 
and region 6 is a single point. For a given eclipse or transit event, the $p$ co-ordinate
is fixed, while the $z$ co-ordinate changes with time.  Therefore,
as an eclipse or transit event is computed in time, the locations
of the $(p,z)$ points ``move'' 
along vertical lines in the plane.
As they do so, these points might pass
through more than one region, and the transitions between the
regions might not be smooth. Depending on where in the plane 
a particular $(p,z)$ point is located,
there are different functions to evaluate the flux fraction
when computing the flux deficit. Note that in our method and the method of Gim\'{e}nez, 
the same functions are evaluated
regardless of where one is in the $(p,z)$ plane.  \\

Consider an eclipse event that is  in region 8  (planet covering
the center and limb of the stellar disk) at
the closest approach of the centers.  Region 7 
(planet's disk touches the stellar center) and region 2 
(planet lies on the limb of the star but does not cover the
center of the stellar disk)
will have to be crossed as the objects move apart.  
When calculating the flux deficit in regions 2 and 8 
one has to evaluate a function, 
$\lambda_1$ (MA2002 Table 1), the last term of which is:
\begin{equation} \label{eq:term_lambda1}
\lambda_1=\frac{1}{9\pi\sqrt{pz}} \bigg[ \ldots -\frac{3q}{a}\Pi\left(\frac{a-1}{a},k\right)\bigg]
\end{equation}
where $a=(z-p)^2$ and $q=p^2-z^2$, and where $\Pi$ is the
complete elliptic integral of the third kind.  Note that
\begin{equation}\label{eq:noteqa}
\frac{q}{a} = \frac{p+z}{p-z}
\end{equation}
so as $z\rightarrow p$, $q/a\rightarrow\infty$. Likewise, the argument
of $\Pi$ also approaches $\infty$ as $z\rightarrow p$.  This probably
means that the complete elliptic integral goes to 0, thus as $z \rightarrow p$, 
the last term of $\lambda_1$ approaches $\infty \times 0$, (which 
is indeterminate). 
Hence, an asymptotic form of the function $\Pi$ would be required 
to determine what should happen at this 
limit.  In any event, the flux values on either side
of the point in time when the planet's limb touches the stellar
center might not line up smoothly.  Indeed, \citet{Ofir2018}
noted difficulties in their models of
transit time variations which require time derivatives of
finely-sampled transit profiles. \\

\begin{figure}[t]
\includegraphics[angle=-90,scale=0.70]{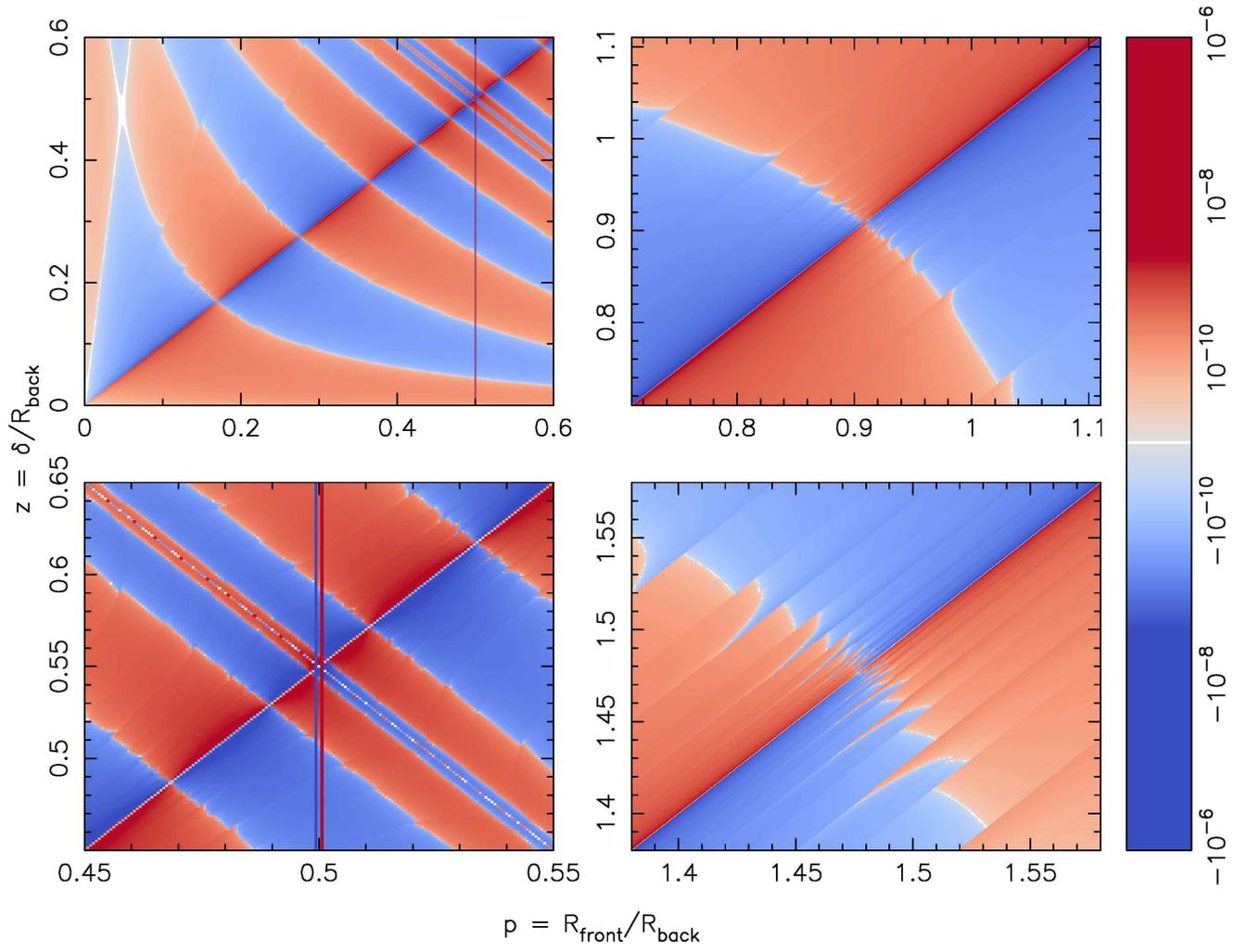}
\caption{Color maps of the MA2002 {\tt occultquad}
quadrature errors in four small regions in the
$(p,z)$ plane.  Red colors indicate a positive
difference ({\tt occultquad} flux is larger than the
$T_{\rm pan}=128$ flux fraction) and blue colors indicate a
negative difference.  Darker colors denote more extreme differences.
The errors are the largest along the lines $z=p$ and
$z=1-p$, and also nearby the vertical line $p=0.5$.
}
\label{fig:MAdiff02}
\end{figure}

Figure \ref{fig:MAdiff02} shows a color map of the
{\tt occultquad} quadrature errors in four small areas of
the $(p,z)$ plane.  The quadrature errors are generally quite small,
except along the lines $z=p$ and  $z=1-p$
(which are boundaries between various regions) and the
vertical line $p=0.5$. 
Given that regions 4, 5, and 7 are lines
in the plane and region 6 is a single point,
these regions will never be reached in an actual calculation.
The boundaries are ``fattened'' inside the {\tt occultquad} routine
by multiplying the terms of $z+p$, $z-p$, and $1-p$ by 
constants equal to $10^{-4}$ or 1.0001.  Large quadrature
errors would occur if a point in the $(p,z)$ plane that is nearby
a boundary is assigned to the wrong region.  \\

In the tests that are summarized in Figure \ref{fig:ploterrortrend1}, 
we have effectively computed eclipse profiles
for an excessively large number of points in time or orbital phase, and with
a large number of points problem areas in the $(p,z)$ plane will  be 
encountered.  In practice, of course,
one would use perhaps a few dozen to a few hundred
points per eclipse or transit event when modeling actual observational data.
So, in cases like these, how likely are we to encounter these
problem areas in the $(p,z)$ plane?  To answer this, we computed
light curves using
the {\sc ELC} code of \citet{Orosz2000}, which has implementations of
all three methods under discussion here.  This
makes it easy for us
to  make direct comparison of light curves computed using 
the various methods for cases that have no more than  two overlapping
bodies at any given time.  \\

\begin{figure}[t]
\plotone{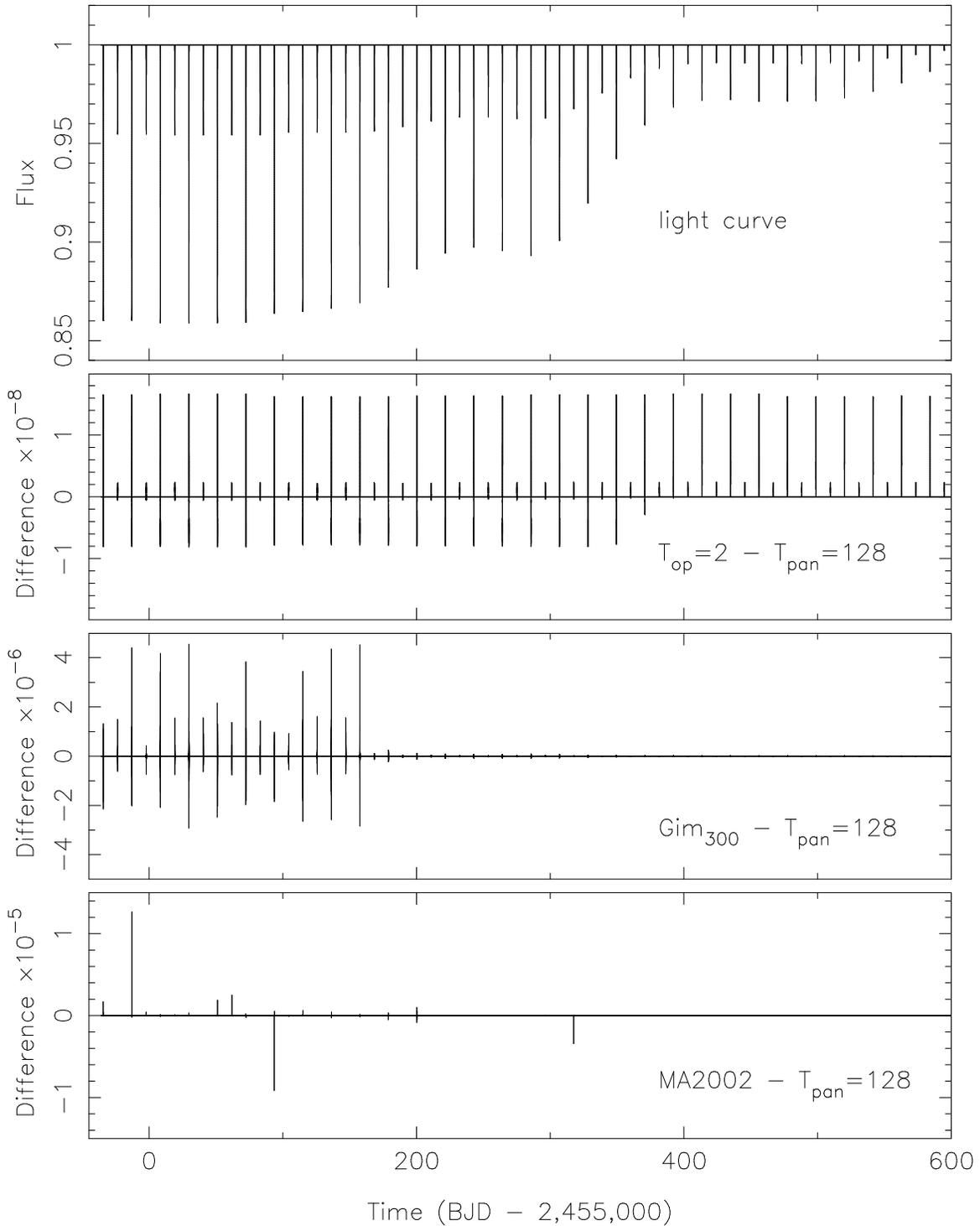}
\caption{
(a): The light curve of
KIC 10319590, computed
using a tolerance of $T_{\rm pan}=128$.  There is a third body in the system
whose dynamical influence causes rapid precession of
the binary, which in turn results in the decreasing
eclipse depths.  (b):
The difference between the light curves
computed with $T_{\rm op}=2$ and $T_{\rm pan}=128$.  The maximum difference is about
$2\times 10^{-8}$.  (c):
The difference between the light curves computed with
$T_{\rm pan}=128$ and the Gim\'{e}nez routine with $N_{\rm terms}=300$.
The maximum differences are about $4\times 10^{-6}$ early on
when the eclipses were deep, and are much smaller thereafter.
(d):  The difference between the light curves computed
with $T_{\rm pan}=128$ and the MA2002  {\tt occultquad} routine.
In this case the maximum difference is about $10^{-5}$.}
\label{fig:difffig1}
\end{figure}

\begin{figure}[t]
\plotone{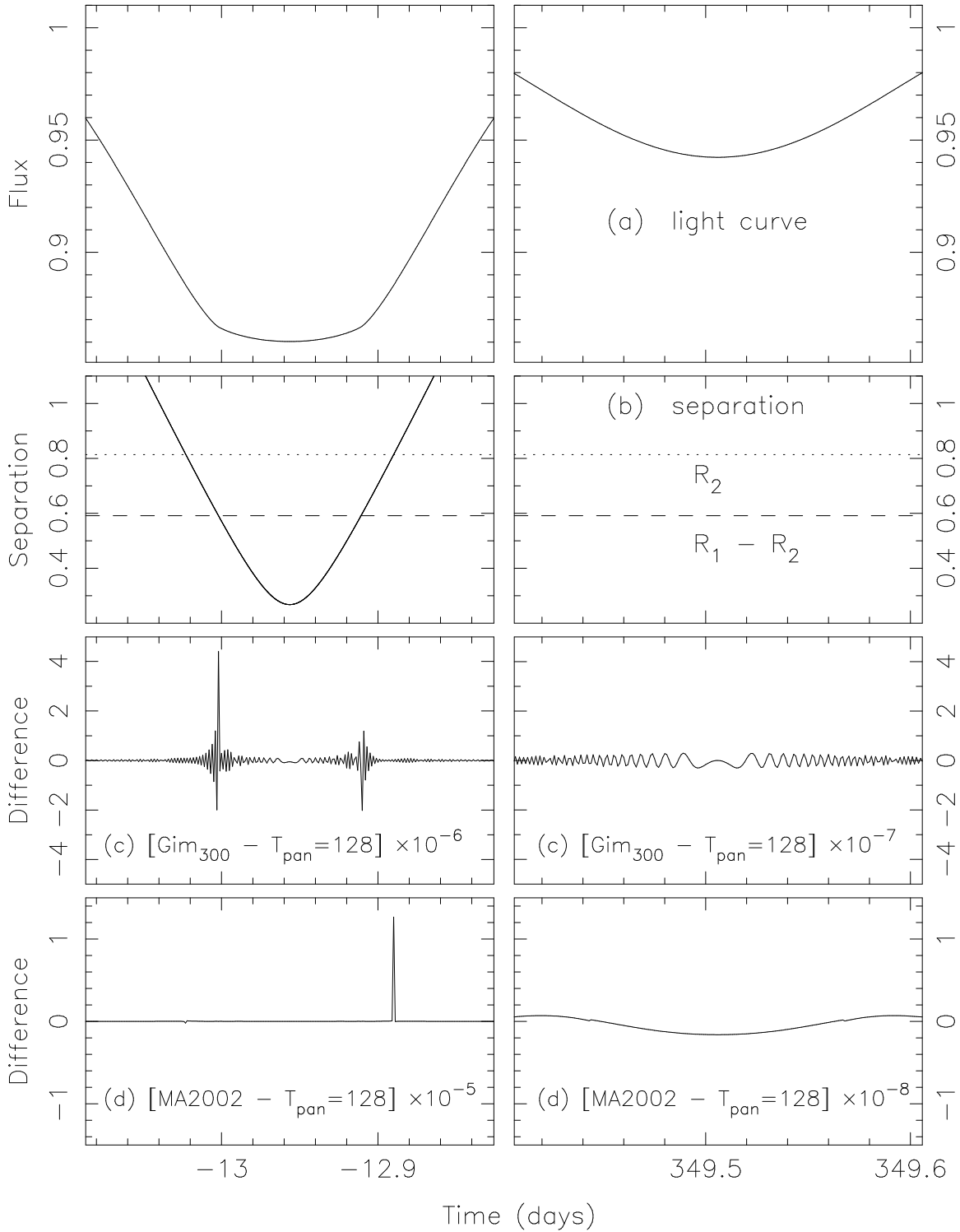}
\caption{
(a): The light curve of
KIC 10319590 from Figure \protect{\ref{fig:difffig1}} near the time
of the first primary eclipse (left) and near the time of a
primary eclipse about 385 days later when the primary
eclipses were much shallower (right). 
(b):  The separation of the centers in
$R_{\odot}$ as a function of
time.  The dotted line denotes the radius of the
secondary, and the intersections of the dotted line
with the solid line mark the times when the secondary
is about to pass over the center of the primary.
Note that $z=p$ at that time.
The dashed line denotes the
difference between the radii, and the intersections of the
dashed line and the solid line marks the times of second
and third contact. 
Note that for the panel
on the right, the separation of centers was much larger so the
solid line does not appear.
(c):  The difference curve between the
$T_{\rm pan}=128$ model and the Gim\'{e}nez model.  For the deeper eclipse near
day $-34.3$ the maximum difference of about $4\times 10^{-6}$ occurs
at the time of the second and third contact points.
(d):  The difference curve between the $T_{\rm pan}=128$ model
and the MA2002 model.  In this case the maximum
difference of $10^{-5}$ for the
deeper eclipse happens near the time when
the secondary star passes over the center of the 
primary star as seen on the sky plane
(this is when $z=p$).  
}
\label{fig:difffig2}
\end{figure}

 We start with the
triple system KIC 10319590 as our first test case.  
This system is one of the more
interesting eclipsing binaries discovered by the {\em Kepler} mission
as the eclipses disappeared over a $\approx 600$ day span near the
start of the mission, owing to the 
influence of a third body companion \citep{Orosz2015}.
Thus we have a convenient way to compute eclipses with different 
impact parameters.  We computed the light
curve over a 686 day span  using the new method with $T_{\rm op}=2$, the
new method with $T_{\rm pan}=128$, the {\tt occultquad} routine, and
our Gim\'{e}nez routine with $N_{\rm terms}=300$.  For this
demonstration we set the initial inclination of the binary
to $89.6^{\circ}$ instead of its best-fitting value of
$88.84^{\circ}$ in order to produce deeper eclipses.
Figure \ref{fig:difffig1} shows the light curve computed with
$T_{\rm pan}=128$, and the difference between that light curve and the three others.
The maximum difference between the $T_{\rm op}=2$ light curve and the
$T_{\rm pan}=128$ light curve is about $2\times 10^{-8}$.
The maximum difference between the $T_{\rm pan}=128$ light curve and the
Gim\'{e}nez light curve is about $4\times10^{-6}$.  Finally,
the maximum difference between the $T_{\rm pan}=128$ light curve and the
{\tt occultquad} light curve is about $10^{-5}$.  Note that the
spikes in that difference curve extend to much later times
than the spikes in the difference curve for the Gim\'{e}nez model
do.
Figure \ref{fig:difffig2} shows a close-up on the difference
curves for the first primary eclipse and a (much shallower) primary
eclipse near day 349.  We see that the maximum 
difference between the $T_{\rm pan}=128$ model and the Gim\'{e}nez model 
of $4\times 10^{-6}$ is near
the times of second and third contact.
The maximum difference between the $T_{\rm pan}=128$ model and the {\tt occultquad} 
model occurs near the time when the secondary star is about to pass
over the center of the primary star
(that is when $z=p$).  Thus there will be transition from one region to
another one in the $(p,z)$ plane..
At some time around
day 320, owing to the precession of the binary caused
by the tertiary companion, the separation of centers on the POS 
has increased to the point where the secondary
star can no longer pass over the center of the primary star (i.e.\
$z>p$).  There is only a single region in the $(p,z)$ plane and
consequently the large spikes in the difference curve for the Mandel \& Agol
model go away at this time. \\

To get a much more thorough idea of the differences between the three
methods and how often one might encounter problem areas in practice, 
we considered five multi-body systems that were discovered by 
{\em Kepler}:
(i) Kepler-16, a circumbinary planetary system
which has relatively deep transits of a planet across
the primary star and total secondary eclipses
\citep{Doyle2011};
(ii) KOI-126, a triple star system
which has transits of two relatively low-mass stars
across a much larger star \citep{Carter2011};
(iii) KIC 8610483, a circumbinary planetary system
which has
$\approx35\%$ deep  primary eclipses and $\approx 13\%$ deep
secondary eclipses;
(iv) KIC 7289157, a multiple star system
which has primary, secondary, and tertiary eclipses
and occultation events, all of which decrease with time \citep{Orosz2015}
(v) KIC 7668648, a triple star system
which has primary, secondary, and tertiary eclipses
and occultation events, all of which increase with time \citep{Orosz2015}.
For each system, we computed 5000 models for these of five cases:
(a) the new method with $T_{\rm pan}=128$;
(b) the new method with $T_{\rm op}=1$;
(c) the new method with $T_{\rm op}=2$;
(d) the Gim\'{e}nez routine with $N_{\rm terms}=300$;
and (e) the MA2002 {\tt occultquad}
routine.  
The parameters for the 5000 models
came from various optimization runs we have done.
We assume the case (a) models (with $T_{\rm pan}=128$) are ``exact'', and compare
the case (b) through (e) models against them.  For each pair
of models, we compute the absolute values of the
difference between the models and record the median 
difference value (excluding differences of zero)
and the maximum difference value.  For each system,
we look at the distributions of the median differences
and the distributions of the maximum differences for the
cases of ``a-b'', ``a-c'', ``a-d'', and ``a-e''.
The typical {\em maximum} difference (in absolute value) is perhaps the more useful statistic
when comparing the different methods, and we refer to this as the error. \\

\begin{figure}[t]
\plotone{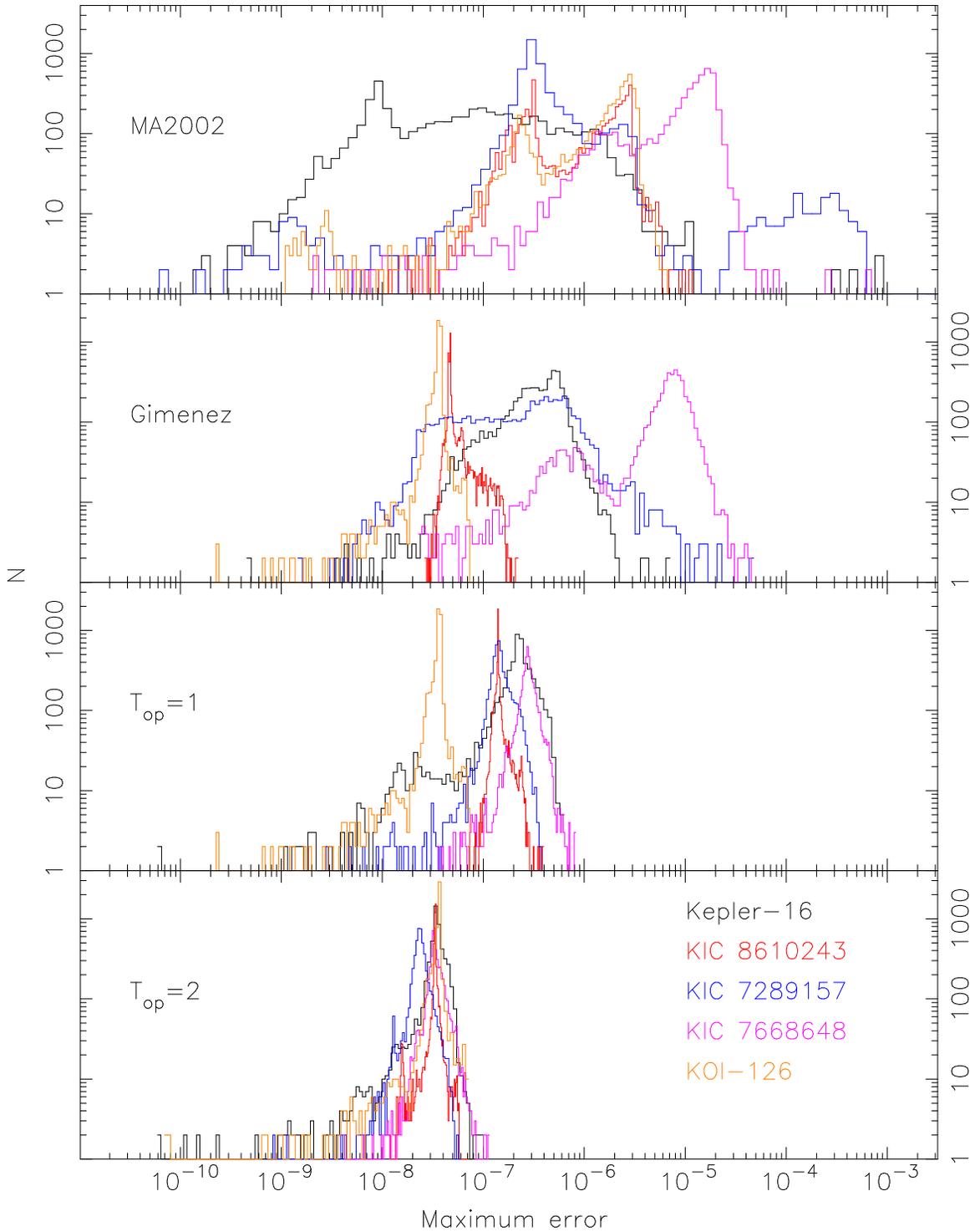}
\caption{The statistics for the comparisons between model light curves
generated using the MA2002 {\tt occultquad} (top), the \citet{Gimenez2006a} routines
(second from top), models with $T_{\rm op}=1$ (second from bottom), 
and models $T_{\rm op}=2$ (bottom).  The histograms show the distribution of
the maximum errors found for 5000 models of Kepler-16
(black lines), KIC 8610243 (red lines), KIC 7289157
(blue lines), KIC 7668648
(cyan lines), and KOI-126 (orange lines).
}
\label{fig:maxerrorhist}
\end{figure}

Figure \ref{fig:maxerrorhist}
shows the distributions for the {\em maximum} differences. 
Over the five sources, {\tt occultquad} gives the largest
spreads in the maximum differences, where the maximum difference in many
of the models can be
as small as $10^{-10}$, or as large as  
$10^{-3}$.   The sources KIC 7668648 and
KIC 7289157 were the most problematic for {\tt occultquad},
as there was a large number of models where the
maximum difference was between $10^{-5}$ and $10^{-3}$.
The Gim\'{e}nez routine gave the next largest spread of
values, where the maximum error encountered was just over
$10^{-5}$. As was the case with
{\tt occultquad}, the sources KIC 7668648 and
KIC 7289157 caused the most difficulties.
The set of models with $T_{\rm op}=1$ and $T_{\rm op}=2$ give the tightest
ranges in the maximum error.  When $T_{\rm op}=1$, the maximum error seen is 
$6\times 10^{-6}$ and when $T_{\rm op}=2$, the maximum error seen is
just over $10^{-7}$, in agreement with the extensive tests performed
above.  \\

\begin{figure}[t]
\plotone{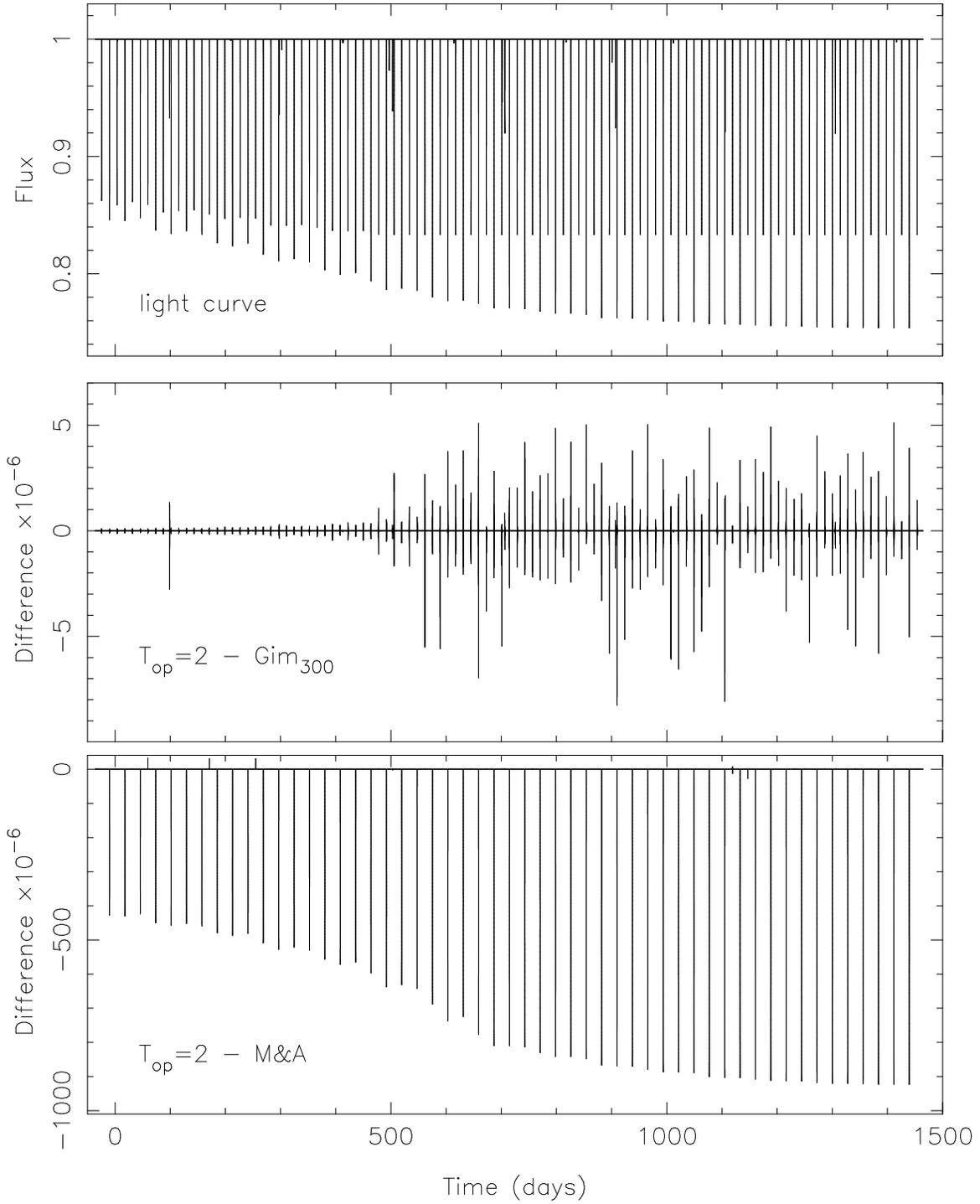}
\caption{Top: The light curve of a system resembling KIC 7668648,
computed with $T_{\rm op}=2$.  Middle: The difference between the $T_{\rm op}=2$ model
and the model computed with the Gim\'{e}nez routine with
$N_{\rm terms}=300$.  The largest differences are about $7\times
10^{-6}$.  Bottom:  The difference between the $T_{\rm op}=2$ model and
the model computed with the MA2002 {\tt occultquad}
routine.  In this case, the differences are systematically large
and approach $10^{-3}$.
}
\label{fig:kid7668diff}
\end{figure}

\begin{figure}[t]
\plotone{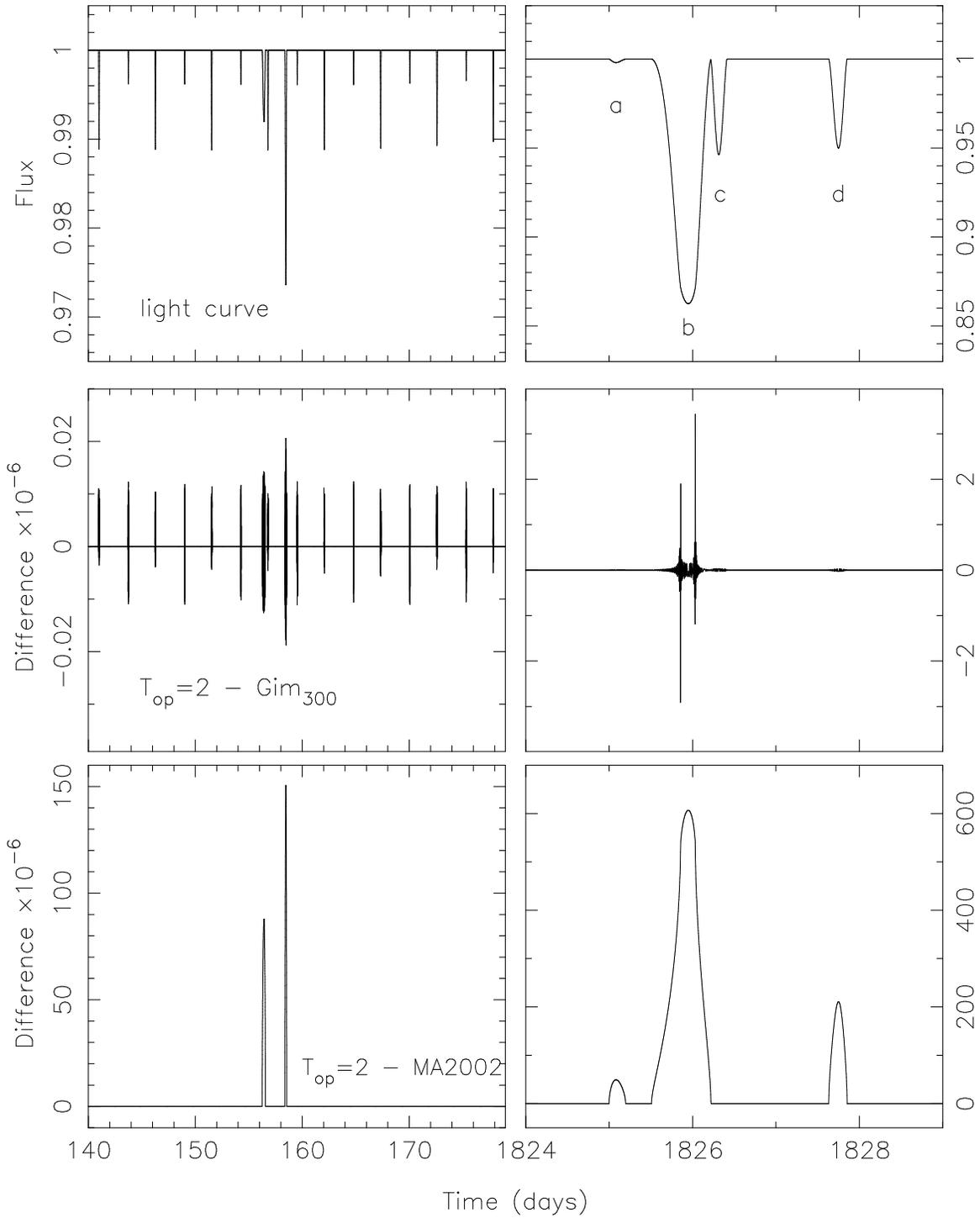}
\caption{Top: The light curve of a system resembling
KIC 7289157, computed with $T_{\rm op}=2$.
On the left the light curves at early times has primary,
secondary, and tertiary eclipses.  On the right the light curve
at late times has only occultation events of the third star by the
stars in the binary.  The event labeled c is an occultation of the third star
by the first star, and the other three events (labeled a, b, and d) are
occultations of the third star by the second star.
Middle: The difference between the model with $T_{\rm op}=2$ and the model
computed with the Gim\'{e}nez routine with $N_{\rm terms}=300.$
The largest deviations are a few parts per million.  Bottom:
The difference between the model with $T_{\rm op}=2$ and the model
computed with the MA2002 {\tt occultquad} routine.
In this case, the differences are systematic and are several hundred parts per million.
}
\label{fig:kid7289diff}
\end{figure}

\begin{figure}[t]
\includegraphics[angle=-90,scale=0.7]{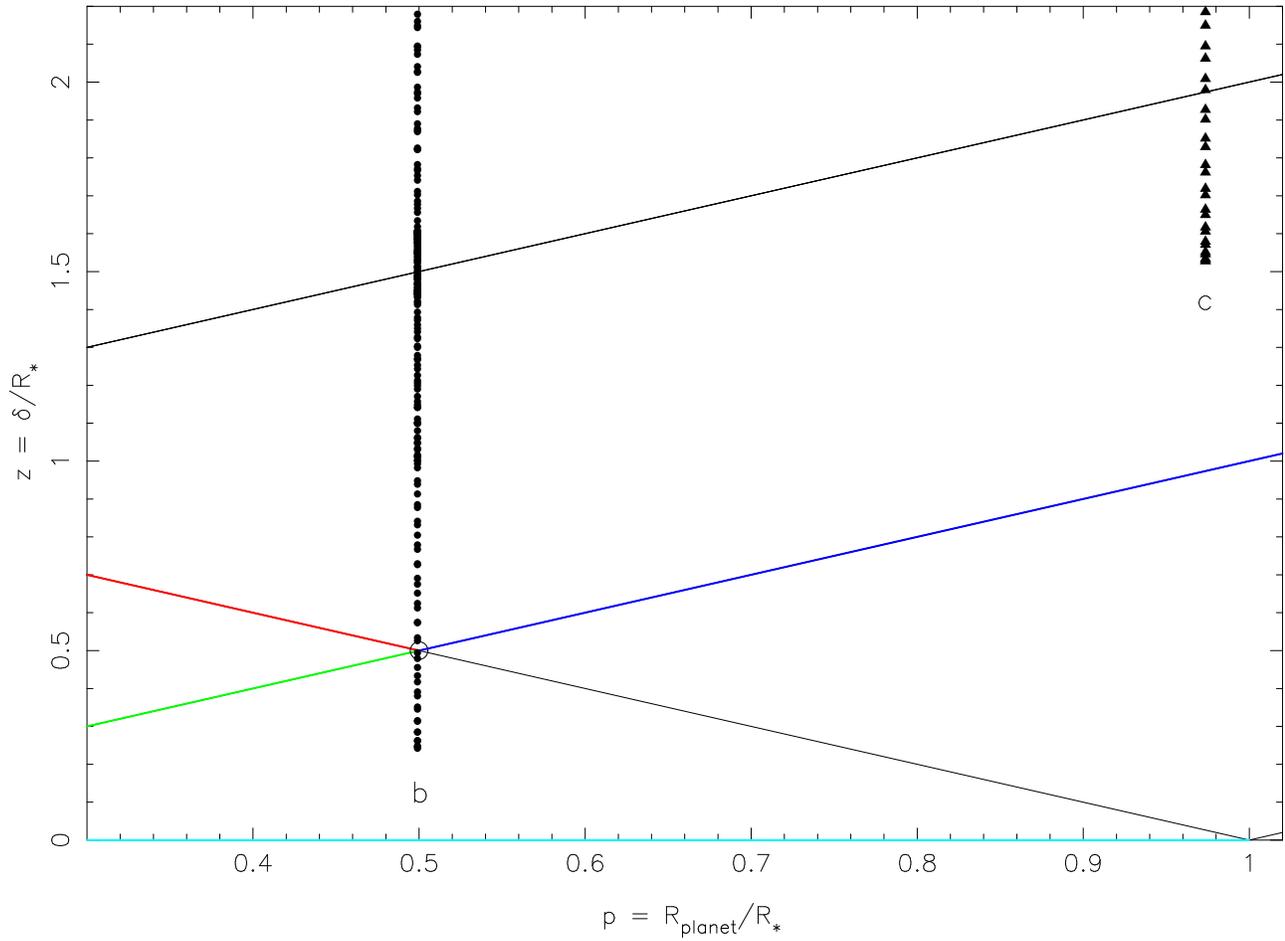}
\caption{Left:
The $(p,z)$ plane showing the various cases used to compute eclipse
events in the MA2002 algorithm.
The filled circles show the location of the points in the event labeled
b in the upper right of Figure \protect{\ref{fig:kid7289diff}}.
The filled triangles show the location of the points in the event
labeled as c in the upper right of 
Figure \protect{\ref{fig:kid7289diff}}.  In the case for event b,
the ratio of radii is 0.49902, so the points pass very close to the
intersections of several regions.
}
\label{fig:drawMAregions1}
\end{figure}

These simulations show that while {\tt occultquad} generates
lightcurves with small errors most of the time, there will be a non-negligible 
number of models where there is a large error somewhere.
In addition,
during the course of these simulations, we found a few
cases where the differences between the models computed with
the new method and models computed with {\tt occultquad}
were very large and systematic (i.e.\ not simply
a ``spike'' in the difference curve).  Figure \ref{fig:kid7668diff} 
shows an example
for KIC 7668648, where the differences
between a model (computed with $T_{\rm op}=2$) and a model computed with {\tt occultquad}
approach one thousand parts per million.
This particular model has $R_1=1.77666 \,R_{\odot}$
and $R_2=0.88963\,R_{\odot}$, which gives $p=0.50073$.
This value of $p$ is very close to $p=0.5$, which is where
several regions intersect
in the $(p,z)$ plane.
Similarly, Figure \ref{fig:kid7289diff} shows an example
for KIC 7289157, where the differences
between a model (computed with $T_{\rm op}=2$) and a  model computed  with {\tt occultquad}
are several hundred parts per million.  Note that in
both cases the differences between our models (with $T_{\rm op}=2$) and
the Gim\'{e}nez models (with $N_{\rm terms}=300$)
are smaller (in absolute value) than about 7 parts per million.
This is shown in the middle panels of each figure.
The upper panel of Figure \ref{fig:kid7289diff} shows occultations
of the third star by the second star
(labeled a, b, and d) in KIC 7289157, and an occultation
of the third star by the first star (labeled c).  Interestingly,
the differences between our method and {\tt occultquad} 
of events a, b, and d shown in the
lower right panel of Figure \ref{fig:kid7289diff} are large,
whereas the differences for event c are small.
In Figure \ref{fig:drawMAregions1}
we plot the locations of these events in the $(p,z)$ plane.
For event b, the $p$ co-ordinate is
0.499012, and there are 
points that fall within {\tt occultquad} regions 9, 3,
and 2.  In addition, the lines denoting regions 5 and 4
are crossed, and a few of the points come very close to the single
point that is region 6.  We noted earlier 
that regions 4 and 5 are lines
in the plane and will never be reached in an actual calculation, so
the boundaries are ``fattened'' inside the {\tt occultquad} routine
by multiplying various factors of $z+p$, $z-p$, and $1-p$ by 
constants equal to $10^{-4}$ or 1.0001, depending on the case.  
Thus it seems likely that for the light curve shown
in Figure \ref{fig:kid7289diff}, incorrect regions in the $(p,z)$
plane were used for the occultation events a, b, and d.
On the other hand,  the eclipse event labeled c falls in the middle of
region 2 far away from the boundaries of other regions
(apart from region 1 where there is no overlap).  As a result the light
curve for this event computed by {\tt occultquad}
closely matches the light curve computed with $T_{\rm op}=2$. \\

\begin{figure}[h]
\includegraphics[angle=-90,scale=0.7]{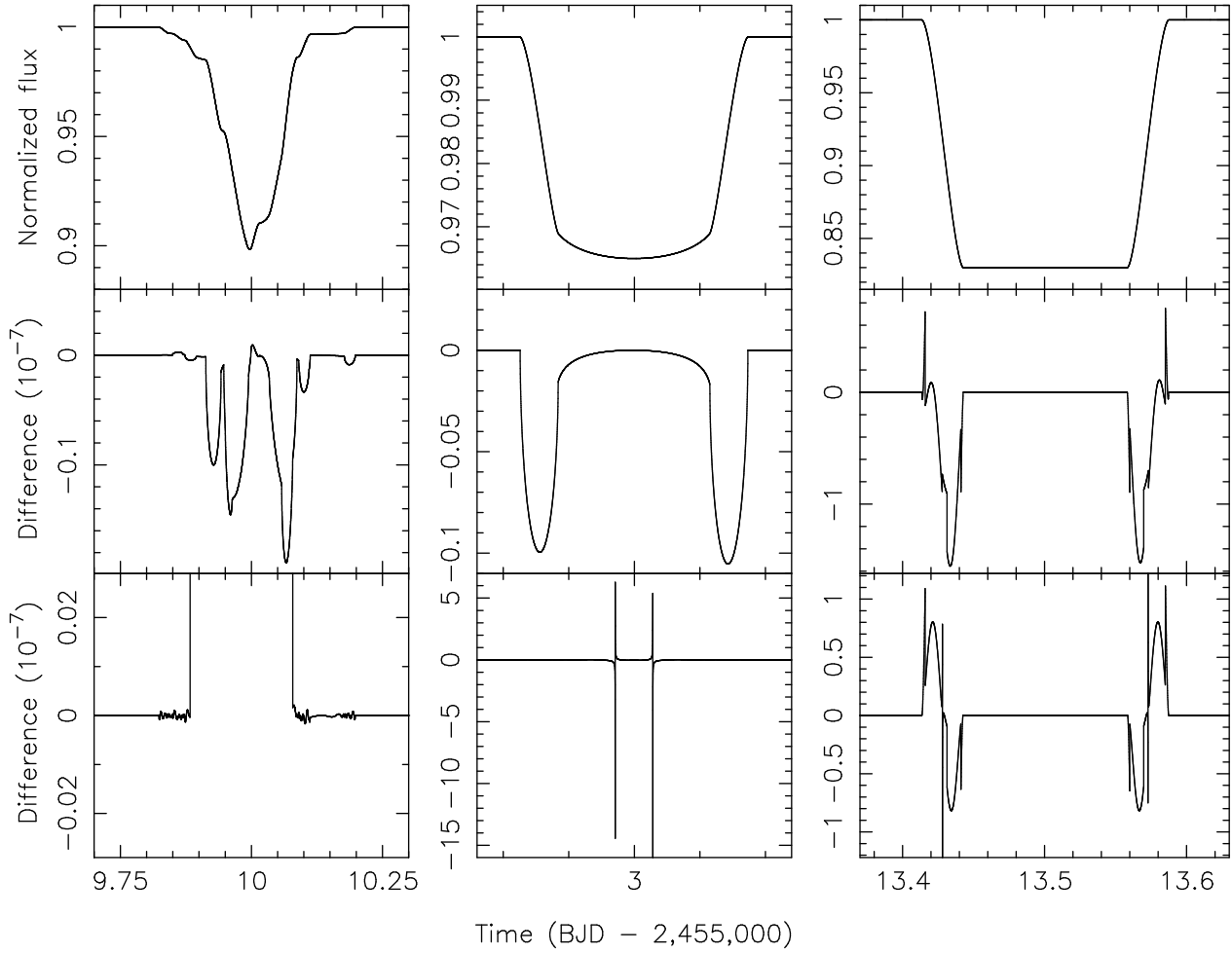}
\caption{Top:
The light curve of the mock 5-body system shown in Figure
\protect{\ref{fig:Jerrys5BodyConfig}}, calculated using
$T_{\rm pan}=128$, during the syzygy
event (left), a primary eclipse (middle), and a total secondary
eclipse (right).  
Middle:  The difference
(in units of $10^{-7}$) between our light curve and the one
calculated with {\sc Photodynam}.  
Bottom:  The difference
(in units of $10^{-7}$) between our light curve
and the one calculated with 
{\tt occultquad} (within {\sc ELC}).  Note the
residuals for each method for the secondary
eclipse are reasonably similar, but not so for the primary eclipse.
}
\label{fig:palcompare}
\end{figure}

It is less straightforward to directly compare our method to that of
\citet{Pal2012} since the subroutines provided by P\'{a}l
are written in {\sc C}, and {\sc ELC} is written in {\sc FORTRAN}.  
In order to make a few direct comparisons between the two methods, we used 
Josh Carter's {\sc Photodynam} code \citep{Carter2011,Doyle2011}, 
which is publicly
available\footnote{https://github.com/dfm/photodynam}. 
{\sc Photodynam} takes initial conditions in the form
of orbital elements and integrates the Newtonian equations of
motion to produce POS co-ordinates and velocities.  When
given POS positions and the radius of each body, it then computes
the flux fractions using P\'{a}l's subroutines  
(which use the quadratic limb darkening law).
We  computed the light curve of the example 5-body system introduced in Section 
\ref{sec:method_conversion} using both codes 
and took the difference between the two (a tolerance of $T_{\rm pan}=128$ was used for the 
light curve computed using our method). 
The results are shown in
 Figure \ref{fig:palcompare}. For the syzygy 
event involving five bodies, and for a primary eclipse
involving two bodies, the maximum difference is on the order of $10^{-8}$.
For a total secondary eclipse, the maximum difference is on the order
of $10^{-7}$.  The difference curves are not quite
symmetric about zero, and we attribute this to small discrepancies  
in the positions of the bodies as a function of time, since
{\sc ELC} and {\sc Photodynam} use different methods to
solve the Newtonian equations of motion (the initial barycentric
co-ordinates agree to better than $3\times 10^{-14}$ AU, and 
the initial POS co-ordinates, corrected for light travel time, 
agree to better than $5\times 10^{-9}$ AU).  \\

When examining the differences between our method and that of
\citet{Pal2012}, it is instructive to look at the difference curve
made from our $T_{\rm pan}=128$ model and the model computed with
{\tt occultquad} within {\sc ELC}.  While {\tt occultquad} cannot
be used for true syzygy events, it works well close to the very
start and the very end of the event near day 10, where the light curve
is the superposition of several individual transits.  
For the total secondary eclipse, the difference curve
between our method and P\'{a}l's method (middle right panel in Figure
\ref{fig:palcompare}) looks broadly similar to the difference curve between our method
and {\tt occultquad} (lower right panel
in Figure \ref{fig:palcompare}).  \citet{Pal2012} noted 
similarities between terms
in his expression for the flux fraction (his Equation 34 for
the linear limb darkening law) and those of MA2002.  
Only one region (Region 8) is required in {\tt occultquad} to compute 
the ingress and egress of the total eclipse, hence only one expression is used 
for the flux fraction. 
In this case, the difference curves look similar.  On the other hand,
several regions are required in {\tt occultquad} to compute
the primary eclipse, so several expressions are pieced together
to produce the transit profile.  In that case, the difference
curves look rather dissimilar and completely different in scale. \\

\begin{figure}[h]
\includegraphics[angle=-90,scale=0.7]{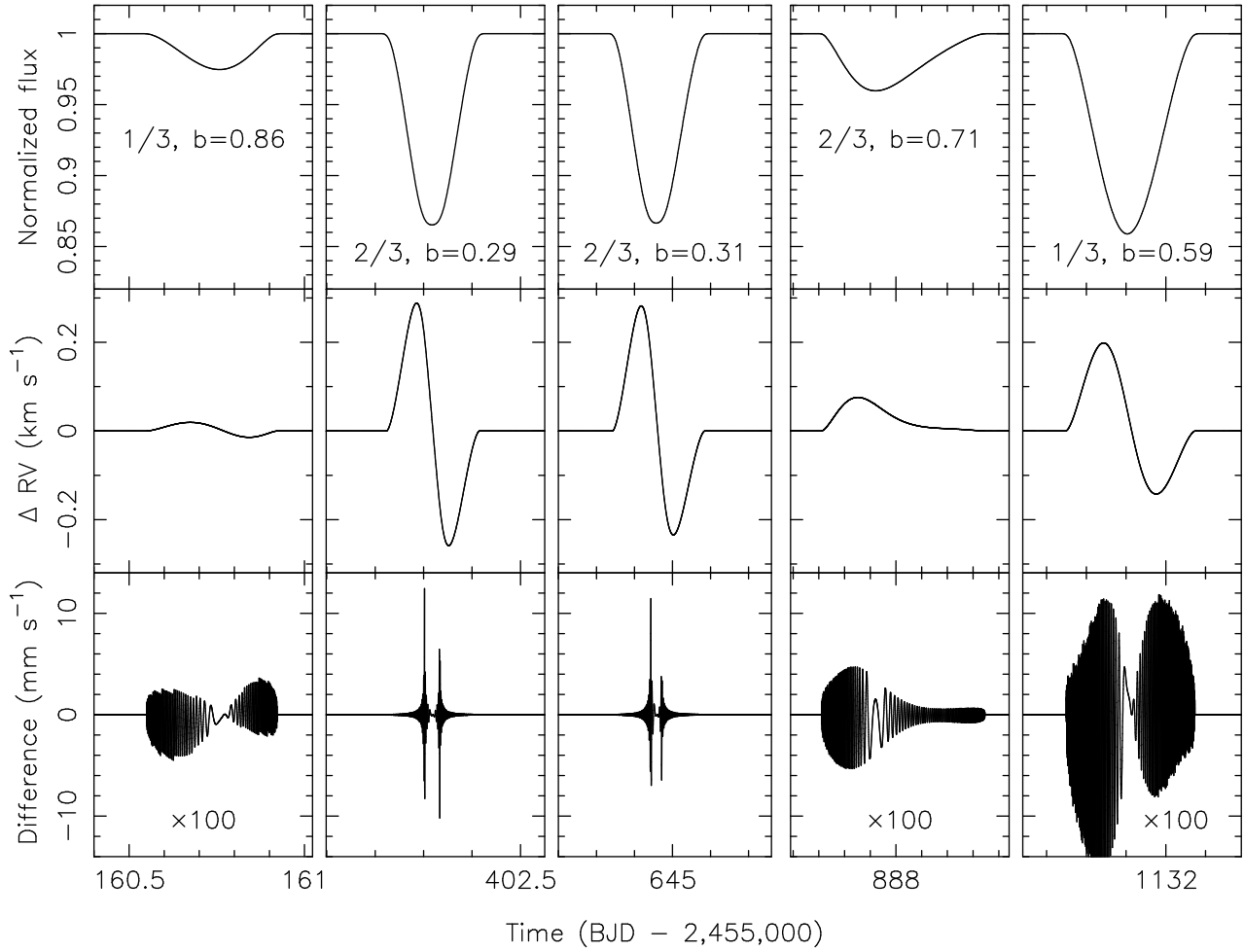}
\caption{Top:
The light curve of five tertiary events in KIC 7289157 where
either the primary of the binary passes in front of
the third star (labeled 1/3) or the secondary of the
binary passes in front of the third star
(labeled 2/3).  The impact parameters
of each event are given.  
Middle:  The R-M curves for each event, where the orbital  
motions have been subtracted.
Bottom: The difference between the R-M curves
computed with our method using $T_{\rm op}=3$ and those computed
using the method of \citet{Gimenez2006b} with $N_{\rm terms}=300$.
Note the vertical scale on the bottom
panels is in ${\rm mm~s^{-1}}$.
}
\label{fig:RVcompare01}
\end{figure}

We can compare the R-M curves computed with our method
to those computed using the method of \citet{Gimenez2006b}.  For the first
comparison we used KIC 7289157.  The third star in this system
(with $R=1.08\,R_{\odot}$), which is 
about as bright as the primary in the binary, is
eclipsed by the primary and secondary stars at a variety of
different impact parameters.  Not knowing the orientation
of the rotation axis of the third star, we assume
$\Phi_{\rm rot}=90^{\circ}$ and $\Theta_{\rm rot}=0$.  These
angles would make the rotation axis of the star slightly
misaligned with its orbital plane about the barycenter.  Furthermore,
we assume this star rotates 10 times faster than its 
orbital period of 243 days.  Figure \ref{fig:RVcompare01}
shows five tertiary events in KIC 7289157, where either
the primary passes in front of the third star (1/3 in the figure), 
or the secondary passes in front of the third star (2/3 in the figure).
The impact parameters are given.  The amplitudes of the R-M
signal range from about 50 m s$^{-1}$ to about 250 m s$^{-1}$.  
Note the variety of shapes. The difference between the R-M curves computed
with our method and those 
computed with the Gim\'{e}nez (2006b) method are shown as well.  
The largest differences in absolute value are about 
$10^{-2} {\rm m~s^{-1}}$ for the second and third events, 
and $\approx 100$ times smaller for the other three.  \\

\begin{figure}[h]
\includegraphics[angle=-90,scale=0.7]{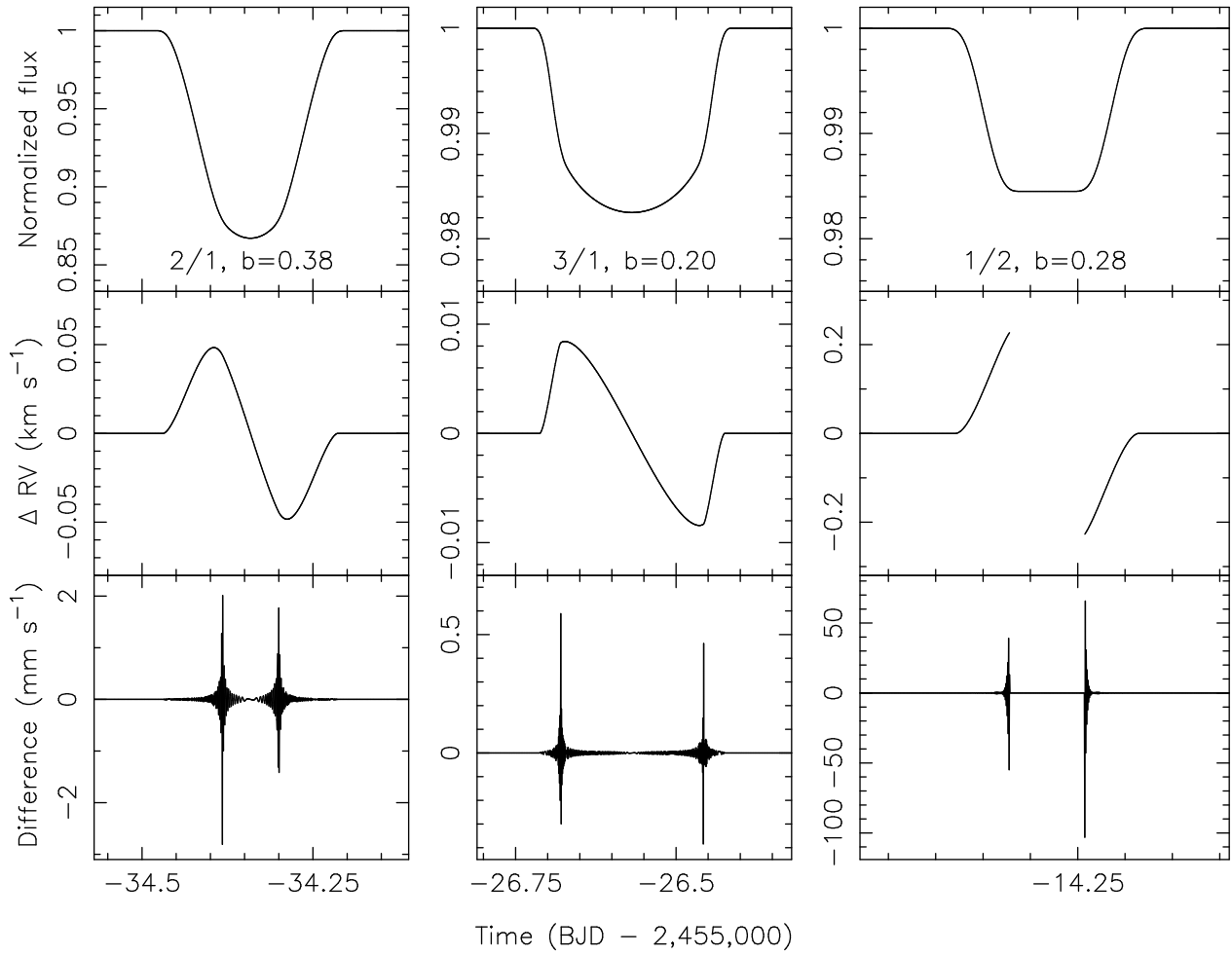}
\caption{Similar to Figure \protect{\ref{fig:RVcompare01}},
but for Kepler-16.  From left to right, we show a primary
eclipse, a planet transit across the primary, and a (total)
secondary eclipse.
}
\label{fig:RVcompare02}
\end{figure}

We also computed R-M curves for Kepler-16, specifically for a primary eclipse, 
for a planet transit of the primary, and for a (total) secondary eclipse.
Figure \ref{fig:RVcompare02} shows the light curves,
the R-M curves, and the difference curves.
The R-M effect for the primary 
star has an amplitude of $\approx 50 ~{\rm m~s^{-1}}$, 
and was observed by \citet{Winn2011}.
The R-M signal for the secondary star is much larger, but given
that the secondary star is only a few percent as bright
as the primary star, the R-M signal for the secondary is probably not 
observable.  Note that the 
R-M signal is not defined during totality since the secondary is
completely hidden by the primary.
The differences in absolute value between the two methods for the primary star events
are a few mm s$^{-1}$, which seems to be sufficiently small.  
For the total secondary eclipse, the absolute value differences are about 100 mm s$^{-1}$ 
near the onset of totality.

\section{Worked Example: KOI-126}\label{sec:example_koi126}

\begin{figure}[h]
\includegraphics[angle=-90,scale=0.8]{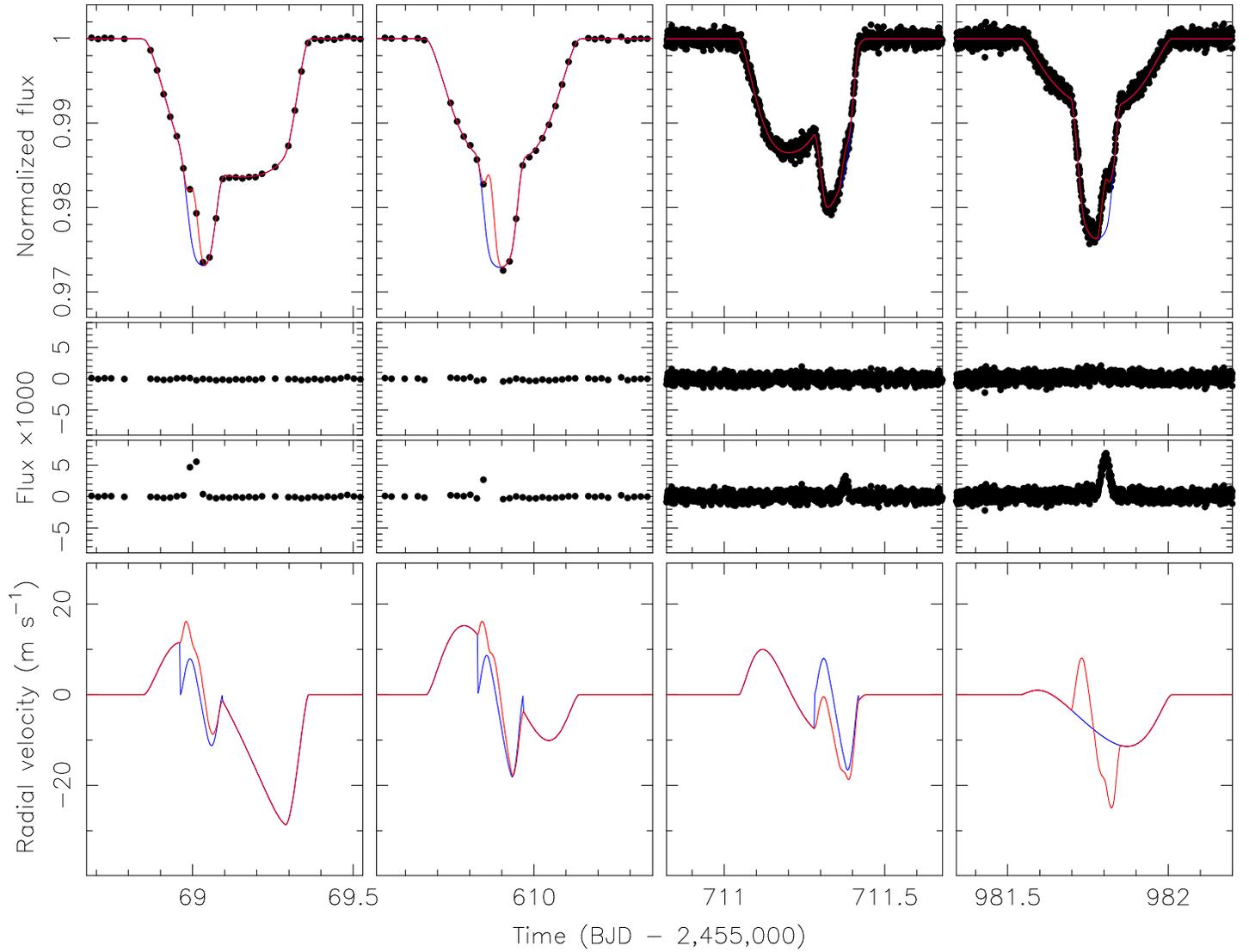}
\caption{Top: The light curves of KOI-126
showing four syzygy events.
The two panels on the left show the {\em Kepler}
long-cadence data, and the other two panels
show {\em Kepler} short cadence data.
The red lines are the model computed with 
our new method, and the blue line shows
the model computed using the \citet{Gimenez2006a}
routine for the same model parameters.
Middle:  The O-C residuals computed
for each model.  
Bottom:  The R-M signal computed for each 
transit event using our new method
(red line) and the \citet{Gimenez2006b} method
(blue line).
}
\label{fig:rossiter_koi}
\end{figure}

Let us return to the system KOI-126 (KIC5897826), a low-mass, short-period binary ($P=1.74$ days) in eccentric 33.5 day orbit about a much larger $1.35M_{\odot}$ F-star star \citep{Carter2011}. 
The system has several distinct syzygies.   
We computed models for these syzygy events using both
our new method and the \citet{Gimenez2006a} method, which will not work well for the syzygies. 
These models are shown in Figure \ref{fig:rossiter_koi}.
In all of the cases shown, our model, which accounts
for the mutual overlap between the three bodies, provides a
significantly better match to the observations.  We also
show in the figure the expected R-M effect signal, where
the orbital motion of the F-star has been removed.  
The signals are complex and have a maximum
velocity displacement on the order of 30 m s$^{-1}$, which
may be detectable. \\

\begin{figure}[ht!]\
\plotone{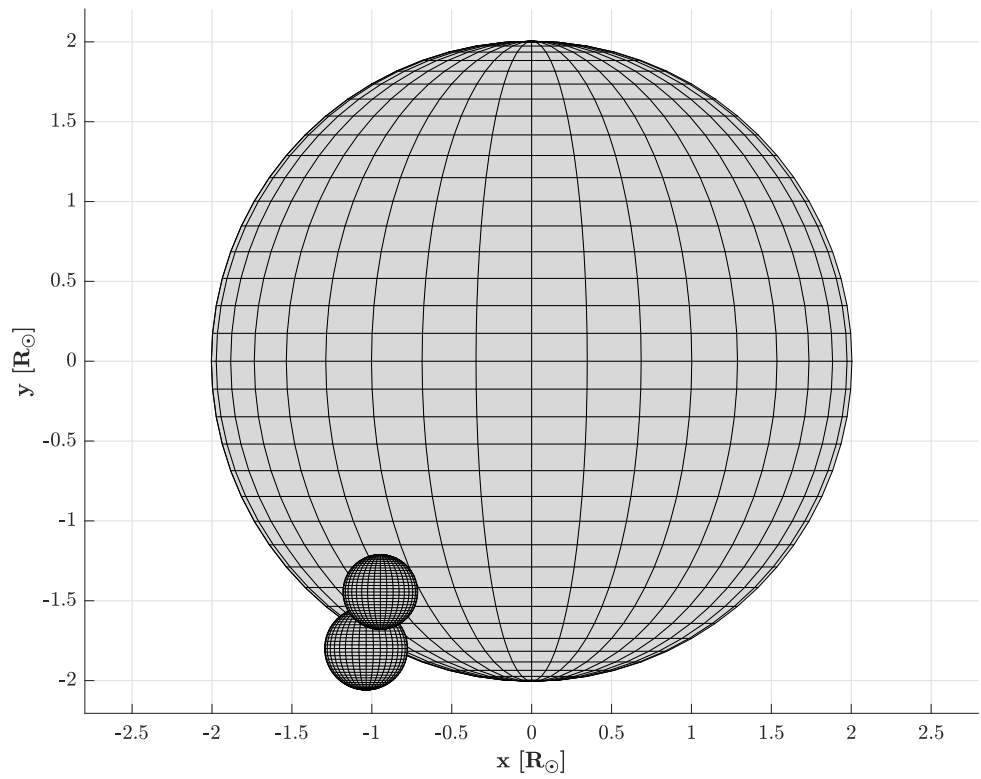}
\caption{The POS view of the three bodies in KOI-126, during the syzygy event at day BJD 2455711.38. The axes are given in Solar radius for convenience. The configuration's origin is the center of the back body, and the foreground binary is moving to the left}
\label{fig:KOI126_3bodycross}
\end{figure}

\begin{figure}[ht!]
\plotone{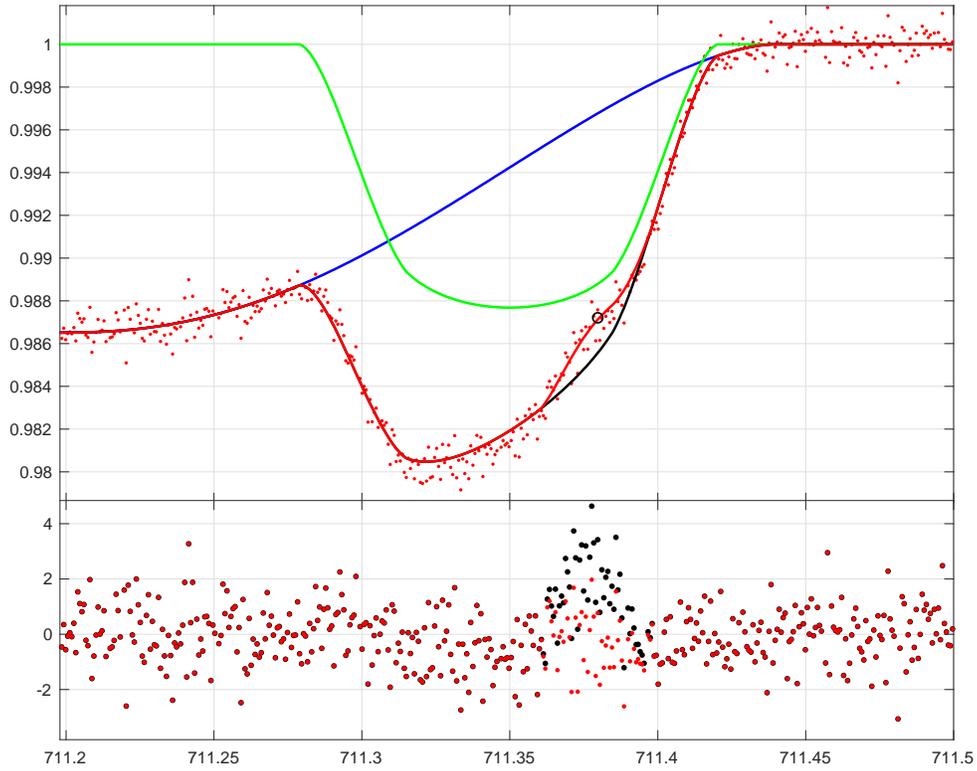} 
\caption{Modeling of a syzygy event in KOI-126, where the open circle marks the time corresponding to the configuration shown in Figure \ref{fig:KOI126_3bodycross} and the example computation we presented in Section \ref{sec:example_koi126}. The red dots show the normalized $Kepler$ short cadence data, where only a small part of the whole transit event is shown (the $x$-axis gives the time in days since BJD 2455000). The red line shows our model that accounts for mutual overlaps of the transiting stars; the blue line shows the expected light curve for the primary star transit of the third star; the green line shows the expected light curve for the secondary star transit of the third star; the black line is a simple combination of the two events, which would ignore overlaps between the primary and the secondary. The bottom panel shows the O-C residuals using the full model (red points) and the model that ignores overlaps (black points). The $\chi^2$ of the fit using the model that accounts for mutual overlaps improves from 976.1 to 849.3. As one might expect from the large feature in the residual near day $711.35$, the improvement in the fit is very significant.}
\label{fig:KOI126_syzygy711_fit}
\end{figure}

To demonstrate the computation of the fractional flux we choose the time in the third syzygy event near 
BJD 2455711.38. Figure \ref{fig:KOI126_3bodycross} shows the POS view of this event. We start this computation with the POS co-ordinates of the three bodies and their radii,  provided in Table \ref{tab:planeskycoord}. In \S\ref{sec:examplearcs} we will show the vector stacks that are the bounding curves, denoted $Arc_{MN}$, for the visible and shadow paths of the three bodies, calculated as explained in \S\ref{sec:boundingcurve}. Note especially our description in \S\ref{sec:boundingcurve} and in Appendix \ref{sec:appendix_workedexample} of the process that allows us to generate the visible path. In \S\ref{sec:exampleintegral} we will describe how the visible path  is integrated and summed over these arcs using Equation (\ref{eq:limbdarkintegral_new}). We will show in detail the calculation of the first and second integrals in this process. Figure \ref{fig:KOI126_syzygy711_fit} shows the resultant synthesized light curve superimposed on the $Kepler$ data with and without modeling the overlap. We note the significant reduction in the $\chi^2$ fit to the data by properly treating the syzygy. \\

\subsection{The Boundary Paths} \label{sec:examplearcs}

Table \ref{tab:planeskycoord} lists the POS co-ordinates. They are arranged from the closest to the farthest along our line of sight. Note that in our co-ordinate system, positive z is toward the observer. \\

Following the method as outlined in \S\ref{sec:boundingcurve}, 
we first need to describe the visible region of each body, beginning with 
the front body, and iteratively determine the visible path for each body as we proceed
along the line of sight. Each step in this process has two parts: updating the shadow path formed by 
all of the preceding bodies, then using that shadow path to determine the visible path 
of the current body.  As described in  \S\ref{sec:boundingcurve}, the visible path and the shadow path 
contain arcs that are stored in a vector stack. See specifically Equation (\ref{eq:Arc_NM}) and Equation (\ref{eq:arc_ij}) for the notation. Since the front body is unobstructed, the bounding arc for it is the entire circumference as is the
shadow that it casts. \\
\begin{align}
\begin{aligned}\label{eq:vispath1_KOI126}
VisiblePath(Body1)=
\begin{bmatrix}
\vspace{0.02in}1 & 1 & 0 & 360 & 360
\end{bmatrix} \\
ShadowPath(Body1)= 
\begin{bmatrix}
\vspace{0.02in} 1 & 1 & 0 & 360 & 360
\end{bmatrix} 
\end{aligned}
\end {align}
where the angles are given in degrees for ease of exposition. \\

Now consider the second body and its relationship to the shadow of the previous body. We see
that the boundary of the second body intersects the shadow arc, hence the visible portion is
described by two arcs (see \S \ref{sec:boundingcurve_example_4bodies}, Equation \ref{eq:vispath1_boundcurve}):
\begin{align}
\begin{aligned}\label{eq:vispath2_KOI126}
VisiblePath(Body2)=
\begin{bmatrix}
\vspace{0.02in} 2 & 1 & 210.87 & 90.103  & 300.97 \\                
\end{bmatrix} \\
\begin{bmatrix}
\vspace{0.02in} 2 & 2 & 115.52  &   280.81 &  396.33 \\
\end{bmatrix} \\
\end{aligned}
\end{align}
The new shadow is also described by two arcs (\S \ref{sec:boundingcurve_example_4bodies}), Equation (\ref{eq:vispath2_boundcurve}):
\begin{align}
\begin{aligned}
ShadowPath(Body2)=
\begin{bmatrix}
\vspace{0.02in} 1 & 1 &  300.97 &  269.90 &  570.87 \\
\end{bmatrix}  \\
\begin{bmatrix}
\vspace{0.02in} 2 & 2 & 115.52 & 280.81 & 396.33 \\
\end{bmatrix}  \\
\end{aligned}
\end{align}

Now consider the third body and its relationship to the shadow of the previous bodies.
Note that the first arc in the shadow is entirely contained within the current body (Body 3) and
that the second arc is only partially contained within Body 3.
\begin{align}
\begin{aligned}\label{eq:vispath3_KOI126}
VisiblePath(Body3)=
\begin{bmatrix}
\vspace{0.02in} 3  &1 & 300.97 & 269.90 & 570.87 \\
\end{bmatrix} \\
\begin{bmatrix}
\vspace{0.02in} 3 & 2 & 349.60 & 46.729 &  396.33 \\
\end{bmatrix} \\
\begin{bmatrix}
\vspace{0.02in} 3 & 2 & 115.52 & 15.045 & 130.56 \\
\end{bmatrix} \\
\begin{bmatrix}
\vspace{0.02in} 3 & 3 & 247.09 & 345.98  & 593.07 \\
\end{bmatrix} \\
\end{aligned}
\end{align}
Although the shadow is not required for this example, it is given for completeness:
\begin{align}
\begin{aligned}
ShadowPath(Body3)=
\begin{bmatrix}
\vspace{0.02in} 2 & 2 & 130.56 & 219.03 & 349.60 \\
\end{bmatrix}  \\
\begin{bmatrix}
\vspace{0.02in}  3 & 3 & 247.09 &  345.98 & 593.07 \\
\end{bmatrix}  \\
\end{aligned}
\end{align}
Note that since Body 1 lies within Body 3, the shadow no longer contains any Body 1 arcs. \\

The flux integral for each body may now be assembled using equations (\ref{eq:xiyi_coord}), (\ref{eq:xiyi_deriv}), and (\ref{eq:limbdarkintegral_new}), noting the value of the orientation given after Equation (\ref{eq:limbdarkintegral_new}). Since the co-ordinate system used to describe the limb darkening law is centered on the body in question having  radius 1,  we must convert all of the arc descriptions to that coordinate system. That is why Equation (\ref{eq:xiyi_coord}) translates the origin of each body's system to the origin of the body in question and scales the co-ordinates to radius 1 for that body. These flux integrals may now be evaluated in closed form or, if needed, numerically. If we normalize 
Equation (\ref{eq:greenapply}) by the integral of the flux over the unobstructed disk, we obtain the flux fraction for the system at the given time.

\subsection{Calculating the Fractional Flux} \label{sec:exampleintegral}

Continuing with the KOI-126 example, in \S\ref{sec:examplearcs} we constructed the bounding arcs for the visible portion of the three stars. Using these visible-path vector stacks, we will implement Equations (\ref{eq:xiyi_coord})-(\ref{eq:deltaRV}) to compute the fractional flux of each body at the syzygy epoch BJD 2455711.38 (refer to Table \ref{tab:planeskycoord}).  For simplicity we will use the linear limb darkening law for all stars, with a coefficient of $0.6$. Repeating Equations \ref{eq:CdotPsi}, \ref{eq:linear_law_dot} we have
\begin{equation} \nonumber
I(\mu)/I_0 = \bm{C} \boldsymbol{\cdot} \bm{\Psi} \text{~where~}
{\bm C}=\big[(1-c_1),c_1\big] \text{~~and~~} \bm{\Psi}=\big[\bm{1},\mu\big] 
\end{equation}
Substituting the exterior anti-derivatives of the functions in the $\Psi$ vector from Table \ref{tab:LimbRM_1Forms} into Equation (\ref{eq:intensity_integral2}) gives:
\begin{equation}
I_{\rm vis}=\bm{C} \boldsymbol{\cdot} \bigg[
\oint\displaylimits_{\partial D_{\rm vis}}\frac{1}{2}\big[-y,x\big] \boldsymbol{\cdot} \big[x',y']d\varphi,\,\,
\oint\displaylimits_{\partial D_{\rm vis}}\frac{1-\mu^3}{3r^2}\big[-y,x\big] \boldsymbol{\cdot} \big[x',y']d\varphi\bigg]
\end{equation}
For each body, in turn, we will evaluate the line integrals in the $\Psi$ vector over the arcs described by that body's visible-path stacks. For Body 1:
\begin{align}
\begin{aligned}\label{eq:vispath1}
VisiblePath(Body1)=
\begin{bmatrix}
\vspace{0.02in} 1 & 1 & 0 & 360 & 360
\end{bmatrix}
\text{~or in radians:~}
\begin{bmatrix}
\vspace{0.02in} 1 & 1 & 0 & 2\pi & 2\pi 
\end{bmatrix} \\
\end{aligned}
\end{align}
This is the front body, which is entirely visible, thus the flux fraction, ${\cal F}=1$. For Body 2:
\begin{align}
\begin{aligned}\label{eq:vispath2}
VisiblePath(Body2)=
\begin{bmatrix}
\vspace{0.02in} 2 & 1 & 210.87 & 90.103 & 300.97 
\end{bmatrix}
\begin{bmatrix}
\vspace{0.02in} 2 & 2 & 115.52 & 280.81 & 396.33 
\end{bmatrix} \\
\text{~or in radians:~}
VisiblePath(Body2)=
\begin{bmatrix}
\vspace{0.02in} 2 & 1 & 3.6804 & 1.5726 & 5.253 
\end{bmatrix}
\begin{bmatrix}
\vspace{0.02in} 2 & 2 & 2.0162 & 4.901 & 6.9172 
\end{bmatrix} \\
\end{aligned}
\end{align}
In this case, the boundary of the visible region is described by two arcs:
\begin{align}\label{KOI126_arc1}
\begin{aligned}
Arc_{MN}=
\begin{bmatrix}
\vspace{0.02in} 2 & 1 & 210.87 & 90.103 & 300.97 
\end{bmatrix}
\end{aligned}
\end{align}
where $N=2$ and $M=1$. The limits of integration are $\varphi_0=3.6804$ and $\varphi_1=5.253$. The parametric equations for this arc are given by Equation (\ref{eq:xiyi_coord}) and the derivatives by Equation (\ref{eq:xiyi_deriv}).
\begin{align}
\begin{aligned}
&x_1=0.90058\cos(\varphi)+0.34217   &y_1&=0.90058\sin(\varphi)+1.3645  \\
&x_1'=-0.90058\sin(\varphi)  &y_1'&=-0.90058\cos(\varphi) 
\end{aligned}
\end{align}
Now assemble the integral $j=1$ (constant function) and $i=1$ (arc $1$) in Equation (\ref{eq:xiyi_deriv}) by substituting the expressions for $x_1$ and $y_1$
\begin{align}\label{eq:Psi11}
\begin{aligned}
\Psi_{1,1}=\oint\displaylimits_{\varphi_0}^{\varphi_1} {\rm Orientation}\times\big[P_1,Q_1\big] \boldsymbol{\cdot} \big[x'_1, y'_1\big] d\varphi=
\int\displaylimits_{\varphi_0}^{\varphi_1}(-1)\frac{1}{2}\big[-y_1,x_1\big] \boldsymbol{\cdot} 
\big[x'_1, y'_1\big]d\varphi=0.25893
\end{aligned}
\end{align}
Since $M < N$, the orientation is $-1$. Next, assemble the integral $j=2$ (the $\mu$ function) and $i=1$ (arc 1) in Equation (\ref{eq:xiyi_deriv}), by substituting the expressions for $x_1$ and $y_1$, and noting the definitions of $r$ and $\mu$:
\begin{align}\label{eq:Psi21}
\begin{aligned}
\Psi_{2,1}=\oint\displaylimits_{\varphi_0}^{\varphi_1}{\rm Orientation}\times\big[P_2,Q_2\big] \boldsymbol{\cdot} \big[x'_1, y'_1\big] d\varphi=
\int\displaylimits_{\varphi_0}^{\varphi_1}(-1)\frac{1-\mu^3}{3r^2}\big[-y_1,x_1\big] \boldsymbol{\cdot} \big[x'_1, y'_1\big]d\varphi=0.22757
\end{aligned}
\end{align}
Since $M < N$, the orientation is $-1$. Moving on to the \textit{second} arc:
\begin{align}\label{KOI126_arc2}
\begin{aligned}
Arc_{MN}=
\begin{bmatrix}
2 & 2 & 2.0162 & 4.901 & 6.9172 
\end{bmatrix}
\end{aligned}
\end{align}
where $N=2$ and $M=2$. The limits of integration are $\varphi_0=2.0162$ and $\varphi_1=6.9172$. The parametric equations for this arc are given by Equation (\ref{eq:xiyi_coord}) and the derivatives are given by Equation (\ref{eq:xiyi_deriv}).
\begin{align}
\begin{aligned}
&x_1=\cos(\varphi)   &y_1&=\sin(\varphi)  \\
&x_1'=-\sin(\varphi)  &y_1'&=\cos(\varphi) 
\end{aligned}
\end{align}
Now assemble the integral $j=1$ (constant function) and $i=2$ (arc 2) in Equation (\ref{eq:limbdarkintegral_new}) by substituting the expressions for $x_1$ and $y_1$: 
\begin{align}\label{eq:Psi12}
\begin{aligned}
\Psi_{1,2}=\oint\displaylimits_{\varphi_0}^{\varphi_1}{\rm Orientation}\times \big[P_1,Q_1\big] \boldsymbol{\cdot} \big[x'_1, y'_1\big] d\varphi=
\int\displaylimits_{\varphi_0}^{\varphi_1}(+1)\frac{1}{2}\big[-y_1,x_1\big] \boldsymbol{\cdot} \big[x'_1, y'_1\big]d\varphi=0.4505
\end{aligned}
\end{align}
Since $M=N$, the orientation is $+1$. Next, assemble the integrand $j=2$ (the $\mu$ function) and $i=2$ (arc 2) in Equation (\ref{eq:limbdarkintegral_new}) by substituting the expressions for $x_1$ and $y_1$ and noting the definitions of $r$ and $\mu$:
\begin{align}\label{eq:Psi22}
\begin{aligned}
\Psi_{2,2}=\oint\displaylimits_{\varphi_0}^{\varphi_1}{\rm Orientation}\times \big[P_2,Q_2\big] \boldsymbol{\cdot} \big[x'_1, y'_1\big] d\varphi=
\int\displaylimits_{\varphi_0}^{\varphi_1}(+1)\frac{1-\mu^3}{3r^2}\big[-y_1,x_1\big] \boldsymbol{\cdot} \big[x'_1, y'_1\big]d\varphi=1.6337.
\end{aligned}
\end{align}
Since $M=N$, the orientation is $+1$. Thus, Equation (\ref{eq:limbdarkintegral_new}) results in
\begin{align}\label{eq:overall_Psi_arc1}
\begin{aligned}
\Psi_1&=\Psi_{1,1}+\Psi_{1,2}=2.7094 \text{~~and~~} \Psi_{2,1}+\Psi_{2,2}=1.8612 \\
\Longrightarrow
I_{\rm vis}&=\bm{C} \boldsymbol{\cdot} \big[2.7094, 1.8612\big]=\big[1-0.6,0.6\big] \boldsymbol{\cdot} \big[2.7094,1.8612\big]=2.2005
\end{aligned}
\end{align}
Following the computation for arc $2$ we can easily compute $I_{\rm total}$, which is the integral over the entire disk. The boundary consists of one arc: $\big[2~~2~~0~~2\pi~~2\pi\big]$:
\begin{align}\label{eq:overall_Itot}
\begin{aligned}
I_{\rm total}&=\bm{C} \boldsymbol{\cdot} \big[3.1416, 2.0944\big]=\big[1-0.6,0.6\big] \boldsymbol{\cdot} \big[3.1416,2.0944\big]=2.5133
\end{aligned}
\end{align}
The flux fraction of Body 2 is then $2.2005/2.5133=0.87556$. Following a similar procedure, we can compute the flux fraction of the third star, which in this case is the bright F-star that dominates the total system light.  As shown in Equation (\ref{eq:vispath3_KOI126}), the visible path of that star has $4$ arcs at this particular time, hence there are $8$ terms in the summation, along with a normalizing term.  The final result is $F_3 = 0.98628$ (we omit the details for brevity).

\section{Summary}\label{sec:summary}

We presented here an efficient method for computing the visible flux for each body during a multi-body eclipsing event for all commonly used limb darkening laws. The method can also calculate the R-M effect. Our approach follows the idea put forth by \citet{Pal2012} to apply Green's Theorem, thus transforming the 2D POS flux integral into a 1D integral over the visible boundary. We executed this idea through an iterative process which combines a fast method for describing the visible boundary of each body with a fast and accurate Gaussian integration scheme to compute the integrals. \\

We first provided the mathematical background for the method, then compare the results of its application to
three other methods currently used, i.e. \ \citet{Mandel2002}, \citet{Gimenez2006a}, and \citet{Pal2012}. Finally, we explained in detail how to implement the technique with the help of several examples and a code which we made available. Specifically, we note that for the two body case, our method compares well in speed with that of \citet{Mandel2002}, and is faster than that of \citet{Gimenez2006a}; the method does not have various distinct cases so it does not produce the occasional spikes and other numerical anomalies caused by crossing boundaries as is seen in the Mandel \& Agol technique; the method uses a tolerance parameter, $T_{\rm op}$, which sets the computational error to the problem being worked.  Most significantly, the method works for any number of eclipsing bodies, and using a multitude of limb darkening laws (including all of the commonly used ones). It can also compute the R-M effect for any of these limb darkening laws. This method can be extended to stars with differential rotation when using the quadratic limb darkening.  \\

We note that in applying Green's Theorem to solve the flux integral, we are not necessarily limited to using spherical bodies. For example, one might wish to compute the transits of an ellipsoidal body with any orientation. 

\appendix 

\section{Constructing the Visible Boundary - a Worked Example} \label{sec:appendix_workedexample}

\subsection{The Special Two Body Case} \label{sec:boundingcurves_2body}

When constructing the bounding curves we employ four operations: we can intersect two different circles (the routine {\tt XCircle} in the code); we can find the complement of an arc in a given circle ({\tt XComp} in the code);  we can intersect two arcs within a given circle ({\tt XSect}); and we can intersect multiple arcs in a given circle ({\tt XSectAll}). In the following sections we show how these routines are used to find the visible paths needed for the fractional flux computation.

\begin{figure}[h]\
\plotone{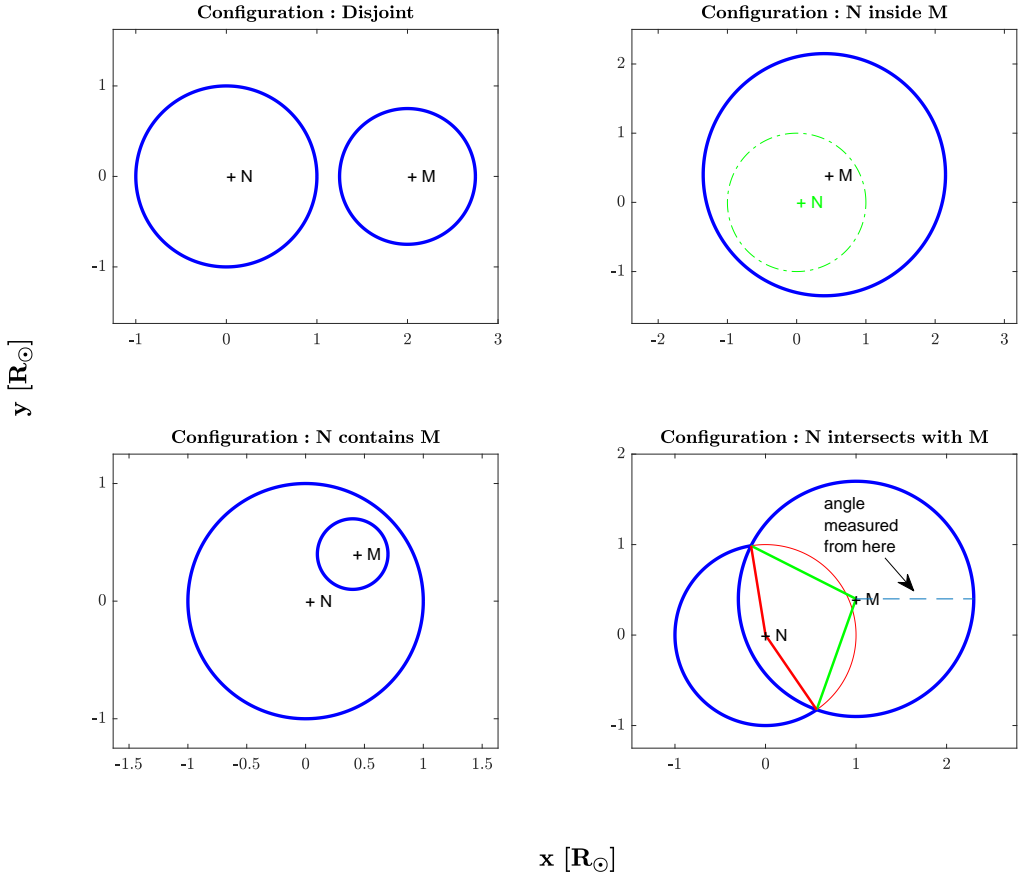}
\caption{The four possible intersections of two circles used when starting the construction of a bounding arc (\S \ref{sec:boundingcurve} and Appendix \ref{sec:boundingcurves_2body}). The body in the back (further from the observer) is always denoted $N$ and the body in the front (closer to the observer) is always denoted $M$. The upper Left panel shows the disjoint case with no arc intersection. The Upper right panel shows the case when the body in the back (labeled $N$) is completely covered by the body in the front (labeled $M$). The lower Left panel shows the case when the smaller body in the front ($M$) is completely within the boundary of the body in the back ($M$). The lower right panel shows the intersection case where the body in front ($M$) partially covers the body in the back ($N$). The visible part of the body in the back is described by these two arcs: $Arc_{NM}= \big[ N~~M~~153.150~~97.301~~250.452\big]$ and $Arc_{NN}= \big[N~~N~~99.198~~205.205~~304.404\big]$, where the angles are measured in degrees.
}\label{fig:ArcConst_CasesExample_angle}
\end{figure}

If we have only two bodies to consider, where Body $M$ is closer to the observer and
body $N$ is further from the observer,  {\tt XCircle} is the only one of the four routines 
mentioned above that we need to use. The operation {\tt XCircle} has one of\
four possible outcomes, as shown in Figure \ref{fig:ArcConst_CasesExample_angle}:  
{\em I}. The two bodies are disjoint and do not intersect.  {\em II}. Body $N$ is contained in Body $M$.
{\em III}. Body $N$ contains body $M$.  {\em IV}. There is partial overlap and the boundaries of each body intersect at two points. In the first three cases, the routine returns special flags to indicate 
which case was encountered.  In the last case, the routine returns the two arcs that make up the visible path of Body $N$. It is important to select the correct arcs, since given two points on the circumference of a circle, one can draw two different arcs.  The first arc can go from $\varphi_0$ to $\varphi_1$ (following the right-hand rule), and the second arc can go from $\varphi_1$ to $\varphi_0$. Which arc do we choose when constructing the visible path on Body $N$ For the arc centered on Body $M$, we want the arc that is {\em closer} to the center of Body $N$.  Likewise, for the arc centered on Body $N$, we want the arc that is {\em furthest} from the center of body $M$. By considering the triangles formed by the two centers and each of the intersection points, and by using the Law of Cosines, we can ensure the correct arcs are chosen. \\

In the discussion below we explicitly give the appropriate visible arc(s) for the four possible outcomes of {\tt XCircle}.  For completeness, we also give the correspondng shadow paths for each case for situations where additional bodies need to be considered (the shadow paths are not
required for the light curve computation, when ${\cal N}=2$). \\

{\em I}. The bodies are \underline{dis}j\underline{oint} \normalsize (top-left panel in Figure \ref{fig:ArcConst_CasesExample_angle}). Since there is no overlap between the two circles, the entire boundary of Body $N$ is visible. If one were to continue to the next stage where there is a third body in the back of these two bodies, the shadow boundary would simply be composed of the outer parts of each circle. Following the notation of Equation (\ref{eq:Arc_NM}), the visible path and shadow path are written as: 
\begin{align}
\begin{aligned}\label{eq:visarc_disjoint}
Visible~path=
\begin{bmatrix}
\vspace{0.02in} N & N & 0 & 360 & 360
\end{bmatrix} \\
Shadow~path= 
\begin{bmatrix}
\vspace{0.02in} N & N & 0 & 360 & 360
\end{bmatrix} \\
\begin{bmatrix} 
\vspace{0.02in} M & M & 0 & 360 & 360
\end{bmatrix} 
\end{aligned}
\end {align} 
where all angles are given in degrees, for the reader's convenience. In this particular case, $N=2$ and $M=1$ following the convention of front-to-back order.  \\

{\em II}. Body $N$ is \underline{contained in} \normalsize Body $M$ (i.e.\ body $N$ is totally eclipsed by body $M$, top-right panel in Figure \ref{fig:ArcConst_CasesExample_angle}). In this case, since Body $N$ is completely obstructed by Body $M$, there is no visible path on Body $N$. Therefore, the arc is of zero length. If one were to continue to an additional third body that is behind these two bodies, one could see that the shadow cast by the two bodies $N$ and $M$ would be bounded by the outer part of body $M$. Thus, following the notation of Equation (\ref{eq:Arc_NM}), the visible path and the shadow path are written as:
\begin{align}
\begin{aligned}\label{eq:visarc_NinsideM}
Visible~path=
\begin{bmatrix}
\vspace{0.02in} N & N & 0 & 0 & 0 
\end{bmatrix} \\
Shadow~path= 
\begin{bmatrix}
\vspace{0.02in} M & M & 0 & 360 & 360
\end{bmatrix} 
\end{aligned}
\end {align}

{\em III}. Body $N$ \underline{contains} \normalsize Body $M$ (i.e.\ body $M$ is transiting body $N$, bottom-left panel in Figure \ref{fig:ArcConst_CasesExample_angle}), Here, the visible path on Body $N$ is an annulus, made of arcs that are two full circles. One of the circles is centered on Body $M$ and the other circle is centered on Body $N$. Since one body is contained within the other, the shadow path is simply the boundary of the larger of the two bodies, Body $N$. Hence, the visible path and the shadow path are written below in the notation of Equation (\ref{eq:Arc_NM}) as:
\begin{align}
\begin{aligned}\label{eq:visarc_NcontainsM}
Visible~path=
\begin{bmatrix}
\vspace{0.02in} N & M & 0 & 360 & 360 \\
\end{bmatrix} \\
\begin{bmatrix}
\vspace{0.02in} N & N & 0 & 360 & 360
\end{bmatrix} & \\
Shadow~path= 
\begin{bmatrix}
\vspace{0.02in} N & N & 0 & 360 & 360 
\end{bmatrix} 
\end{aligned}
\end {align} \\
 
 {\em IV}. The boundary of Body $N$ and the boundary of Body $M$ \underline{intersect} \normalsize (bottom-right panel in Figure \ref{fig:ArcConst_CasesExample_angle}). In this case, the visible path on Body $N$ will have two arcs, one centered on Body $M$, and the other centered on Body $N$. These arcs are output by the {\tt XCircle} routine.  In a similar manner, the boundary of the shadow cast by these two bodies will consist of two arcs. The first arc, centered on Body $M$, will be the {\em complement} of the arc (centered on Body $M$) that appeared in the visible path on Body $N$. The second arc, centered on Body $N$, will be the same arc centered on Body $N$ that was in the visible path.  In the notation of Equation (\ref{eq:Arc_NM}), the visible path and shadow path are therefore written as:
\begin{align}
\begin{aligned}\label{eq:visarc_intersect}
Visible~path=
\begin{bmatrix}
\vspace{0.02in} N & M & \varphi_0 & \Delta \varphi & \varphi_1
\end{bmatrix} \\
\begin{bmatrix}
\vspace{0.02in} N & N & \psi_0 & \Delta \psi & \psi_1
\end{bmatrix} \\
Shadow~path= 
\begin{bmatrix}
\vspace{0.02in} M & M & \rm{complement}(\varphi_0,\varphi_1)
\end{bmatrix} \\
\begin{bmatrix}
\vspace{0.02in} N & N & \psi_0 & \Delta \psi & \psi_1 
\end{bmatrix} 
\end{aligned}
\end {align}
The ``$ \rm{complement}(\varphi_0,\varphi_1)$'' in Equation (\ref{eq:visarc_intersect}) is the complementary arc to
$(\varphi_0,\varphi_1)$, and is obtained with the routine {\tt XComp} in the code:
\begin{align}
\begin{aligned}\label{eq:visarc_xcomp}
{\tt XComp}\bigg(
\begin{bmatrix}
\vspace{0.02in} N & M & \varphi_0 & \Delta \varphi & \varphi_1
\end{bmatrix} 
\bigg)=
\begin{bmatrix}
\vspace{0.02in} N & M & (\varphi_1-2\pi) & (2\pi- \Delta \varphi) & \varphi_0
\end{bmatrix} \quad \varphi_1 \geq 2\pi \\
\begin{bmatrix}
\vspace{0.02in} N & M & \varphi_1 & (2\pi- \Delta \varphi) & (2\pi+\varphi_0)
\end{bmatrix} \quad \varphi_1 < 2\pi \\
\end{aligned}
\end {align}

\subsection{The Case of Three or More Bodies}\label{sec:boundingcurve_example_3more_bodies}

When there are three or more bodies, we can find the visible path on the third star (which
is behind the other two) as follows. The shadow path of the first two bodies 
(which are closer to the observer) needs to be found as outlined above. That shadow path
can be the outer boundary of either star (in the case 
of an occultation or transit), it can be the outer boundaries of both
stars (in the case of no overlap), or it can resemble the outside of a``dumbell'' (in the case
of partial overlap). This shadow could cover parts of the third star. Therefore, 
the visible path on the third star (the one in the back) could consist of arcs on its own boundary and arcs from the shadow path of the first two stars. To do that, we use {\tt XCircle} on Body 1 and Body 3
(the ordering that the subroutine uses is front, back), and also on 
Body 2 and Body 3. This results in two arcs on the perimeter of Body 3 that are 
potentially visible. These arcs could overlap, and we want to find their {\em intersection}.
In addition, there might be arcs on the perimeter of Body 1 and on the perimeter of Body 2 
that are potentially visible, and these arcs may overlap with arcs in the
shadow path of the first two bodies.  As before, we need
to find their intersections.  The routines {\tt XSect} and {\tt XSectAll} 
are used to find those. \\

If there is a fourth star, the shadow path of the three stars is required. The shadow path of the three bodies is formed by taking the union of the shadow path formed by the two front bodies with the visible path of the third body and then removing the extraneous interior arcs. This is done as follows: for each of the front bodies, we look at that body's contribution to the visible path on the back body, we find the complimentary arcs to that contribution, and intersect them (using the routines {\tt XSect} and {\tt XSectAll})  with the arcs in the shadow path contributed by that front body. Finally, we add all of the visible arcs on the perimeter of the back body.\\

\begin{figure}[ht!]\
\plotone{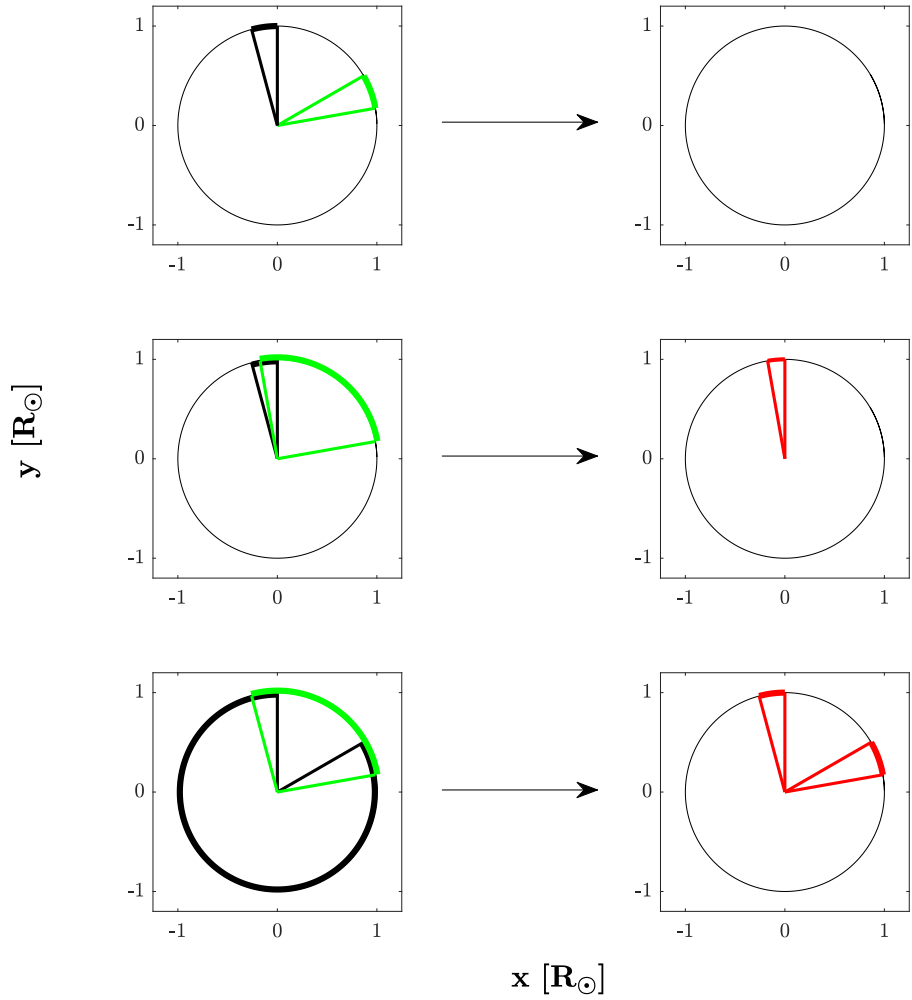}
\caption{The three possibilities of intersecting two arcs, as demonstrated in \S\ref{sec:boundingcurve} and Appendix \ref{sec:appendix_workedexample}. The panels on the left show the two arcs to be intersected, and the panels on the right show the resulting arc. The two arcs in the top panels are disjoint, so the intersection is empty (Equation \ref{eq:visarc_overlapempty}). The arcs in the middle panels intersect so that only a part of the smaller arc is preserved (Equation \ref{eq:visarc_overlap1}). The two arcs in the bottom panels intersect into two different, disjoint, arcs (Equation \ref{eq:visarc_overlap2}). In the code, this is done with the {\tt XSect} routine.} 
\label{fig:GraphXSect}
\end{figure}

In the following section we present an example of a system with four overlapping bodies, and explain step-by-step how to find the visible paths and the shadow boundaries. But first we show below how the routines {\tt XSect} and {\tt XSectAll} work.  Consider the intersection of bounding arcs for a given disk. The simplest case is the intersection of two arcs ({\tt XSect}). There are three possibilities, as shown by the following example (see Figure \ref{fig:GraphXSect}): \\
\begin{align}
\begin{aligned}\label{eq:visarc_overlapempty}
\text{No Overlap:}
\begin{bmatrix}
\vspace{0.02in} N & M & 10 & 20 & 30
\end{bmatrix} \cap
\begin{bmatrix}
\vspace{0.02in} N & M & 90 & 15 & 105
\end{bmatrix}  \Longrightarrow ~Empty
\end{aligned}
\end {align} 
\begin{align}
\begin{aligned}\label{eq:visarc_overlap1}
\text{Overlap (1 arc)}=
\begin{bmatrix}
\vspace{0.02in} N & M & 10 & 90 & 100
\end{bmatrix} \cap
\begin{bmatrix}
\vspace{0.02in} N & M & 90 & 15 & 105
\end{bmatrix}  \Longrightarrow
\begin{bmatrix}
\vspace{0.02in} N & M & 90 & 10 & 100
\end{bmatrix} 
\end{aligned}
\end {align}
\begin{align}
\begin{aligned}\label{eq:visarc_overlap2}
\text{Overlap (2 arcs)}=
\begin{bmatrix}
\vspace{0.02in} N & M & 10 & 95 & 105
\end{bmatrix} \cap
\begin{bmatrix}
\vspace{0.02in} N & M & 90 & 300 & 390 
\end{bmatrix} \Longrightarrow
\begin{bmatrix}
\vspace{0.02in} N & M & 90 & 15 & 105
\end{bmatrix}  \\
\begin{bmatrix} 
\vspace{0.02in} N & M & 10 & 20 & 30
\end{bmatrix} \\
\end{aligned}
\end {align} 
If there are more than two arcs to be intersected, then we proceed iteratively, starting with the intersection of the first two, and intersecting the remaining arcs, in turn, with the intersection of the previous arcs (see the {\tt XSectAll} routine). However, if at some step, the intersection, {\tt XSect}, yields two arcs, the process bifurcates, as in Equation (\ref{eq:visarc_overlap2}). Those 2 arcs must be intersected individually with the remaining arcs. For example, consider the following three arcs:
\begin{align}
\begin{aligned}\label{eq:visarc_3arcs}
\begin{bmatrix}
\vspace{0.02in} N & M & 10 & 95 & 105
\end{bmatrix} \\
\begin{bmatrix}
\vspace{0.02in} N & M & 90 & 300 & 390
\end{bmatrix} \\
\begin{bmatrix}
\vspace{0.02in} N & M & 25 & 70 & 95 
\end{bmatrix}
\end{aligned}
\end {align} 
Intersect the first two:
\begin{align}
\begin{aligned}\label{eq:visarc_3arcs_first2}
\begin{bmatrix} 
\vspace{0.02in} N & M & 10 & 95 & 105
\end{bmatrix} \cap
\begin{bmatrix}
\vspace{0.02in} N & M & 90 & 300 & 390 
\end{bmatrix} \Longrightarrow
\begin{bmatrix}
\vspace{0.02in} N & M & 90 & 15 & 105
\end{bmatrix} \\
\begin{bmatrix} 
\vspace{0.02in} N & M & 10 & 20 & 30
\end{bmatrix} \\
\end{aligned}
\end {align}
Bifurcation yields two intersections, each with the remaining arc $\big[N ~M ~25 ~70 ~95\big]$:
\begin{align}
\begin{aligned}\label{eq:visarc_3arcs_bifur1}
\begin{bmatrix} 
\vspace{0.02in} N & M & 90 & 15 & 105
\end{bmatrix} \cap
\begin{bmatrix}
\vspace{0.02in} N & M & 25 & 70 & 95 
\end{bmatrix} \Longrightarrow
\begin{bmatrix}
\vspace{0.02in} N & M & 90 & 15 & 105
\end{bmatrix} \\ 
\end{aligned}
\end {align} 
and
\begin{align}
\begin{aligned}\label{eq:visarc_3arcs_bifur2}
\begin{bmatrix}
\vspace{0.02in} N & M & 10 & 20 & 30
\end{bmatrix} \cap
\begin{bmatrix}
\vspace{0.02in} N & M & 25 & 70 & 95 
\end{bmatrix} \Longrightarrow
\begin{bmatrix}
\vspace{0.02in} N & M & 25 & 5 & 30
\end{bmatrix} \\
\end{aligned}
\end {align}
Thus, the result of intersecting the three arcs in (\ref{eq:visarc_3arcs}) is two disjoint arcs (\ref{eq:visarc_3arcs_bifur1}) and (\ref{eq:visarc_3arcs_bifur2}). \\

We are now ready to present a worked example with four overlapping bodies

\subsection{Worked Example: a Four Body System}\label{sec:boundingcurve_example_4bodies}

\begin{figure}[ht!]\
\plotone{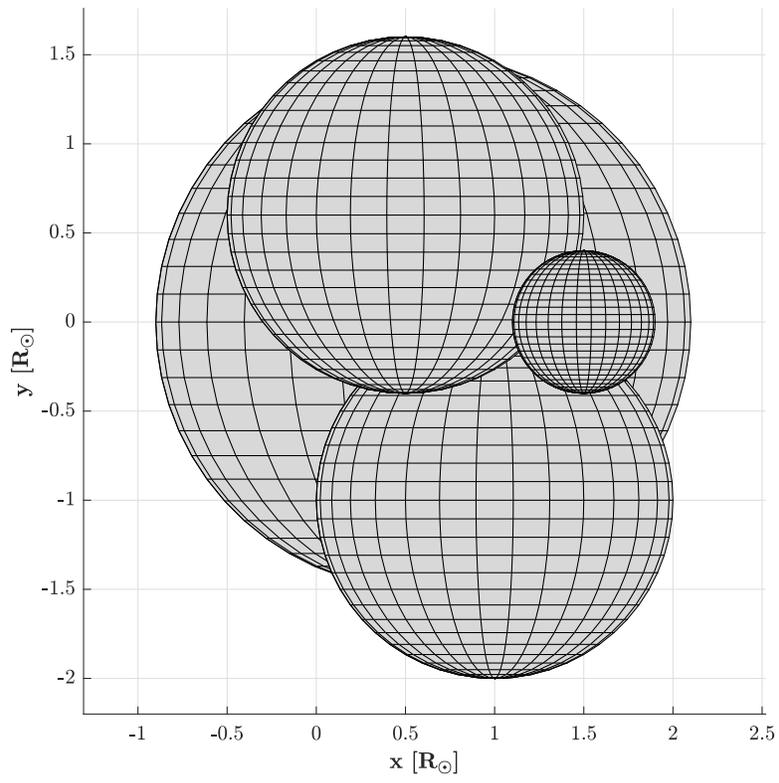}
\caption{The event of 4 bodies crossing, the arcs of which are constructed in Figure \ref{fig:ArcConst_BoundingCurve}}
\label{fig:ArcConst_SphereExample}
\end{figure}

To fully understand the details of this process, consider the following $4$-body illustrative example based on the plane-of-the-sky co-ordinates given in Table \ref{tab:planeskycoord_4body}, corresponding to the 4-body crossing event shown in Figure \ref{fig:ArcConst_SphereExample}. \\

We describe the boundary of the visible path of a particular body, say Body 3, as the {\it stack} of arcs denoted, for clarity, by $VisiblePath(Body3)$. In the code, this stack is held in the variable {\tt Path}. Similarly, the variable $ShadowPath(Body3)$ refers to the shadow formed at the stage where Body 3 is the back body.
In the code, this stack is held in the variable {\tt Bdr}. Both {\tt Path} and {\tt Bdr} are assigned by the function {\tt IntegrationPaths}. We occasionally refer to specific arcs. In those instances, we use the notation $(Body3,Arc2)$ referring to the second arc in the stack, centered on Body 3. \\

\begin{figure}[ht!]\
\plotone{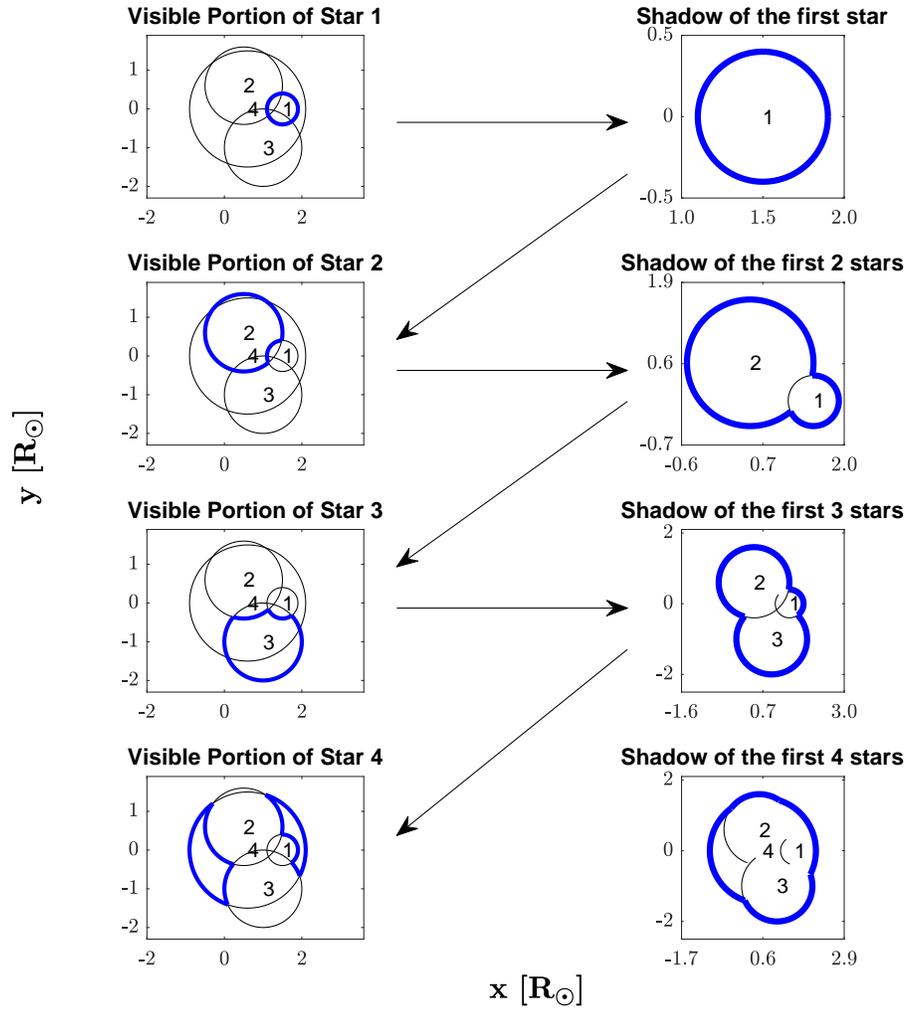}
\caption{Construction of the Bounding Curve in the 4 body example given in \S \ref{sec:boundingcurve_example_4bodies} for the configuration shown in Figure \ref{fig:ArcConst_SphereExample}. The panels on the left show the progression of the visible path of arcs, and those on the right show the shadow path of arcs resulting from each previous visible path. Both the visible path and the shadow path arcs are described as vector stacks in Equations (\ref{eq:vispath1_boundcurve}) -  (\ref{eq:shadpath4_boundcurve}). The arrows indicate the progression of the process where each visible path yields the shadow path which, in turn, is used to generate the next visible path. The process ends on the visible path of Star 4, but we show the resulting shadow path of Star 4 for completion.}
\label{fig:ArcConst_BoundingCurve}
\end{figure}

To initialize the process, set $VisiblePath(Body1)$ and $ShadowPath(Body1)$ to the boundary of Body 1. This is shown in the upper two panels in Figure \ref{fig:ArcConst_BoundingCurve}. The visible path and the shadow path are:
\begin{align}
\begin{aligned}\label{eq:vispath1_boundcurve}
Visible\-Path(Body1)=
\begin{bmatrix}
\vspace{0.02in} 1 & 1 & 0 & 360 & 360
\end{bmatrix} \\
ShadowPath(Body1)= 
\begin{bmatrix}
\vspace{0.02in} 1 & 1 & 0 & 360 & 360
\end{bmatrix} 
\end{aligned}
\end {align}
where, again, all angles are given in degrees, for the reader's convenience.\\

Now consider Body 2 and the shadow of Body 1 (see Figure \ref{fig:ArcConst_BoundingCurve}). 
Using {\tt XCircle}(Body1,Body2), two arcs are produced, which describe the visible portion of Body 2: 
\begin{align}
\begin{aligned}\label{eq:vispath2_boundcurve}
VisiblePath(Body2)=
\begin{bmatrix}
\vspace{0.02in} 2 & 1 & 92.910 & 112.25 & 205.16
\end{bmatrix} \\
\begin{bmatrix}
\vspace{0.02in} 2 & 2 & 348.43 & 321.21 & 669.64
\end{bmatrix} \\
\end{aligned}
\end{align}
The first arc is centered on Body 1 and the second arc is centered on Body2.  We now compute the combined shadow of Body 1 and Body 2.  Note that this requires the complement of the first arc, together with the second arc.
\begin{align}
\begin{aligned}\label{eq:shadpath2_boundcurve}
ShadowPath(Body2)=
\begin{bmatrix}
\vspace{0.02in} 1 & 1 & 205.16 & 247.75 & 452.91
\end{bmatrix} \\
\begin{bmatrix}
\vspace{0.02in} 2 & 2 & 348.43 & 321.21 & 669.64
\end{bmatrix} \\
\end{aligned}
\end{align} 

Now consider Body 3. Here, for the first time, we see the complication of the general case (see Figure \ref{fig:ArcConst_BoundingCurve}). We begin with the first arc of the shadow. The arc is centered on Body 1, so  we use {\tt XCircle} to find the two arcs detailing the portion of Body 3 that is not occluded by Body 1:
\begin{align}
\begin{aligned}\label{eq:XCircle_Body3Body1}
{\tt XCircle}(Body1,Body3)=
\begin{bmatrix}
\vspace{0.02in} 3 & 1 & 180.72 & 125.43 & 306.15
\end{bmatrix} \\
\begin{bmatrix}
\vspace{0.02in} 3 & 3 & 84.259  & 318.35 & 402.61
\end{bmatrix} \\
\end{aligned}
\end{align}
Note that the resulting arc, centered on Body 1, has a smaller starting angle than the shadow arc $(Body2,Arc1)$, while the shadow arc $(Body2,Arc1)$ has a larger ending angle than the arc from {\tt XCircle}. Thus we must take the intersection of these arcs to obtain the visible arc $(Body3,Arc1)$. A similar intersection is required for Body 2: 
\begin{align}
\begin{aligned}\label{eq:XCircle_Body3Body2}
{\tt XCircle}(Body2,Body3)=
\begin{bmatrix}
\vspace{0.02in} 3 & 2 & 254.3 & 66.109 & 320.41
\end{bmatrix} \\
\begin{bmatrix}
\vspace{0.02in} 3 & 3 & 140.41  & 293.89 & 434.3
\end{bmatrix} \\
\end{aligned}
\end{align}
Finally, for Body 3, the two arcs centered on Body 3 produced by {\tt XCircle} must also be intersected, 
resulting in the following description of the boundary of Body 3:
\begin{align}
\begin{aligned}\label{eq:vispath3_boundcurve}
VisiblePath(Body3)=
\begin{bmatrix}
\vspace{0.02in} 3 & 1 & 205.16 & 100.99 & 306.15
\end{bmatrix} \\
\begin{bmatrix}
\vspace{0.02in} 3 & 2 & 254.30 & 55.340 & 309.64
\end{bmatrix} \\
\begin{bmatrix}
\vspace{0.02in} 3 & 3 & 140.41 & 262.20 & 402.61
\end{bmatrix} \\
\end{aligned}
\end{align}
Given  $VisiblePath(Body3)$, we intersect the complement of the paths centered at Body 1 with 
those from the shadow centered at Body 1, do so for Body 2, and add in the Body 3 paths. This results
in updating the shadow to that of Body 1, Body 2, and Body 3.
\begin{align}
\begin{aligned}\label{eq:shadpath3_boundcurve}
ShadowPath(Body3)=
\begin{bmatrix}
\vspace{0.02in} 1 & 1 & 306.15 & 146.76 & 452.91
\end{bmatrix} \\
\begin{bmatrix}
\vspace{0.02in} 2 & 2 & 348.43 & 265.87 & 614.30
\end{bmatrix} \\
\begin{bmatrix}
\vspace{0.02in} 3 & 3 & 140.41 & 262.20 & 402.61
\end{bmatrix} \\
\end{aligned}
\end{align}

We now repeat the process to get the visible portion of Body 4. The first arc in $ShadowPath(Body3)$ is centered at Body 1. We, therefore, use {\tt XCircle}
to compute the arcs between Body 1 and Body 4.
\begin{align}
\begin{aligned}\label{eq:XCircle_Body4Body1}
{\tt XCircle}(Body1,Body4)=
\begin{bmatrix}
\vspace{0.02in} 4 & 1 & 0 & 360 &360
\end{bmatrix} \\
\begin{bmatrix}
\vspace{0.02in} 4 & 4 & 0  & 360  & 360
\end{bmatrix} \\
\end{aligned}
\end{align}
Note that Body 1 is contained in Body 4, but we still get two arcs, $\big[4~1~0~360~360\big]$, and
$\big[4~4~0~360~360\big]$. The first must be intersected with the shadow arc $(Body3,Arc1)$, giving
$\big[4~1~306.15~146.76~452.91\big]$, and the second must be collected for intersection with all
of the other $\big[4~4~x~x~x\big]$ arcs obtained during this step. Continuing the process of generating arcs, Body 2 and Body 3 are present in the shadow, which prompts the use of {\tt XCircle} to compute the arcs between Body 4 and Body 2, and between Body 4 and Body 3.
\begin{align}
\begin{aligned}\label{eq:XCircle_Body4Body2}
{\tt XCircle}(Body2,Body4)=
\begin{bmatrix}
\vspace{0.02in} 4 & 2 & 143.13 & 272.66 & 415.79
\end{bmatrix} \\
\begin{bmatrix}
\vspace{0.02in} 4 & 4 & 126.87  & 305.18 & 432.05
\end{bmatrix} \\
\end{aligned}
\end{align}
\begin{align}
\begin{aligned}\label{eq:XCircle_Body4Body3}
{\tt XCircle}(Body3,Body4)=
\begin{bmatrix}
\vspace{0.02in} 4 & 3 & 19.407 & 184.79 & 204.20
\end{bmatrix} \\
\begin{bmatrix}
\vspace{0.02in} 4 & 4 & 333.57  & 276.47 & 610.04
\end{bmatrix} \\
\end{aligned}
\end{align}
As before, we intersect the visible arc $(Body4,Arc1)$ with the shadow arc $(Body1,Arc1)$. The next intersection of the visible arc $(Body4,Arc2)$ and shadow arc $(Body2,Arc2)$ brings a new complication. The intersection has two components:
\begin{align}
\begin{aligned}\label{eq:XCircle_Body4Body2_components}
\begin{bmatrix}
\vspace{0.02in} 4 & 2 & 348.43 & 67.362 & 415.79
\end{bmatrix} \\
\begin{bmatrix}
\vspace{0.02in} 4 & 2 & 143.13  & 111.17 & 254.30
\end{bmatrix} \\
\end{aligned}
\end{align}
Likewise, the intersection of visible arc $(Body4,Arc3)$ and the shadow arc $(Body3,Arc3)$ has two components. The Visible Path is then given by:
\begin{align}
\begin{aligned}\label{eq:vispath4_boundcurve}
VisiblePath(Body4)=
\begin{bmatrix}
\vspace{0.02in} 4 & 1 & 306.15 & 146.76 & 452.91
\end{bmatrix} \\
\begin{bmatrix}
\vspace{0.02in} 4 & 2 & 348.43 & 67.362 & 415.79
\end{bmatrix} \\
\begin{bmatrix}
\vspace{0.02in} 4 & 2 & 143.13 & 111.17 & 254.30
\end{bmatrix} \\
\begin{bmatrix}
\vspace{0.02in} 4 & 3 & 140.41 & 63.788 & 204.20
\end{bmatrix} \\
\begin{bmatrix}
\vspace{0.02in} 4 & 3 & 19.407 & 23.204 & 42.611
\end{bmatrix} \\
\begin{bmatrix}
\vspace{0.02in} 4 & 4 & 333.57 & 98.488 & 432.05
\end{bmatrix} \\
\begin{bmatrix}
\vspace{0.02in} 4 & 4 & 126.87 & 123.17 & 250.04
\end{bmatrix} \\
\end{aligned}
\end{align} 
Even though the shadow boundary is not required, it is given by:
\begin{align}
\begin{aligned}\label{eq:shadpath4_boundcurve}
ShadowPath(Body4)=
\begin{bmatrix}
\vspace{0.02in} 2 & 2 & 55.795 & 87.336 & 143.13
\end{bmatrix} \\
\begin{bmatrix}
\vspace{0.02in} 3 & 3 & 204.20 & 175.21 & 379.41
\end{bmatrix} \\
\begin{bmatrix}
\vspace{0.02in} 4 & 4 & 333.57 & 98.488 & 432.05
\end{bmatrix} \\
\begin{bmatrix}
\vspace{0.02in} 4 & 4 & 126.87 & 123.17 & 250.04
\end{bmatrix} \\
\end{aligned}
\end{align}
Note, for this particular example, Body 1 is completely contained within Body 4. The complement of the first arc in ${\tt XCircle}(Body1,Body4)$ is empty, and therefore does not contribute to $ShadowPath(Body4)$ from the first body. \\

We provide a {\sc FORTRAN} code to both compute and plot the arcs on the visible path and shadow path boundaries for ${\cal N}$ bodies when given their radii, the POS co-ordinates of their centers, and their ordering in terms of their distance from the observer. In addition, we provide a {\sc Matlab} code to produce this example.

\section{Exterior Derivatives}\label{sec:appendix_exderiv}

Every term in the limb darkening laws is of the form $f=f(r^2)$, a continuous function on the unit disk.
Since the 2-form $f(x,y)dxdy$ is a closed form on the unit disk, Poincar\'{e}'s Lemma asserts that there
exists a 1-form $P(x,y)dx+Q(x,y)dy$ such that 
\begin{equation}\label{eq:DPQpartial}
d\wedge\big[P,Q\big]=\frac{\partial Q}{\partial x}-\frac{\partial P}{\partial y}=f(x,y)
\end{equation}
Try $P(x,y)=-yF(r^2),~~Q(x,y)=xF(r^2)$ for some smooth function $F$ on the unit disk. Then
\begin{align}\label{eq:DPQpartial_or}
\begin{aligned}
&\frac{\partial Q}{\partial x}-\frac{\partial P}{\partial y}=F(r^2)+2x^2\frac{dF(r^2)}{dr^2}+F(r^2)+2y^2\frac{dF(r^2)}{dr^2} \\
&\rm{or}\\
&2F(r^2)+2r^2\frac{dF(r^2)}{dr^2}=f(r^2) \\
&{\rm or}\\
&\frac{2d\big(r^2F(r^2)\big)}{dr^2}=f(r^2)
\end{aligned}
\end{align}
integrating $f=f(z)$ with respect to $z$ where $z=r^2$, from $0$ to $r^2$, gives
\begin{align}\label{eq:integrateFr2}
\begin{aligned}
2r^2F(r^2)=\int\displaylimits_0^{r^2}f(z)dz, \text{ hence} \\
F(r^2)=\frac{\mathlarger{\int}\displaylimits_0^{r^2}f(z)dz}{2r^2}
\end{aligned}
\end{align}
and $P=-yF(r^2),~~Q=xF(r^2)$ is the 1-form, the exterior derivative of which is $f(r^2)$. \\

In the case of the R-M effect, all of the terms in Equation (\ref{eq:deltaRV}) (Equation 1 in \citealt{Gimenez2006b}) using any of the limb darkening laws have the 2-form $g=xf(r^2)$ or $h=yf(r^2)$ on the unit disk.
Try
\begin{align}\label{eq:P1P2Q1Q2}
\begin{aligned}
&P_1(x,y)=0 &Q_1(x,y)&=G(r^2) \\
&P_2(x,y)=G(r^2) &Q_2(x,y)&=0 \\
\end{aligned}
\end{align}
Evaluating the partial derivatives for the first case only (the second case is very similar) yields:
\begin{align} 
\begin{aligned} \nonumber
&\frac{\partial P_1}{\partial y}=0 \\
&\frac{\partial Q_1}{\partial x}=\frac{dG(r^2)}{dr^2}(2x)
=2x\frac{dG(r^2)}{dr^2}=g(x,y)=xf(r^2) \\
&{\rm or}~ \frac{dG(r^2)}{dr^2}=\frac{1}{2}f(r^2) \\
\end{aligned}
\end{align}
which we then integrate:
\begin{align} \nonumber
\begin{aligned}
&G(r^2)=\frac{1}{2}\int\displaylimits_0^{r^2}f(z)dz \\
\end{aligned}
\end{align}
Hence
\begin{align} \nonumber 
\begin{aligned}
&d\wedge\big[0,G(r^2)\big]=xf(r^2) ~\rm{and}\\
&d\wedge\big[-G(r^2),0\big]=yf(r^2)\\
\end{aligned}
\end{align}
In summary,
\begin{align}\label{eq:RMgfunc_integral}
\begin{aligned}
G(r^2)=\frac{1}{2}\int\displaylimits_0^{r^2}f(z)dz \\ 
\end{aligned}
\end{align}
where $f=f(r^2)$  is a continuous function on the unit disk. Then,
\begin{align}\label{eq:RMgfunc_summary}
\begin{aligned}
&d\wedge\bigg[-y\frac{G(r^2)}{r^2},x\frac{G(r^2)}{r^2}\bigg]=f(r^2) \\
&d\wedge\bigg[0,G(r^2)\bigg]=xf(r^2) \\
&d\wedge\bigg[-G(r^2),0\bigg]=yf(r^2) \\
\end{aligned}
\end{align} \\
We note, again, that Equations (\ref{eq:RMgfunc_summary}) explicitly gives the 1-forms $[P, Q]$ for the flux calculation for any limb darkening given by
\begin{align} \nonumber
\begin{aligned}
I(\mu)/I_0=f(r^2)=f(1-\mu^2)
\end{aligned}
\end{align}
(where $f$ is a continuous function on the disk $D$), and for the 1-forms $[P, Q]$ for the R-M effect based on that limb darkening. The specific form of the integral $G$ (Equation \ref{eq:RMgfunc_integral}) will determine if $[P,Q]$ can be expressed in closed form, by a special function, or will require numerical evaluation. \\

Consider two special cases, $\mu^{p}$, and, $\mu \log(\mu)$.
\begin{align}\label{eq:Gexpression}
\begin{aligned}
&\rm{Let} ~~f(r^2)=\big(1-r^2\big)^{\slantfrac{p}{2}}=\mu^{p}~~~(p\not=-2) \\
&G_{\rm pow}(r^2)=\frac{1}{2}\int\displaylimits_0^{r^2}(1-z)^{\frac{p}{2}} dz \\
&{\rm if} ~w=1-z,~~dw=-dz \\
&G_{\rm pow }(r^2)=\frac{1}{2}\int\displaylimits_{1-r^2}^{1}w^{\frac{p}{2}} dw=
\frac{1}{2}\frac{w^{\slantfrac{p}{2}+1}}{(\frac{p}{2}+1)}\bigg|_{1-r^2}^1=\frac{1}{p+2}\bigg[1-\big(1-r^2\big)^{\slantfrac{p+2}{2}}\bigg] \\
&~~~~~~~~~~~~=\frac{1}{p+2}\big[1-\mu^{p+2}\big]~~~(p\not=-2) \\
&\\
&\rm{Let}~~f(r^2)=\sqrt{1-r^2}\log\big(\sqrt{1-r^2}\big)=\mu\log\mu \\
&G_{\rm log}(r^2)=-\frac{1}{9}-\frac{\big(1-r^2\big)^{\slantfrac{3}{2}}}{3}\bigg[\log\big(\sqrt{1-r^2}\big)-\frac{1}{3}\bigg]\\
\end{aligned}
\end{align}
From the closed form expressions for $G$, we obtain closed form expressions for all of the 1-forms needed to efficiently apply Green's Theorem to the computation of the flux fraction and the R-M effect (Equations \ref{eq:linear_law_dot}-\ref{eq:log_law_dot}). In Appendix \ref{sec:appendix_rm} we extend this analysis to the R-M effect with differential stellar rotation. Note that Equation (\ref{eq:Gexpression}) provides the closed form of the integral $G$ (Equation \ref{eq:RMgfunc_integral}) needed to extend the computation of the flux fraction and the R-M effect to the power-2 law \citep{Hestroffer1997,Maxted2018}.

\section{Rossiter-McLaughlin Effect with Differential Stellar Rotation}\label{sec:appendix_rm}

For a star with differential rotation, the latitude variation in angular velocity will be modeled by the simple solar approximation 
\begin{equation}
\omega=\omega_e\big[1-\varepsilon \cos^2(\varphi)\big]
\end{equation}
where $\omega_e$ is the equatorial angular velocity, $\varepsilon=\big(\omega_e-\omega_{\rm{pole}}\big)/\omega_e$, and $\varphi$ is the co-latitude with respect to the axis of rotation. From \S\ref{sec:LimbRM1forms},
 the R-M effect is the product of a limb darkening law and the radial rotational velocity function. For this
 exposition, the quadratic law will be used. Thus, the integrand $F$ over the unit disk is given by
\begin{align}\label{eq:Fover_unitdisk}
\begin{aligned}
&F(x,y)=\big[1-c_1(1-\mu)-c_2(1-\mu)^2\big]\big[Ax+By\big]\omega \\
&\rm{or}\\
&F(x,y)=\big[1-(c_1+c_2)+(c_1+2c_2)\mu-c_2\mu^2\big]\big[Ax+By\big]\omega \\ \\
\end{aligned}
\end{align} 
where $A$ and $B$ are defined in Equation (\ref{eq:defAB}). The dynamical co-ordinates $(x,y,z)$ for the axis of rotation, $\Psi$, are given by
\begin{align}\label{eq:dynam_coord}
\begin{aligned}
& \Psi=\big[-B,A,\cos(\Phi_{\rm rot})\big] ~~\rm{and} \\
& \cos(\varphi)=\Psi \boldsymbol{\cdot}[x,y,\mu], ~\rm{or} \\
&\cos(\varphi)=-Bx+Ay+\mu\cos(\Phi_{\rm rot})  \\
&\Longrightarrow \omega=\omega_e\bigg[1-\varepsilon\bigg(-Bx+Ay+\mu\cos(\Phi_{\rm rot})\bigg)^2\bigg]
\end{aligned}
\end{align} \\

To compute the exterior anti-derivative of $F$, as in the previous cases of the computation of the flux fraction and the R-M effect, we will find the exterior anti-derivative for each simple term, compute the path integrals, apply the coefficients, and sum them up. The problem then becomes one of detailed bookkeeping. \\

For the product of the quadratic law and the rotational radial velocity field, the simple terms are 
\begin{align}
\begin{aligned} \nonumber
\big[x, x\mu,  x\mu^2, y, y\mu, y\mu^2\big]
\end{aligned}
\end{align}
and for the latitude effect, they are
\begin{align}
\begin{aligned}\label{eq:simplelat_terms}
\big[1, x^2, y^2, \mu^2, xy, x\mu, y\mu\big]
\end{aligned}
\end{align}
The product of these terms forms a $7\times6$ array, $\rm{Terms_{i,j}}$, shown in Table \ref{tab:termproducts}, with a corresponding $7\times6$  array of coefficients:
\begin{equation}\label{eq:coeff_DiEj}
\rm{Coeff_{i,j}}=D_iE_j
\end{equation}
where the coefficients of the simple terms of the quadratic law and the rotational radial velocity
field are given by
\begin{align}\label{eq:alphabeta_E}
\begin{aligned}
&\alpha=c_1+c_2,~\beta=c_1+2c_2 \\
&E=\big[A(1-\alpha),A\beta,-c_2A,B(1-\alpha),B\beta,-c_2B\big] \\
\end{aligned}
\end{align}
and the coefficients of the simple terms of the Latitude effect are given by
\begin{align}\label{eq:alphabeta_D}
\begin{aligned}
&D_1=1 \\
&D_2=-eB^2 \\ 
&D_3=-eA^2 \\ 
&D_4=-e\cos^2(\Phi_{\rm rot}) \\
&D_5=2eAB \\
&D_6=2eB\cos(\Phi_{\rm rot}) \\
&D_7=-2eA\cos(\Phi_{\rm rot}) \\
\end{aligned}
\end{align}
Thus, 
\begin{align}\label{eq:Fxy_Coeff_Terms}
\begin{aligned}
F(x,y)=\omega \sum_{i\&j} \rm{Coeff_{ij}} \cdot {\rm Terms_{ij}}
\end{aligned}
\end{align}
Table \ref{tab:RMwLatitude_1Forms} has the 1-forms for all of the simple terms in ${\rm Terms_{ij}}$ contained in Equation \ref{eq:Fxy_Coeff_Terms}. Note that in the case of no latitude variation, the vector $D$ consists of just
the first term, $D_1$, which then simplifies the coefficient matrix and Equation (\ref{eq:Fxy_Coeff_Terms}).

\section{Numerical Properties of the Integrands of the Path Integral}\label{sec:appendix_numproperties}

To illuminate the numerical properties of the path integral's integrand (Equation \ref{eq:limbdarkintegral_new}), we will substitute the general form (Equation \ref{eq:RMgfunc_summary}) for $\big[P,Q\big]$
into a typical term of Equation (\ref{eq:limbdarkintegral_new}). Note that Equation (\ref{eq:xiyi_coord}) defines $(x,y)$ and Equation (\ref{eq:xiyi_deriv}) defines $(x',y')$.
\begin{align}\label{eq:general_into_typical}
\begin{aligned}
&\int\displaylimits_{\varphi_1}^{\varphi_2}\big[P,Q\big] \boldsymbol{\cdot} \big[x', y'\big] d\varphi \\
=&\int\displaylimits_{\varphi_1}^{\varphi_2}\bigg[-y\frac{G(r^2)}{r^2},x\frac{G(r^2)}{r^2}\bigg]
\boldsymbol{\cdot}\big[x', y'\big]d\varphi \\
=&\int\displaylimits_{\varphi_1}^{\varphi_2}\frac{G(r^2)}{r^2}\big[-y, x\big]\boldsymbol{\cdot}
\big[x', y'\big]d\varphi \\
=&\int\displaylimits_{\varphi_1}^{\varphi_2}\frac{G(r^2)}{r^2}
det\begin{vmatrix}
x & y \\
x' & y' 
\vspace{0.04in}
\end{vmatrix}
d\varphi
\end{aligned}
\end{align}
the determinant of a $2 \times 2$ matrix is the signed area of the parallelogram defined by the row vectors.
Therefore, the integral can be written as:
\begin{align} \label{eq:signed area}
\begin{aligned}
&\int\displaylimits_{\varphi_1}^{\varphi_2}\frac{G(r^2)}{r^2}\lVert(x,y)\rVert \cdot \lVert(x',y')\rVert 
\sin\theta d\varphi
\end{aligned}
\end{align}
where $\theta$ is the signed angle from $(x,y)$ to $(x',y')$ with a right hand orientation. Consider the signed area term.
\begin{align}\label{eq:signed_area2}
\begin{aligned}
 \lVert(x',y')\rVert  & = \frac{\text{Radius of Front Body}}{\text{Radius of Back Body}} \\
 \lVert(x,y)\rVert & \leq 1
\end{aligned}
\end{align}
since the point $(x,y)$ lies in the unit disk. Thus 
\begin{align}\label{eq:signed_area3}
\begin{aligned}
\lvert\text{signed area}\rvert \leq \frac{\text{Radius of Front Body}}{\text{Radius of Back Body}}\lVert(x,y)\rVert
\end{aligned}
\end{align}
Now consider the first term.
\begin{align}\nonumber
\begin{aligned}
G(r^2)=\frac{1}{2}\int\displaylimits_0^{r^2}f(z)dz \text{~~~~~(Equation \ref{eq:RMgfunc_integral})}
\end{aligned}
\end{align}
where $f$ is a continuous function on the unit disk (limb darkening). Or:
\begin{align}\label{eq:average_of_f}
\begin{aligned}
\frac{G(r^2)}{r^2}=\frac{1}{2}\frac{\mathlarger\int\displaylimits_0^{r^2}f(z)dz}{r^2}=
\frac{1}{2}\bigg(\text{Average of } f \text{ on } \big[0,r^2\big]\bigg)
\end{aligned}
\end{align}
Thus
\begin{align}\label{eq:lim_G_1}
\begin{aligned}
\lim_{r \to 0}\frac{G(r^2)}{r^2}=\frac{1}{2}f(0)
\end{aligned}
\end{align}
Now compute the derivative of Equation (\ref{eq:RMgfunc_integral}):
\begin{align}\label{eq:dGdr}
\begin{aligned}
\frac{d}{dr}G(r^2)=rf(r^2)
\end{aligned}
\end{align}
This implies that the derivative order of $G$ is one more than that of $f$ and that 
\begin{align}\label{eq:lim_dGdr}
\begin{aligned}
\lim_{r \to 0}\frac{d}{dr}\frac{G(r^2)}{r}=f(0)
\end{aligned}
\end{align} \\

Suppose that the code traps for $r=0$ in the evaluation of the integral integrand. That is, if $\lvert r \rvert < 10^{-\mathlarger\epsilon}$ then set the integrand to zero.
What is the error induced? It is
\begin{align}\label{eq:error_induced}
\begin{aligned}
&\bigg\lvert \frac{G(r^2)}{r^2} \cdot \text{signed area} \bigg\rvert \\
\leq~ &\bigg\lvert \frac{G(r^2)}{r^2} \bigg\rvert \cdot \bigg\lvert \text{signed area} \bigg\rvert \\
\leq~ & \frac{1}{2}\bigg(\text{Average of } f \text{ on } \big[0,10^{-2\mathlarger\epsilon}\big]\bigg)
\frac{\text{Radius of Front Body}}{\text{Radius of Back Body}} 10^{-\mathlarger\epsilon}
\end{aligned}
\end{align} \\
If f is decreasing then the largest value is the left end point, which is 
\begin{align}\nonumber
\begin{aligned}
\leq \frac{1}{2}f(0) \frac{\text{Radius of Front Body}}{\text{Radius of Back Body}} 10^{-\mathlarger\epsilon}
\end{aligned}
\end{align} \\

\section{Closed Form Expressions}\label{sec:appendix_closedform}

The computation of the light loss and the R-M effect through a transit
involves the sum of integrals of simple functions (Equation
\ref{eq:limbdarkintegral_new}).  
Several of these definite integrals can be evaluated in closed form
using simple functions.  We provide the closed form expressions for these integrals,
which were found using the {\sc MATLAB}
symbolic toolbox. \\

The expression for the integral of the constant term for the flux fraction
($\mu^0$, see Table \ref{tab:LimbRM_1Forms}) is
\begin{align}\label{eq:symbolic_forms_costerm}
\begin{aligned}
&\int\displaylimits_{\rm arc}\big[P(\mu^0),Q(\mu^0)\big]\big[x',y'\big]d\varphi=
\int\displaylimits_{\rm arc}\frac{1}{2}\big(-yx'+xy'\big)d\varphi=\\
&{\frac{p}{2}}\bigg\{p(\varphi_1-\varphi_0)-\xi_Y\big[\cos(\varphi_1)-\cos(\varphi_0)\big]
+\xi_X\big[\sin(\varphi_1)-\sin(\varphi_0)\big]\bigg\}
\end{aligned}
\end {align}
with $\xi_X=\xi_{M,x}-\xi_{N,x}$ and $\xi_Y=\xi_{M,y}-\xi_{N,y}$ being
the differences in the center of mass positions on the POS (see
Equation \ref{eq:xiyi_coord}), and where $p=R_M/R_N$ is the ratio
of the front body's radius to the back body's radius.
Note that the vector $[\xi_X,\xi_Y]$ 
goes from the center of the back body to the 
center of the front body in units of the back body radius.
We define the length of this vector as
$z=\sqrt{\xi_X^2+\xi_Y^2}$.   \\

The expression for the integral of the 
Quadratic term  for the flux fraction
($\mu^2$, see Table \ref{tab:LimbRM_1Forms}) 
is written using the following constants:
\begin{align}\label{eq:symbolic_forms_constants}
\begin{aligned}
&A=\frac{p}{8}\bigg(2\xi_X^2\xi_Y+2\xi_Y^3+6p\xi_X-4\xi_Y\bigg)\\
&B=\frac{1}{4}p^2\xi_X \xi_Y  \\
&C=-\frac{p}{8}\bigg(p\xi_X^2-p\xi_Y^2\bigg) \\
&D=-\frac{p}{8}\bigg[2\xi_X^3+2\xi_X\xi_Y^2+6p^2\xi_X-4\xi_X\bigg] \\
&E=-\frac{p^2}{8}\bigg[4z^2+2p^2-4\bigg]
\end{aligned} 
\end{align}
The Quadratic term integral is then given by:
\begin{align}\label{eq:symbolic_forms_quadterm}
\begin{aligned}
&\int\displaylimits_{\rm arc}\big[P(\mu^2),Q(\mu^2)\big]\big[x',y'\big]d\varphi=
\int\displaylimits_{\rm arc}\bigg[-yx'\bigg(\frac{1}{2}-\frac{r^2}{4}   \bigg)
+xy'\bigg(\frac{1}{2}-\frac{r^2}{4}   \bigg) \bigg] d\varphi= \\
&\quad A\big[\cos(\varphi_1)-\cos(\varphi_0)\big]+B\big[\cos(2\varphi_1)-\cos(2\varphi_0)\big]+C\big[\sin(2\varphi_1)-\sin(2\varphi_0)\big]\\
&+D\big[\sin(\varphi_1)-\sin(\varphi_0)\big]+E\big(\varphi_1-\varphi_0\big)
\end{aligned}
\end {align}

For $M=N$, all of the integrals for the flux fraction
for each limb darkening law can be evaluated in 
closed form.  These laws can then be written as:
\begin{align}
\begin{aligned}
&\text{linear: }  && \bigg[\frac{\Delta\varphi}{2}, \frac{\Delta\varphi}{3}\bigg] \\
&\text{quadratic: } && \bigg[\frac{\Delta\varphi}{2}, \frac{\Delta\varphi}{3}, \frac{\Delta\varphi}{4}\bigg] \\
&\text{square root: } && \bigg[\frac{\Delta\varphi}{2}, \frac{2\Delta\varphi}{5}, \frac{\Delta\varphi}{3}\bigg] \\
&\text{logarithmic: } && \bigg[\frac{\Delta\varphi}{2}, \frac{\Delta\varphi}{3}, -\frac{\Delta\varphi}{9}\bigg] \\
&\text{Claret (4 parameter nonlinear): } && \bigg[\frac{\Delta\varphi}{2}, \frac{2\Delta\varphi}{5}, \frac{\Delta\varphi}{3},
 \frac{2\Delta\varphi}{7}, \frac{\Delta\varphi}{4}\bigg]\\
\end{aligned}
\end{align}
where $\Delta\varphi$ is the fourth element in the vector 
$Arc_{NM}$ in Equation (\ref{eq:arc_ij}). \\

When computing the R-M effect using the quadratic limb darkening
law
and no differential rotation, 
four out of the six terms result in 
exterior anti-derivative integrands which can 
be integrated in terms of simple functions, when
$M<N$. \\
  
For the $x$ term,
which is index (1,1) in Table 6, we have the following constants
\begin{align}\label{eq:RM1constants}
\begin{aligned}
&D=p\xi_X \\
&C=\frac{p^2}{12}  \\
&A=\xi_X^2 + 9C \\
&B=\frac{D}{2}
\end{aligned} 
\end{align}
The integral is then given by:
\begin{align}\label{eq:symbolic_forms_RM1}
\begin{aligned}
&\int\displaylimits_{\rm arc}\big[P_{\rm RM}(x),Q_{\rm RM}(x)\big]\big[x',y'\big]d\varphi =
\int\displaylimits_{\rm arc}\left(\frac{x^2y'}{2}   \right)     d\varphi= \\
& \frac{p}{2}
\bigg\{A\big[\sin(\varphi_1)-\sin(\varphi_0)\big]
+B\big[\sin(2\varphi_1)-\sin(2\varphi_0)\big] + C\big[\sin(3\varphi_1)
-\sin(3\varphi_0)\big] +D(\varphi_1-\varphi_0) \bigg\}
\end{aligned}
\end {align} \\

For the $\mu^2x$ term,
which is index (1,3) in Table 6, we have the following constants
\begin{align}\label{eq:RM2constants}
\begin{aligned}
&A=p^2\xi_X\xi_Y \\
&B=\frac{p}{2}\xi_Y\left[z^2+p^2 -1  \right] \\
&C=\frac{A}{3} \\
&D=\frac{1}{2}\left[-\xi_X^4-2\xi_X^2\xi_Y^2 -\xi_Y^4
 + 2z^2 + p^2
\left(2-5\xi_X^2 - 3\xi_Y^2 -p^2      \right)-1     \right] \\
&E= \frac{p}{2}\xi_X\left[1-z^2 -p^2 \right] \\
&F=\frac{p^2}{6}(\xi_Y^2-\xi_X^2) \\
&G=2E
\end{aligned} 
\end{align}
The integral is then given by:
\begin{align}\label{eq:symbolic_forms_RM2}
\begin{aligned}
&\int\displaylimits_{\rm arc}\big[P_{\rm RM}(\mu^2x),Q_{\rm RM}(\mu^2x)\big]\big[x',y'\big]d\varphi =
\int\displaylimits_{\rm arc}\left(-\frac{\mu^4y'}{4}   \right)     d\varphi = \\
&\frac{p}{2}
\bigg\{A\big[\cos(\varphi_1)-\cos(\varphi_0)\big]
+B\big[\cos(2\varphi_1)-\cos(2\varphi_0)\big] + 
C\big[\cos(3\varphi_1)-\cos(3\varphi_0)\big] \\
& +D\big[\sin(\varphi_1)-\sin(\varphi_0)\big] 
+E\big[\sin(2\varphi_1)-\sin(2\varphi_0)\big]
+F\big[\sin(3\varphi_1)-\sin(3\varphi_0)\big] + G(\varphi_1 - \varphi_0)   \bigg\}
\end{aligned}
\end {align}  \\

For the $y$ term,
which is index (1,4) in Table 6, we have the following constants
\begin{align}\label{eq:RM3constants}
\begin{aligned}
&A=-\left(p^2+\xi_Y^2   \right) \\
&B=\frac{p^2}{3} \\
&C=-\frac{p}{2}\xi_Y  \\
&D=-2C
\end{aligned} 
\end{align}
The integral is then given by:
\begin{align}\label{eq:symbolic_forms_RM3}
\begin{aligned}
&\int\displaylimits_{\rm arc}\big[P_{\rm RM}(y),Q_{\rm RM}(y)\big]\big[x',y'\big]d\varphi
= \int\displaylimits_{\rm arc}\left(-\frac{y^2x'}{2}   \right)     d\varphi = \\
&\frac{p}{2}
\bigg\{A\big[\cos(\varphi_1)-\cos(\varphi_0)\big]
+B\big[\cos^3(\varphi_1)-\cos^3(\varphi_0)\big] + 
C\big[\sin(2\varphi_1)-\sin(2\varphi_0)\big]
+ D(\varphi_1 - \varphi_0) \bigg\}
\end{aligned}
\end {align}  \\

For the $\mu^2y$ term,
which is index (1,6) in Table 6, we have the following constants
\begin{align}\label{eq:RM4constants}
\begin{aligned}
&A=\frac{1}{4}\bigg[\xi_X^2\big(\xi_X^2+2\xi_Y^2 +  3p^2-2\big)
+\xi_Y^2\big(\xi_Y^2 +  5p^2-2\big)+p^2
\big(p^2-2   \big)+1\big] \\
&B=\frac{p}{4}\xi_X\big[z^2+p^2 - 1 \big] \\
&C=\frac{p^2}{12}  (\xi_X^2-\xi_Y^2) \\
&D=-\frac{p^2}{2}\xi_X\xi_Y \\
&E= \frac{p}{4}\xi_Y\big[z^2 +p^2 - 1\big] \\
&F=\frac{D}{3} \\
&G=-2E
\end{aligned} 
\end{align}
The integral is then given by:
\begin{align}\label{eq:symbolic_forms_RM4}
\begin{aligned}
&\int\displaylimits_{\rm arc}\big[P_{\rm RM}(\mu^2y),Q_{\rm RM}(\mu^2y)\big]\big[x',y'\big]d\varphi
=\int\displaylimits_{\rm arc}\bigg(\frac{\mu^2x'}{4} \bigg)     d\varphi = \\
&p
\bigg\{A\big[\cos(\varphi_1)-\cos(\varphi_0)\big]
+B\big[\cos(2\varphi_1)-\cos(2\varphi_0)\big] + 
C\big[\cos(3\varphi_1)-\cos(3\varphi_0)\big] \\
& +D\big[\sin(\varphi_1)-\sin(\varphi_0)\big] 
+E\big[\sin(2\varphi_1)-\sin(2\varphi_0)\big]
+F\big[\sin(3\varphi_1)-\sin(3\varphi_0)\big] + G(\varphi_1 - \varphi_0)   \bigg\}
\end{aligned}
\end {align}  \\

When $M=N$ the R-M integrals are along the boundary of $N$, 
and hence simplify greatly.  
All six of these integrals can be evaluated in closed 
form in terms of simple functions.  
The path being on the boundary of $N$ 
implies that $\mu = 0$ and hence many of these integrals are zero.  
For the $x$ term,
which is index (1,1) in Table 6, 
the integral is evaluated using:
\begin{align}\label{eq:symbolic_forms_RM1closed}
\begin{aligned}
&\int\displaylimits_{\rm arc}\big[P_{\rm RM}(x),Q_{\rm RM}(x)\big]\big[x',y'\big]d\varphi =
\int\displaylimits_{\rm arc}\bigg(\frac{x^2y'}{2}   \bigg)     d\varphi= 
\frac{1}{2}\sin\varphi_1 -\frac{1}{6}(\sin\varphi_1)^3 -
\frac{1}{2}\sin\varphi_0 + \frac{1}{6}(\sin\varphi_0)^3
\end{aligned}
\end {align} \\

For the $y$ term,
which is index (1,4) in Table 6,
the integral is evaluated using:
\begin{align}\label{eq:symbolic_forms_RM3closed}
\begin{aligned}
&\int\displaylimits_{\rm arc}\big[P_{\rm RM}(y),Q_{\rm RM}(y)\big]\big[x',y'\big]d\varphi
= \int\displaylimits_{\rm arc}\bigg(-\frac{y^2x'}{2} \bigg)     d\varphi = 
\frac{1}{6}\bigg\{\cos\varphi_1\bigg[\big(\cos\varphi_1)^2-3\bigg]
-\cos\varphi_0\bigg[(\cos\varphi_0)^2-3\bigg]  \bigg\}
\end{aligned}
\end {align}  \\
The integrals for the $\mu x$ term (index (1,2) in Table 6),
the $\mu^2x$ term (index (1,3) in Table 6),
the $\mu y$ term (index (1,5) in Table 6), and $\mu^2y$
term (index (1,6) in Table 6) all evaluate to zero.

\acknowledgments.

We thank the anonymous referee for comments and suggestions that have significantly improved the paper.
We gratefully acknowledge funding from the National Science Foundation through Award AST-1617004, and from NASA via grant NNX14AB91G. We are also deeply grateful to John Hood, Jr. for his generous support of exoplanet research at San Diego State University.


\startlongtable
\begin{deluxetable*}{lcc}
\tablefontsize{\large}
\tablenum{1}
\tablecaption{\large Flux fraction and rigid-body R-M effect limb darkening 1-Forms}\label{tab:LimbRM_1Forms}
\tablewidth{0pt}
\tablehead{}
\startdata
Flux Fraction\\
LD Law Terms & $P_{FF}$ & $Q_{FF}$ \\
\\
\hline
$\mu^0=1$                                            & $-y/2$                                & $x/2$  \\
$\mu^{\frac{1}{2}}=\big(1-r^2\big)^{\frac{1}{4}}$ & $-y\big[1-\mu^{\frac{5}{2}}\big]\big/\big(\frac{5}{2}r^2\big)$ & $x\big[1-\mu^{\frac{5}{2}}\big]\big/\big(\frac{5}{2}r^2\big)$ \\
$\mu^{1}=\sqrt{1-r^2}$ & $-y\big[1-\mu^{3}\big]\big/\big(3r^2\big)$ & $x\big[1-\mu^{3}\big]\big/\big(3r^2\big)$ \\ 
$\mu^{\frac{3}{2}}=\big(1-r^2\big)^{\frac{3}{4}}$ & $-y\big[1-\mu^{\frac{7}{2}}\big]\big/\big(\frac{7}{2}r^2\big)$ & $x\big[1-\mu^{\frac{7}{2}}\big]\big/ \big(\frac{7}{2}r^2\big)$ \\
$\mu^{2}=1-r^2$ & $-y\big[\frac{1}{2}-r^2/4\big]$ & $x\big[\frac{1}{2}-r^2/4\big]$ \\ 
$r^2$ & $-yr^2/4$ & $xr^2/4$ \\ 
$\mu \log(\mu)$ 
& $-y\bigg\{-\frac{1}{9}-\mu^3\big[\log(\mu)-\frac{1}{3}\big]/3\bigg\}\big/r^2$ 
& $x\bigg\{-\frac{1}{9}-\mu^3\big[\log(\mu)-\frac{1}{3}\big]/3\bigg\}\big/r^2$  \\
\\
\hline
R-M\\
LD Law Terms & $P_{RM}$  & $Q_{RM}$ \\
\\
\hline
$x$ & $0$ & $x^2/2$ \\
$y$ & $-y^2/2$ & $0$ \\
$x\mu^{\frac{1}{2}}$ & $0$ & $-2\mu^{\frac{5}{2}}/5 \tablenotemark{~$\dagger$}$\\
$y\mu^{\frac{1}{2}}$ & $2\mu^{\frac{5}{2}}/5$ & $0$ \\
$x\mu$ & $0$ & $-\mu^3/3$ \\
$y\mu$ & $\mu^3/3$ & $0$ \\
$x\mu^{\frac{3}{2}}$ & $0$ & $-2\mu^{\frac{7}{2}}/7$ \\
$y\mu^{\frac{3}{2}}$ & $2\mu^{\frac{7}{2}}/7$ & $0$ \\
$x\mu^2$ & $0$ & $-\mu^4/4$ \\
$y\mu^2$ & $\mu^4/4$ & $0$ \\
$xr^2$ & $0$ & $r^4/4$ \\
$yr^2$ & $-r^4/4$ & $0$ \\
$x\mu\log(\mu)$ & $0$ & $-\mu^3\big[\log(\mu)-\frac{1}{3}\big]\big/3$\\
$y\mu\log(\mu)$ & $\mu^3\big[\log(\mu)-\frac{1}{3}\big]\big/3$ & $0$ \\
\\
\enddata
\tablenotetext{^\dagger}{\large $G(r^2)=2\big[1-\mu^{\frac{5}{2}}\big]\big/5=\frac{2}{5}-2\mu^{\frac{5}{2}}/5$ 
~However, $P$ and $Q$ are determined up to a constant, hence we may set $Q=-2\mu^{\frac{5}{2}}/5$, omitting the constant terms.}
\end{deluxetable*}

\begin{deluxetable*}{lcccc}[b!]
\tablenum{2}
\tablecaption{\large Error analysis in the 4-Body Example}\label{tab:example_4body_error}
\tablewidth{0pt}
\tablehead{ & Tolerance ($T_{\rm pan}$) & Body-2 & Body-3 & Body-4 }
\startdata
Flux & 2 & 0.971859521247178 &  0.899899635157968 & 0.233698941823416 \\
& 4 & 0.971859485798051 & 0.899899574207016 & 0.233699053887767  \\
& 8 & 0.971859478886569 & 0.899899562225870 & 0.233699073758843 \\
& 16 & 0.971859477607772 & 0.899899559999681 & 0.233699077276634 \\
& 32 & 0.971859477376681 & 0.899899559596524 & 0.233699077898928 \\
& 64 & 0.971859477335384 & 0.899899559524400 & 0.233699078008974 \\
& 128 & 0.971859477328045 & 0.899899559511577 & 0.233699078028432 \\
\hline
Flux Difference\tablenotemark{a} & 4 & $-3.54\times10^{-8}$ & $-6.01\times10^{-8}$ & $1.12\times 10^{-7}$ \\
& 8 & $-6.91\times 10^{-9}$ & $-1.20\times 10^{-8}$ &	1.99$\times 10^{-8}$ \\
& 16 & $-1.28\times 10^{-9}$ & $-2.23\times 10^{-9}$ & 3.52$\times 10^{-9}$ \\
& 32 & $-2.31\times 10^{-10}$ & $-4.03\times 10^{-10}$ & 6.22$\times 10^{-10}$ \\
& 64 & $-4.13\times 10^{-11}$ & $-7.21\times 10^{-11}$ & 1.10$\times 10^{-10}$ \\
& 128 & $-7.34\times 10^{-12}$ &	$-1.28\times 10^{-11}$ & 1.95$\times 10^{-11}$ \\
\hline
Difference Ratio\tablenotemark{b} & 8 & 1.95$\times 10^{-1}$ & 1.97$\times 10^{-1}$ & 1.77$\times 10^{-1}$ \\
& 16 & 1.85$\times 10^{-1}$ & 1.86$\times 10^{-1}$ & 1.77$\times 10^{-1}$ \\
& 32 & 1.81$\times 10^{-1}$ & 1.81$\times 10^{-1}$ & 1.77$\times 10^{-1}$ \\
& 64 & 1.79$\times 10^{-1}$ & 1.79$\times 10^{-1}$ & 1.77$\times 10^{-1}$ \\
&128 & 1.78$\times 10^{-1}$ & 1.78$\times 10^{-1}$ & 1.77$\times 10^{-1}$ \\
\hline
Extrapolated Flux\tablenotemark{c} & 8 & 0.971859477212689 & 0.899899559294515 & 0.233699078041798 \\
& 16 & 0.971859477317445 & 0.899899559491640 & 0.233699078033354 \\
& 32 & 0.971859477325710 & 0.899899559507367 & 0.233699078032670 \\
& 64 & 0.971859477326398 & 0.899899559508686 & 0.233699078032615 \\
& 128 & 0.971859477326459 & 0.899899559508804 & 0.233699078032612 \\
\hline
Flux Error Estimate\tablenotemark{d} & 8 & 1.67 $\times 10^{-9}$ & 2.93 $\times 10^{-9}$ & -4.28 $\times 10^{-9}$ \\ 
& 16 & 2.90 $\times 10^{-10}$ & 5.08 $\times 10^{-10}$ &	-7.57$\times 10^{-10}$ \\
& 32 & 5.10$\times 10^{-11}$ & 8.92 $\times 10^{-11}$ & -1.34$\times 10^{-10}$ \\
& 64 & 8.99 $\times 10^{-12}$ & 1.57$\times 10^{-11}$ &	-2.36$\times 10^{-11}$ \\
& 128 & 1.59$\times 10^{-12}$ & 2.77$\times 10^{-12}$ & $-4.18 \times 10^{-12}$ \\
\enddata
\tablenotetext{a}{For each body, this is the difference in a given tolerance between the flux value for that tolerance and the flux value for the tolerance in the line above it; e.g.  the first line is 
Difference$_{T_{\rm pan}=4}$ = Flux$_{T_{\rm pan}=4}$ $-$ Flux$_{T_{\rm pan}=2}$}
\tablenotetext{b}{For each body, this is the ratio in a given tolerance between the difference value for that tolerance and the difference value for the tolerance in the line above it}
\tablenotetext{c}{This is the Aitken extrapolation method for linearly convergent sequences \citep{Atkinson1989}}
\tablenotetext{d}{The error estimate is ${\rm Flux}-{\rm Extrapolated~Flux}$, based on the asymptotic error formula
for Composite Gaussian Integration}
\end{deluxetable*}

\begin{deluxetable*}{lcccc}[b!]
\tablenum{3}
\tablecaption{\large POS Co-ordinates for the KOI-126 Syzygy}\label{tab:planeskycoord}
\tablewidth{0pt}
\tablehead{
     &   & co-ordinates at epoch 2455711.38 & &\\
Star & $x~\rm{(AU)}$ & $y~\rm{(AU)}$  & $z~\rm{(AU)}$  & Radius \rm{(AU)}
}
\startdata
Body 1 (Star B) &  -0.003241   &  -0.004790 & ~0.1428 &  0.001087 \\    
Body 2 (Star A) & -0.003654  &   -0.006437 & ~0.1211  & 0.001207  \\  
Body 3 (Star C) & ~0.001161   & ~0.001930 & ~~-0.04473 & 0.009320  \\
\enddata
\end{deluxetable*}

\begin{deluxetable*}{cccc}[htb!]
\tablefontsize{\large}
\tablenum{4}
\tablecaption{\normalsize POS Co-ordinates (4-body example)}\label{tab:planeskycoord_4body}
\tablewidth{0pt}
\tablehead{
  \#   & x $[R_\odot]$  & y $[R_\odot]$   & Radius $[R_\odot]$
}
\startdata
Body 1  &  1.5  & 0.0 &  0.4 \\    
Body 2  &  0.5  & 0.6  & 1.0  \\  
Body 3  & 1.0   & -1.0 & 1.0  \\
Body 4  & 0.6   & 0.0  & 1.5 \\ 
\enddata
\end{deluxetable*}

\begin{deluxetable}{c|c|cccccc}
\tablenum{5}
\tablecolumns{7}
\tablewidth{1.0\columnwidth} 
\tablecaption{\large Product of the Terms} \label{tab:termproducts}
\tablehead{ & & $1$ & $2$ & $3$ & $4$ & $5$ & $6$ \\
                   \hline
              ~~ & &  $x$ &  $x\mu$ &  $x\mu^2$ & $y$ & $y\mu$ & $y\mu^2$}
\startdata
$1$ & $1$         & $x$          & $\mu x$         & $\mu^2x$       & $y$              & $\mu y$       & $\mu^2y$ \\
$2$ & $x^2$     & $x^3$       & $\mu x^3$     & $\mu^2x^3 $   & $x^2y $       & $\mu x^2y$  & $\mu^2x^2y$ \\
$3$ & $y^2$     & $xy^2$     & $\mu xy^2 $   & $\mu^2xy^2 $ & $y^3 $        & $\mu y^3 $   & $\mu^2y^3 $ \\
$4$ & $\mu^2$ & $\mu^2x$ & $\mu^3x $      & $\mu^4x $      & $\mu^2y $  & $\mu^3y $    & $\mu^4y $ \\
$5$ & $xy$       & $x^2y$     & $\mu x^2y $   & $\mu^2x^2y $ & $xy^2 $      & $\mu xy^2 $  & $\mu^2xy^2 $ \\
$6$ & $x\mu$  & $\mu x^2$ & $\mu^2x^2 $  & $\mu^3x^2 $  & $\mu xy $   & $\mu^2xy $   & $\mu^3xy $ \\
$7$ & $y\mu$  & $\mu xy$  & $\mu^2 xy $    & $\mu^3 xy $   & $\mu y^2 $ & $\mu^2y^2 $ & $\mu^3y^2 $ \\
\enddata  
\end{deluxetable}

\clearpage
\startlongtable
\begin{deluxetable*}{cllcc}
\tablefontsize{\large}
\tablenum{6}
\tablecaption{\large Quad Law 1-forms for R-M effect with differential rotation}\label{tab:RMwLatitude_1Forms}
\tablewidth{0pt}
\tablehead{
Term & Index  & Redundant & $P_{RM}$ & $Q_{RM}$ \\
         & in array\tablenotemark{a}           & index\tablenotemark{b}          &        &   
}
\startdata
\hline
$x$ & $(1,1)$ & & $0$ & $x^2/2$  \\
$\mu x$ & $(1,2)$ & & $0$ & $-\mu^3/3$ \\
$\mu^2x$ & $(1,3)$ & & $0$ & $-\mu^4/4$ \\
$y$ & $(1,4)$ & & $-y^2/2$ & $0$ \\
$\mu y$ & $(1,5)$ & & $\mu^3/3$ & $0$ \\
$\mu^2y$ & $(1,6)$ & & $\mu^4/4$ & $0$ \\
$x^3$ & $(2,1)$ & & $0$ & $x^4/4$ \\
$\mu x^3$ & $(2,2)$ & & $-x^3H(x,y,\mu)$  \tablenotemark{$\dagger$} & $0$ \\
$\mu^2x^3$ & $(2,3)$ & & $0$ & $x^4\mu^2/4+x^6/12$ \\
$x^2y$ & $(2,4)$ & & $-x^2y^2/2$ & $0$ \\
$\mu x^2y$ & $(2,5)$ & & $\mu^3x^2/3$ & $0$ \\
$\mu^2x^2y$ & $(2,6)$ & & $\mu^4x^2/4$ & $0$ \\
$xy^2$ & $(3,1)$ & & $0$ & $x^2y^2/2$ \\
$\mu xy^2$ & $(3,2)$ & & $0$ & $-\mu^3y^2/3$ \\
$\mu^2xy^2$ & $(3,3)$ & & $0$ & $-\mu^4y^2/4$ \\
$y^3$ & $(3,4)$ & & $-y^4/4$ & $0$ \\
$\mu y^3$ & $(3,5)$ & & $0$ & $y^3H(y,x,\mu)$  \tablenotemark{$\dagger$}   \\
$\mu^2y^3$ & $(3,6)$ & & $-y^4\mu^2/4-y^6/12$ & $0$ \\
$\mu^2x$ & $(4,1)$ &$(1,3)$ & $0$ & $-\mu^4/4$ \\
$\mu^3x$ & $(4,2)$ & & $0$ & $-\mu^5/5$ \\
$\mu^4x$ & $(4,3)$ & & $0$ & $-\mu^6/6$ \\
$\mu^2y$ & $(4,4)$ &$(1,6)$ & $\mu^4/4$ & $0$ \\
$\mu^3y$ & $(4,5)$ & & $\mu^5/5$ & $0$ \\
$\mu^4y$ & $(4,6)$ & & $\mu^6/6$ & $0$ \\
$x^2y$ & $(5,1)$ & $(2,4)$ & $-x^2y^2/2$ & $0$ \\
$\mu x^2y$ & $(5,2)$ & $(2,5)$ & $\mu^3x^2/3$ & $0$ \\
$\mu^2x^2y$ & $(5,3)$ & $(2,6)$ & $\mu^4x^2/4$ & $0$ \\
$xy^2$ & $(5,4)$ &(3,1) & $0$ & $x^2y^2/2$ \\
$\mu xy^2$ & $(5,5)$ & $(3,2)$ & $0$ & $-\mu^3y^2/3$ \\
$\mu^2xy^2$ & $(5,6)$ & $(3,3)$ & $0$ & $-\mu^4y^2/4$ \\
$\mu x^2$ & $(6,1)$ & & $-x^2H(x,y,\mu)$  \tablenotemark{$\dagger$} & $0$ \\
$\mu^2x^2$ & $(6,2)$ & & $y^3x^2/3$ & $x^3/3-x^5/5$ \\
$\mu^3x^2$ & $(6,3)$ & & $-x^2(1-x^2)^2L(x,y,\mu)$ \tablenotemark{$\dagger\dagger$} & $0$ \\
$\mu xy$ & $(6,4)$ & & $x\mu^3/6$ & $-y\mu^3/6$ \\
$\mu^2xy$ & $(6,5)$ & & $x\mu^4/8$ & $-y\mu^4/8$ \\
$\mu^3xy$ & $(6,6)$ & & $x\mu^5/10$ & $-y\mu^5/10$ \\
$\mu xy$ & $(7,1)$ & $(6,4)$ & $x\mu^3/6$ & $-y\mu^3/6$ \\
$\mu^2xy$ & $(7,2)$ & $(6,5)$ & $x\mu^4/8$ & $-y\mu^4/8$ \\
$\mu^3xy$ & $(7,3)$ & $(6,6)$ & $x\mu^5/10$ & $-y\mu^5/10$ \\
$\mu y^2$ & $(7,4)$ & & $0$ & $y^2H(y,x,\mu)$  \tablenotemark{$\dagger$} \\
$\mu^2y^2$ & $(7,5)$ & & $y^5/5-y^3/3$ & $-x^3y^2/3$ \\
$\mu^3y^2$ & $(7,6)$ & & $0$ & $-y^2(1-y^2)^2L(y,x,\mu)$ \tablenotemark{$\dagger\dagger$} \\
\\
\enddata
\tablenotetext{a}{Index position in the $7\times6$ array of simple terms given in Table \ref{tab:termproducts}, with the matching array of constants given in Equation (\ref{eq:alphabeta_E})}
\tablenotetext{b}{for terms re-appearing in the $7\times6$ array of simple terms.}
\tablenotetext{^\dagger}{\large where 
                       $H(u,v,\mu)=\bigg[(1-u^2)\arcsin\big(\frac{v}{\sqrt{1-u^2}}\big)+v\mu\bigg]\big/2$}
\tablenotetext{^{\dagger\dagger}}{\large where 
                       $L(u,v,\mu)=\big[3\arcsin(g)+5gw-2g^3w\big]\big/8$, \large
                       $w=\mu/\sqrt{1-u^2}$, and $g=v/\sqrt{1-u^2}$}
\end{deluxetable*}


\newpage

\clearpage

\newpage

\end{document}